\documentclass[a4paper,12pt]{article}
\usepackage{jheppub}
\usepackage{amsmath,epsfig}
\usepackage{amssymb,amsfonts}
\usepackage{latexsym}
\usepackage[latin1]{inputenc}
\usepackage{slashed}
\usepackage{empheq}
\numberwithin{equation}{section}
\usepackage{cancel}
\usepackage{overpic}
\usepackage{subcaption}
\usepackage{subeqnarray}
\usepackage{xcolor}

\usepackage{longtable}
\usepackage{color}
\usepackage{multirow}
\usepackage{epstopdf}
\usepackage{caption,graphicx,color}
\usepackage{amsmath,epsfig}
\usepackage{amssymb,amsfonts}
\usepackage{latexsym}
\usepackage[latin1]{inputenc}
\usepackage{float}
\usepackage{overpic}
\usepackage{slashed}
\usepackage{xcolor}

\usepackage{graphicx}
\usepackage{longtable}

\relax
\renewcommand{\theequation}{\arabic{section}.\arabic{equation}}

\def\be{\begin{equation}}
	\def\ee{\end{equation}}

\newcommand{\de}{\partial}
\newcommand{\bear}{\begin{eqnarray}}
	\newcommand{\bea}{\begin{eqnarray}}
		\newcommand{\eear}{\end{eqnarray}}
	\newcommand{\eea}{\end{eqnarray}}

\def\bsq{\begin{subequations}}
	\def\esq{\end{subequations}}
\def\hri#1#2{\href{http://arxiv.org/abs/#1}{[ArXiv:#1]#2}}
\def\hre#1#2{\href{http://arxiv.org/abs/#1/#2}{[ArXiv:#1/#2]}}

\def\hrj#1#2{\href{https://doi.org/#1}{#2}}

\def\x{X_{\infty}}
\def\y{Y_{\infty}}
\def\v{V_{\infty}}
\def\si{S_{\infty}}
\def\ti{T_{\infty}}

\def\xx{X^{(1)}_{\infty}}
\def\ss{S^{(1)}_{\infty}}
\def\Rz{R^{(\z)}}

\newbox\pippobox

\def\II{\relax{\rm I\kern-.18em I}}

\def\tu{{\tilde{u}}}

\def\m{\mu}
\def\n{\nu}

\def\g{\gamma}

\def\pa{\partial}

\def\sp{\;\;\;,\;\;\;}

\def\f{\varphi}
\def\z{\zeta}
\def\a{\alpha}
\def\b{\beta}

\def\t{\tau}

\def\nn{\nonumber}

\preprint{\\
	\hspace*{12cm} CCTP-2026-6\\
	\hspace*{12.1cm} ITCP-2026/6\\ }

\title{The phase diagram of confining holographic theories on constant curvature manifolds in the presence of a $\theta$-angle. }

\author[a]{Ahmad Ghodsi}
\author[b,c]{, Elias Kiritsis}
\author[b]{and Francesco Nitti}
\affiliation[a]{
	Department of Physics, Faculty of Science,	Ferdowsi University of Mashhad,  Mashhad, Iran.}
\affiliation[b]{	\href{http://www.apc.univ-paris7.fr}{Universit\'e Paris Cit\' e, CNRS, Astroparticule et Cosmologie}, F-75013 Paris, France.}
\affiliation[c]{\href{http://hep.physics.uoc.gr}{Crete Center for Theoretical Physics},\\ Institute for Theoretical and Computational Physics,
	Department of Physics, \\
	University of Crete, Heraklion, Greece.}

\abstract{Large families of confining holographic QFTs, described by Einstein-Dilaton gravity, are considered on constant-curvature manifolds in the presence of a $\theta$-angle. The space of ground states of such theories is explored as a function of the UV parameters, namely the dimensionless curvature and the $\theta$ angle. The free energy is computed, and the phase structure is determined. For constant negative curvature manifolds, we find solutions dual to single QFTs as well as solutions describing interfaces.
	The single QFTs exhibit an infinite family of saddle points, with the leading one dominating the gravitational path integral and no phase transitions present.
	For constant positive curvature manifolds, like de Sitter, the ($\theta$-angle, curvature) phase diagram exhibits both first and second order phase transitions, as a function of the class of theories considered. We also show that when $\theta=0$, a holographic Vafa-Witten-like theorem can be proven.}


\begin{document}
	\maketitle
	
	\section{Introduction} \label{INT}
	The gauge/gravity duality provides a tool to study
	Large-$N$, strongly coupled quantum field theories (QFTs) in both flat and
	curved space-time. In particular, QFTs which are confining in flat space may display new features when set on a curved
	background. This situation has
	been recently discussed in a very general holographic context in
	\cite{Ghodsi:2024jxe}, which dealt with holographic QFTs on
	(Euclidean) AdS, and
	\cite{Kastikainen:2025eys}, which focused on the QFTs defined
	on QFTs on dS (or spheres, in the
	Euclidean case). In both these articles, the playground was
	Einstein-Dilaton gravity, with an appropriate dilaton potential that
	assures a mass gap in the infrared region of the dual geometry
	\cite{Gursoy:2007er}. In this type of model, when defined on a flat manifold, the corresponding dual QFT displays  confinement, a discrete spectrum of gauge-invariant excitation (``glueballs''), and a finite-temperature, first-order phase transition \cite{Gursoy:2008za}.
	
	In this paper, we add a new ingredient to the setup studied in
	\cite{Ghodsi:2024jxe, Kastikainen:2025eys}:  a bulk axion field. By
	this, we mean a bulk scalar which enjoys an exact shift symmetry (at
	least in the semi-classical gravity approximation). Such
	a field is expected to exist generically in holographic duals to gauge
	theories, as it is dual to the gauge-invariant instanton density
	operator $Tr F\tilde F$. Its boundary value corresponds to a
	$\theta$-angle.  If we limit ourselves to the gauge-sector, the
	metric-axion-dilaton system is enough to describe all the lowest-dimension gauge-invariant operators holographically.
	
	In top-down string theory, holographic
	models, it usually stems from a Ramond-Ramond form, possibly upon
	dimensional reduction (e.g., it may be the RR axion of type IIB string
	theory, or a higher form reduced to an internal cycle, as in the Witten
	D4-brane model, \cite{Witten:1998uka}).
	
	The history of axion-related studies in holography is long, starting
	with \cite{Witten:1998uka}. It is an important ingredient in holographic
	phenomenology, as the axial sector gives rise to various observables
	which may be computed on the lattice (topological susceptibility,
	CP-odd particle state masses). Moreover, at  large-$N$, axion
	fields can give rise to  discrete vacua,
	a feature that has been exploited in BSM model-building in
	\cite{relaxion} and adapted in holography in \cite{Hamada:2020bbf}.
	
	Finally, axial-sector observables are
	important from the theoretical perspective because they can serve as
	order parameters between different phases of the theory. In
	particular, it is known that the
	topological susceptibility in Yang-Mills is non-zero in the confined
	phase, but vanishes above the deconfinement temperature.  We shall be particularly interested in this later aspect of holographic axion physics: as we discuss below, the axion will provide an order
	parameter to distinguish between different phases of holographic
	confining theories when set on a positive curvature manifold.
	
	In this paper, we study holographic {\em axion RG-flows} for confining
	theories on curved space-time. The ground state of such theories is described by gravitational solutions of
	$d+1$-dimensional Einstein-axion-dilaton theory, where
	
	1) The bulk is sliced by a constant-curvature $d$-dimensional manifold, whose
	geometry can be mapped to that where the dual field theory is defined;
	
	2) Both the axion and the dilaton have a non-trivial profile in the bulk, whose full back reaction is taken into
	account.
	
	This may be seen as a non-trivial running of both  couplings:
	the relevant coupling associated with the dilaton, and the effective
	$\theta$-angle associated with the axion.
	
	Such solutions may be parameterized (up to diffeomorphisms) by the metric scale
	factor, $A(u)$, and the dilaton and axion profiles, $\f(u)$ and $a(u)$,
	\be \label{intro1}
	ds^2 = du^2 + e^{2A(u)} \zeta_{\mu\nu} dx^\mu dx^\nu \sp \f =
	\f(u) \sp a = a(u)\,,
	\ee
	where $u$ is the holographic transverse coordinate and
	$\zeta_{\mu\nu}$ is  a $u$-independent $d$-dimensional metric,  which we take to have constant curvature.
	We shall always assume that the metric asymptotes to an AdS$_{d+1}$ conformal boundary as $u\to -\infty$,
	which corresponds to  the dual field theory reaching a conformal fixed point in the UV\footnote{In the case where the slice manifold has negative constant curvature,
		there may be two AdS$_{d+1}$ conformal boundaries, the other being at $u\to +\infty$. The holographic interpretation of such solutions is that of a conformal interface.}.
	The scalar field $\varphi$ corresponds to a relevant deformation away from the fixed point.
	
	As the bulk axion is massless, it has a schematic, near-boundary expansion of the form,
	\be \label{intro1-ax}
	a(u) = a_{\textrm{UV}} + Q e^{d u/\ell} + \ldots \sp u\to -\infty\,.
	\ee
	The leading therm $a_{\textrm{UV}}$ is interpreted as the $\theta$-angle of the boundary theory
	\be
	a_{UV}=w\frac{\theta+2\pi\; k}{N_c}\sp k\in \mathbb{Z}\,,
	\ee
	where $w$ is an $O(1)$ numerical factor that depends on the precise identification of the axion, $a$, with the dual instanton density.
	
	The subleading parameter $Q$ is proportional to the vev of the topological density operator, $Q\sim \langle Tr F\tilde{F}\rangle$. We shall refer to it sometimes as the {\em axion vev} parameter.  Due to the axion shift symmetry,
	$Q=0$ implies $a(u)$ is constant throughout the bulk. Therefore,  any non-trivial solution with $a'\neq 0$ requires a non-zero axion vev parameter.
	The {\em topological susceptibility} is given by,
	\be \label{intro3}
	\chi_{top} = \frac{d Q }{ d a_{\textrm{UV}}}\,.
	\ee
	For a detailed description of these features and properties in flat space, the reader is referred to \cite{Gursoy:2007er, Gursoy:2012bt}.

	When the dual field theory is  defined  on flat space-time, i.e., when $\zeta_{\mu\nu}$ is the Minkowski metric,
	axion RG-flows in Einstein-Dilaton gravity were  studied in \cite{Ha1,Ha2}; Curved
	RG-flows with trivial axion $a(u) = 0$, were the subject of a series of
	papers which systematically analyzed these solutions in both ungapped \cite{Ghosh:2017big, Ghodsi:2022umc}
	and gapped theories \cite{Ghodsi:2024jxe,Jani,Raymond}.
	
	The goal of this paper is to extend those works in the presence of a bulk axion.  As the axion leading term closes
	to the AdS boundary is dual to the field theory $\theta$-angle,  the resulting phase diagram will be a function of two independent
	parameters: the  dimensionless curvature ${\cal R}$ and the UV value of the axion $a_{\textrm{UV}})$.
	In the present work, we study the space of solutions,  the resulting phase diagram, and phase transitions as a function of the two-dimensional parameter space $({\cal R}, a_{\textrm{UV}})$.
	
	One of the main results of our analysis is that, in the axial sector, one can find an order parameter with a clear interpretation on the QFT side (namely the {\em topological susceptibility}) which can tell apart phases with different geometric features in the bulk interior. Such an order parameter was missing in the Einstein-Dilaton system studied in earlier work.
	
	In the rest of this introduction, we briefly summarize our findings.
	
	\subsection{Summary of results}
	
	Our setup consists of Einstein-Dilaton-Axion gravity in $d+1$ dimensions, with a scalar potential such that, for a flat  $d$-dimensional metric $\zeta_{\mu\nu} =\eta_{\mu\nu}$ in (\ref{intro1}), the ground state (Poincaré-invariant) solution corresponds to a  confining dual QFT (by which we mean that the QFT has mass gap and discrete spectrum of gauge-invariant excitation).  For this, we require that the dilaton potential has an exponential asymptotic at large $\varphi$,
	\be \label{intro4}
	V(\varphi) \sim -V_\infty e^{2 b \varphi} \sp \varphi \to +\infty\,,
	\ee
	where the  exponent $b$ satisfies:
	\be \label{intro5}
	\sqrt{\frac{1}{2(d-1)}} < b <  \sqrt{\frac{d}{2(d-1)}} \,.
	\ee
	The lower bound, known as the ``confinement bound",  corresponds to requiring  IR confinement, \cite{Gursoy:2007er,Gursoy:2008za}; the upper bound (also known as the {\em Gubser bound})
	ensures the existence of ``holographically acceptable'' solutions,\cite{good}. An extended discussion can be found in  \cite{Jani}.
	
	We always consider solutions which have (at least) one asymptotic boundary region, where the geometry is close to the large-volume region of  AdS, and the scalar field approaches an extremum of $V(\varphi)$.
	
	Unlike the dilaton, the axion field cannot have a potential. Its dynamics is governed by its kinetic action, which takes the general form:
	\be \label{intro6}
	S_{axion} = \frac12 \int \sqrt{g} \,Y(\varphi) (\de a)^2 \,,
	\ee
	where $Y(\f)$ is  an {\em a priori} generic  function of the dilaton. We assume that $Y(\f)$ approaches a constant close to the  UV fixed point, and we parameterize it by an exponential at large $\varphi$, like we have done for $V(\f)$,
	\be \label{intro7}
	Y(\varphi) \simeq Y_\infty e^{\gamma \varphi} \sp \f \to +\infty\,.
	\ee
	for some real parameter $\gamma$. In this work we concentrate on a specific value of $\gamma$ which is motivated by dimensional uplift of the large--$\f$ regime: namely $\gamma = 1/b$ where $b$ is the parameter governing the  IR asymptotic of the potential, see equation  (\ref{intro4}).
	In Appendix \ref{IRdetails} we give a more general account of the IR solutions if we relax this assumption.

	\paragraph*{Classification of the solutions:} Solutions of the form (\ref{intro1}) are characterized by two dimensionless  parameters which can be identified in the near-boundary expansion, and can be thought of as UV boundary conditions:
	\begin{itemize}
		\item The dimensionless curvature parameter defined by
		\be \label{intro2}
		{\cal R} \equiv  \frac{R^{\textrm{uv}} }{ \varphi_-^{2/(d-\Delta)}}\,,
		\ee
		where $R^{\textrm{uv}}$ is the curvature of the metric on which the dual field theory is defined, and  $\varphi_-$ is the (relevant or marginally relevant) coupling that drives the theory away from the UV fixed point;   	
		\item The  UV axion source term $a_{\textrm{UV}}$  defined in \ref{intro1-ax}.
	\end{itemize}	
	
	Since the bulk axion field has no potential,  and appears in the field equations only through its derivative, for any value  $a_{\textrm{UV}}$ one can find a solution with constant axion field $a(u) = a_{\textrm{UV}}$,  which trivially extend any of the solutions of the Einstein-Dilaton system. Solutions with non-trivial axion, with $\dot{a}(u)\neq 0$, on the other hand, have a non-trivial back reaction on the metric and dilaton.
	
	It is useful to label non-trivial solutions by a dimensionless parameter $Q$, which parameterizes the axion subleading term in the near-boundary expansion (in QCD, this would be the vev of $Tr F\tilde{F}$) and which vanishes for the trivial (constant axion) solution.
	
	Some of the solutions of the Einstein-Dilaton theory, in particular those which reach $\varphi\to +\infty$,  are singular in the IR. There are various criteria one may use to decide whether such singularities may be acceptable in holography or not, for which we refer the reader to \cite{Jani}.  In the presence of a bulk axion, new criteria arise for accepting or rejecting a solution. We identify three such criteria:
	\begin{enumerate}
		
		\item{\bf Strict regularity:} The axion back reaction should not drive to infinite curvature an otherwise regular Einstein-Dilaton solution.
		
		\item {\bf Consistency with Dimensional uplift:} Einstein-Dilaton theories with an exponential dilaton potential may be understood in terms of generalized dimensional reduction \cite{GK}. In this context,  the axion field arises as the reduction of a higher-dimensional form potential over the internal manifold. For the solution to be acceptable, we require that, whenever the internal manifold shrinks to zero size, the corresponding form is regular.
		
		\item{\bf Holographic Consistency:} This criterion was  introduced in \cite{Ha1},  specifically for singular solutions.  In holographic theories, typically the subleading term in a near-boundary expansion (vev) is fixed by regularity {\em and} by the value of the leading term (source). This is because one can interpret the vev as the response to turning on a source. If, on the other hand, the vev is fixed {\em independently} of the source, this signals an inconsistency (one can interpret this as having an infinite susceptibility at vanishing source). We apply this criterion in particular to the bulk axion: we discard all singular solutions where the axion vev parameter $Q$ is fixed by the IR  independently of the UV source $\theta_{\textrm{uv}}$.

	\end{enumerate}

	Based on these criteria, we identify three kinds of acceptable  IR asymptotics. The same kinds of IR asymptotics
	for the metric and dilaton were found in \cite{Jani}, where they were labeled  Type I, II, and III, and here we follow the same labeling:
	\begin{itemize}
		\item {\bf Type I IR:} These solutions extend to the field space asymptotic region $\varphi \to +\infty$.
		The metric scale factor behaves asymptotically as in the flat-sliced solutions with an acceptable singularity.
		The solution admits no small deformation, i.e., they exist only for a fixed value of the curvature and, importantly,
		they require a vanishing axion. In particular, they only exist at $a_{\textrm{UV}}=0$.

		\item {\bf Type II IR:} These are similar to Type I, but come in continuous families: they admit a non-trivial axion
		vev (therefore a non-trivial axion profile) and the curvature parameter ${\cal R}$   can vary over a continuous range.
		As ${\cal R} \to 0$,  they connect to the flat-slicing solution.  The IR singularity is of the same type as the one encountered in the flat-slicing case.
		
		\item {\bf Type III IR:} These solutions only exist for positive curvature, and only for ${\cal R}$ above a
		certain threshold. They display a regular IR endpoint, at which the (Euclidean) solution shrinks to a point.
		The value $\varphi_0$ taken on by the dilaton at the endpoint may be used as a parameterization of these solutions.
		Regularity further demands that the axion be constant and equal to the UV value, i.e., the vev parameter $Q=0$.
		In particular, this implies that the topological susceptibility, equation (\ref{intro1-ax}), vanishes.
		This is unlike solutions with   Type II-IR, for which $\chi_{top}\neq 0$.
		
		\item{\bf Solutions with no IR endpoints:} All the above solutions have a UV boundary and an IR endpoint.
		When the slice metric $\zeta_{\mu\nu}$ has negative curvature, it is possible to find solutions with two AdS UV
		boundaries and no IR endpoints, which are similar to Janus solutions. These solutions are non-singular, and they
		correspond to interfaces between CFTs \cite{Ghodsi:2022umc}. As such, they are characterized by two sets of UV parameters
		(curvature and $\theta$-angle).
	\end{itemize}

	\paragraph*{Phase diagram at positive curvature:}
	Guided by the classification of regular, or acceptable, IR geometries described above, we proceed to explore the space of solutions as a function of the two UV parameters ${(\cal R}, \theta_{\textrm{UV}})$. When more than one solution exists at a given point in parameter space, we compare their free energies to establish which one is the leading saddle point. In this way, we build the phase diagram of the Einstein-axion-dilaton system as a function of curvature and $\theta$-angle.

\begin{figure}[htbp]
		\centering
		\includegraphics[width=.7\linewidth]{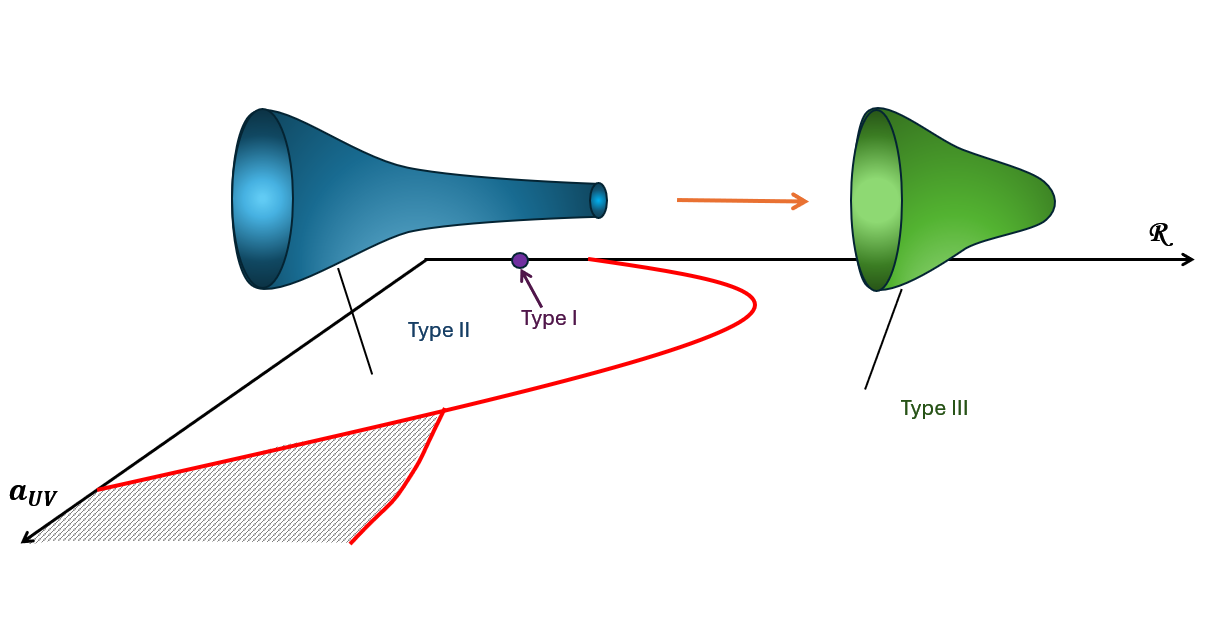}
		\caption{\footnotesize{Phase diagram with $b$ above the Efimov bound (\ref{intro-ef}) . The Schematic depiction
				of the phase structure in the $(a_{\textrm{UV}}, \mathcal{R})$ plane shows a transition from Type II to Type III
				solutions. Type I solution presents only at $a_{\textrm{UV}} = 0$ and exists prior to the transition point. No solutions exist within the shaded regions.
		}}
		\label{phase1}
	\end{figure}

\begin{figure}[htbp]
		\centering
		\includegraphics[width=.7\linewidth]{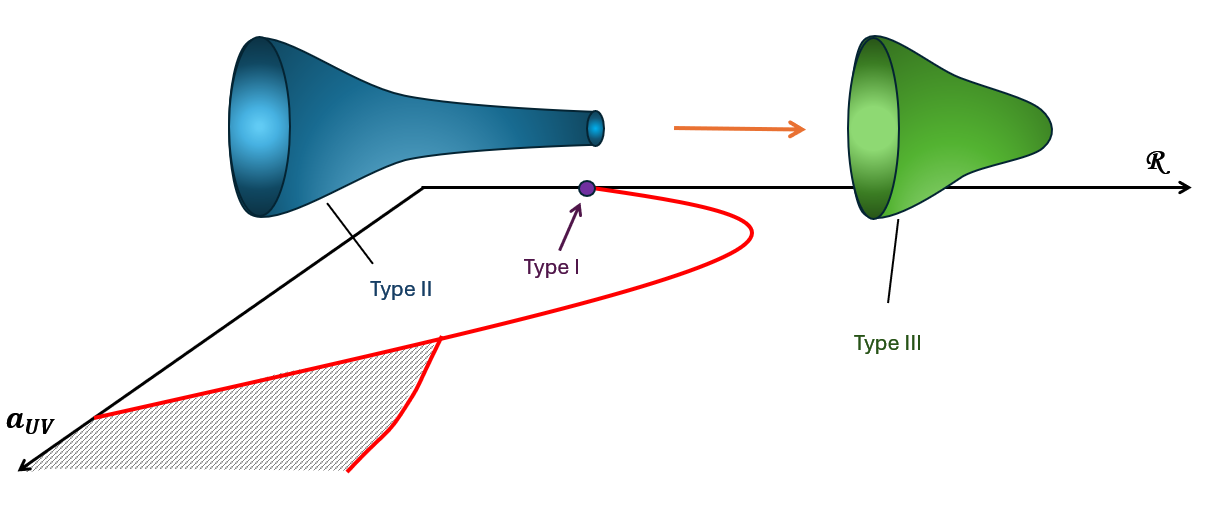}
		\caption{\footnotesize{Phase diagram below the Efimov bound  (\ref{intro-ef}). The transition line ends on the Type I solution at $a_{\textrm{UV}} = 0$, which corresponds to a higher order phase transition.
		}}
		\label{phase2}
	\end{figure}

	We find that in this two-dimensional phase space characterized by the coordinates $a_{\textrm{UV}}$ and $\mathcal{R}$, a phase transition occurs from Type II to Type III solutions.
	We identify two situations, which are illustrated schematically in figures \ref{phase1} and \ref{phase2}.
	In these figures, the red lines separate phases with  IR geometries in different classes (Type II or III).
	The transition across the red line is generically first order, except at the point corresponding to Type I, when
	it is second order.  In the shaded region of the diagram,  we could not find any acceptable solution.
	We comment on this point in the concluding section.
	\begin{enumerate}
		\item In the situation described in  Figure \ref{phase1}, the phase transition line ends, at $a_{\textrm{UV}}=0$,
		on a Type II solution (whose curvature is higher than that of the Type I solution).  This case corresponds
		to the first-order phase transitions found in \cite{Kastikainen:2025eys} at $a_{\textrm{UV}}=0$. The phase diagram
		in figure \ref{phase1} is an extension of that result into the plane with $a_{\textrm{UV}}\neq 0$.
		\item Figure \ref{phase2} corresponds to the case where the phase transition line ends, at $a_{\textrm{UV}}=0$,
		exactly on the Type I solution. In this case, the zero-axion phase transition found in \cite{Kastikainen:2025eys} was second order.
	\end{enumerate}
	Whether the phase diagram is of one or the other type is determined by the parameter $b$ governing the potential
	asymptotics in (\ref{intro4}).  The critical value that separates the two types of diagrams is the {\em Efimov bound}, defined as
	\be  \label{intro-ef}
	b_E \equiv \frac{2}{\sqrt{(d-1)(9-d)}}\,.
	\ee
	This value can be understood in terms of the higher-dimensional origin of the potentials of type (\ref{intro4}) \cite{Kastikainen:2025eys}.

	Our findings reduce to previous results in the absence of an axion, i.e., on the $a_{\textrm{UV}}=0$ axis of figures \ref{phase1} and \ref{phase2}.
	In that case,  a phase transition at positive curvature between two different bulk geometries (with  Type II and Type III IR, respectively)  was found in
	\cite{Kastikainen:2025eys}\footnote{See also \cite{Marolf:2010tg,Blackman:2011in} for a similar phenomenon in a different
		setup}.
	Although the gravity solutions are clearly different in the Einstein-dilaton theory,  there was
	no obvious order parameter on the QFT  side to clearly distinguish the two
	phases.
	
	One of the main results  of this paper is that the axial sector  provides such a  QFT order parameter:  as discussed above,  the topological susceptibility vanishes in the high-curvature phase with Type III IR)
	but is non-zero in the low-curvature phase with Type II IR. This is
	similar (and has the same geometrical origin) to the behavior of
	topological susceptibility in finite-temperature holography, where the
	topological susceptibility vanishes in the high-temperature phase
	described by a bulk black hole.
	
	We also find that for some of the confining theories, there are missing solutions, as in some cases, we have a phase transition with a discontinuous free energy. The same effect is signaled also by the fact that, as was already shown in \cite{Ha1} for the flat-sliced case, the space of axion sources is a finite interval instead of the expected real line. We have commented on these issues both in section \ref{VW}, where we show a holographic version of the Vafa-Witten theorem, \cite{vw}, as well as in the last section, \ref{last}.

	\paragraph*{Solutions at negative curvature}
	
	When the slice curvature is negative, the presence of an axion field with a non-trivial bulk profile does not significantly modify the space of solutions, whose qualitative features are similar to those found in \cite{Ghodsi:2024jxe} and that we summarize below.
	
	If the slice metric $\zeta_{\mu\nu}$ in (\ref{intro1}) has negative curvature, then the Type III IR (regular endpoint at finite  $\varphi$) cannot occur. On the other hand, it is possible that the scale factor {\em bounces}, in which case the solution connects to two UV-like boundaries \cite{Ghosh:2017big}. Therefore, there are two possible types of solutions at negative curvature:
	\begin{enumerate}
		\item {\bf UV-UV:} These are solutions with two asymptotically AdS boundaries, each one having its own boundary
		sources ${\cal R}$ and $a_{\textrm{UV}}$. The geometry corresponds to an interface between two UV CFTs with generically distinct $\theta$-angles.
		
		\item {\bf UV-IR:} These are single-boundary solutions connecting a UV  and a Type II endpoint (like the ones discussed above for positive curvature) and may have a more traditional interpretation in terms of holographic RG flows for field theories on negatively curved manifolds.  These solutions exist for any value of the curvature parameter, and the solution with no bounces connects at ${\cal R}=0$ with the flat-sliced one-boundary confining solution.
	\end{enumerate}
	
	In both cases, solutions exist that display any number of $\varphi$-bounces (points where the dilaton profile in the holographic direction turns around)  and $A$-bounces (points where the scale factor turns around). The axion field, on the other hand, is always monotonic.
	
	Due to the possibility of multiple bounces,  there is a discrete infinity of solutions with the same boundary data (curvature parameter and axion source). We compare the free energy of solutions with the same boundary data but with a different number of bounces. As in \cite{Ghodsi:2024jxe},  we find that the dominant solution is always the one with the least number of bounces (for the UV-IR solution, this has a monotonic scale factor and dilaton).
	
	We can also comment on the expectation mentioned in \cite{CW}, concerning the nature of the instanton gas of gauge theories on negative curvature manifolds.
	It was argued in \cite{CW} that in a confining gauge theory with electric boundary conditions on AdS, one would expect an instanton gas instead of an instanton liquid.
	We do find the analogue of an instanton liquid like the flat-sliced case, and this is correlated with the fact that the confining holographic theories we study seem to be of the magnetic type, \cite{Ghodsi:2024jxe}.

	\subsection*{The structure of the paper}
	
	In section 2, we develop the holographic Einstein-axion-dilaton framework in $(d+1)$ dimensions, starting from a general two-derivative bulk action with shift-symmetric axions. We derive the equations of motion for the metric, dilaton, and axion and specialize them to a holographic ansatz where all fields depend only on the radial coordinate. We establish a connection to higher-dimensional theories via dimensional reduction, showing how the lower-dimensional axion and dilaton can emerge from compactifications on curved or flat internal spaces, leading to confining or deconfining effective potentials. Finally, we recast the formalism in a first-order framework, introducing functions that simplify the equations, elucidate holographic RG flows, and allow the Ricci scalar and axion field to be expressed in terms of these first-order quantities.
	
	Section 3 analyzes the asymptotic behavior of holographic solutions near both the UV boundary and IR end-points. Near the UV, the geometry approaches an asymptotically AdS region, facilitating the identification of sources and vacuum expectation values of the dual operators. The IR expansions determine the termination of the flow and indicate whether singularities or confinement-like behavior arise. We classify the IR solutions into three types: Type I, with zero axion value and shrinking internal spheres; Type II, with a non-trivial axion and smooth shrinking of the internal $S^n$; and Type III, with a shrinking $d$-dimensional Einstein space and constant axion. Regularity of the uplifted higher-dimensional geometry constrains the IR data, particularly fixing the axion to vanish at the endpoint, providing a physical interpretation of lower-dimensional singularities through smooth higher-dimensional completions.
	
	In section 4, we present a holographic analogue of the Vafa-Witten theorem. We begin by reviewing the field theory argument and its implications for parity-odd operators. We then translate this reasoning into the holographic framework, focusing first on the axion and its dual operator, and show how IR regularity and monotonic radial evolution enforce the vanishing of the vev in the absence of sources. We subsequently generalize the argument to other massless RR forms. We discuss how the structure of the effective action ensures that the on-shell action is an even function of the sources. Finally, we comment on possible non-perturbative effects, such as D-instanton contributions, and their expected role in the holographic picture.
	
	Section 5 investigates holographic solutions with positively curved spatial slices in the presence of an axion field. Among the three IR-regular solution types, only Type II exhibits a non-constant axion. By examining the scalar potential and axion kinetic function, we study UV and IR data, including the dimensionless curvature $\mathcal{R}$, the vacuum expectation value $C$, and the UV axion value $a_{\textrm{UV}}$. The solutions display distinct behaviors above and below the Efimov bound, with oscillatory Efimov spirals appearing for $b > b_E$ and gradually disappearing for $b < b_E$. Holographic free energy analysis identifies regions in the $(\mathcal{R}, a_{\textrm{UV}})$ plane where Type II and Type III solutions coexist or dominate, revealing a first-order phase transition between these branches, characterized by a discontinuity in the free energy derivative and a latent heat, with the unique Type I solution serving as a limiting connecting case.
	
	Section 6 focuses on axionic solutions in negatively curved backgrounds, distinguishing UV-UV and IR-UV configurations. UV-UV solutions, connecting two UV fixed points, require at least one A-bounce and form a three-parameter family, potentially describing holographic interfaces or wormholes depending on slice topology. IR-UV solutions extend from a regular IR endpoint to a UV fixed point, forming a two-parameter family corresponding to single holographic theories on negatively curved manifolds. We analyze the parameter spaces, field profiles, and axion behavior for both classes, and examine holographic UV data, including sources and vevs. Free energy computations show that one-boundary solutions are smooth and dominant, whereas connected two-boundary solutions are subdominant, with the axion affecting free energy only at subleading order.
	
	We have presented the details of the computations in appendix \ref{IRdetails} to \ref{FE}.

\section{The holographic theory and its equations}\label{AEA}

Throughout this paper, we work with an  Einstein-axion-dilaton setup in a $(d+1)$-dimensional bulk space-time, characterized by the coordinates $x^A \equiv (u, x^{\mu})$. The most general two-derivative action compatible with the axionic shift symmetry can be written as
\begin{equation}\label{axact}
	S=M_p^{d-1}\int d^{d+1}x\sqrt{-g}\left[R-\frac{1}{2}g^{AB}\partial_A\varphi\partial_B\varphi-\frac{1}{2}Y(\varphi)g^{AB}\partial_A a\partial_B a-V(\varphi)\right]\,,
\end{equation}
where $M_{p}$ denotes the bulk Planck scale, and $g_{AB}$ represents the bulk metric with $R$ being its corresponding Ricci scalar. The function $V(\varphi)$ is the bulk scalar potential, while $Y(\varphi)$ is a positive function that determines the axionic kinetic term. The bulk field equations are then given by
\begin{equation}\label{axeq1}
	R_{AB}-\frac{1}{2}g_{AB}R=\frac{1}{2}\partial_{A}\varphi\partial_{B}\varphi+\frac{Y}{2}\partial_{A}a\partial_{B}a-\frac{1}{2}g_{AB}\left(\frac{1}{2}(\partial\varphi)^2+\frac{Y}{2}(\partial a)^{2}+V\right)=0\,,
\end{equation}

\begin{equation}\label{axeq2}
	\frac{1}{\sqrt{-g}}\partial_{A}\left(\sqrt{-g}g^{AB}\partial_{B}\varphi\right)-\frac{dV}{d\varphi}-\frac{1}{2}\frac{dY}{d\varphi}(\partial a)^{2}=0\,,
\end{equation}

\begin{equation}\label{axeq3}
	\partial_A\left(\sqrt{-g}\,Y g^{AB}\partial_B  a\right)=0\,.
\end{equation}

The  solution corresponding to the ground state of the holographically dual  theory on a constant curvature manifold is captured by an ansatz where the scale factor, dilaton, and axion depend only on the holographic coordinate $u$:
\begin{equation}\label{axanz}
	ds^2=du^2+e^{2A(u)}\zeta_{\mu\nu}dx^\mu dx^\nu \sp
	\varphi=\varphi(u) \sp
	a=a(u)\,.
\end{equation}
Above,  $\zeta_{\mu\nu}$ denotes the metric of a $d$-dimensional space-time with constant positive or negative curvature characterized by a length scale $\a$,  satisfying:
\begin{equation}\label{axsym}
	R_{\mu\nu}^{(\zeta)}=\kappa\zeta_{\mu\nu}\sp R^{(\zeta)}=d\kappa ,\quad\mathrm{with}\quad\kappa=\pm \frac{(d-1)}{\alpha^2} \,.
\end{equation}
In what follows, we use a shorthand notation in which differentiation with respect to $u$ is denoted by a dot, while differentiation with respect to $\varphi$ is indicated by a prime
\begin{equation}\label{axdef1}
	\dot{f}(u)\equiv\frac{df(u)}{du}
	\sp g'(\varphi)\equiv\frac{dg(\varphi)}{d\varphi} \,.
\end{equation}
The bulk field equations for the ansatz \eqref{axanz} from relations \eqref{axeq1}--\eqref{axeq3} are
\be\label{eom1}
2(d-1)\ddot{A}+\frac{2}{d}R^{(\zeta)} e^{-2A}+\dot{\varphi}^{2}+Y\dot{a}^{2}=0 \,,
\ee

\be\label{eom2}
d(d-1)\dot{A}^{2}-R^{(\zeta)} e^{-2A}-\frac{1}{2}\dot{\varphi}^{2}-\frac{1}{2}Y\dot{a}^{2}+V=0 \,,
\ee

\be\label{eom3}
\ddot{\varphi}+d\dot{A}\dot{\varphi}-V^{\prime}-\frac{1}{2}Y^{\prime}\dot{a}^{2}=0 \,,
\ee

\be \label{eom4}
\partial_{u}\left(Y e^{dA} \dot{a}\right)=0 \,.
\ee
Integrating the last equation yields
\be \label{Q}
\dot{a}=\frac{Q}{Y e^{dA}}\,,
\ee
where $Q$ is an integration constant, interpreted as the conserved axionic charge associated with the shift symmetry.

\subsection{Connection to generalized compactifications/uplift}\label{CHD}

The Einstein-axion-dilaton system described above can be understood in terms of a higher-dimensional origin \cite{GK}. Consider a $(d+n+1)$-dimensional gravitational theory coupled to a scalar field $\chi$ and an $n$-form gauge potential $C$, with  action given by:
\be
S_{d+n+1}=M_p^{d+n-1} \int d^{d+n+1} \tilde{x} \sqrt{-\tilde g} \left[\tilde R - \frac12(\pa{\chi})^2- \tilde V({\chi})-\frac{Z(\chi)}{2(n+1)!} (dC)^2\right]\,.\label{co1}
\ee
This setup provides a natural framework for dimensional reduction, where the lower-dimensional axion and dilaton fields can emerge from the compactification of the higher-dimensional metric and form fields.
To see this, we consider  a  background within this higher-dimensional theory, described by the metric ansatz
\be
ds^2=\tilde{g}_{AB} d\tilde{x}^A d\tilde{x}^B=d{\tu}^2+ e^{2A_1({\tu})}\zeta^{(1)}_{\a\b}d\tilde{x}^{\a}d\tilde{x}^{\b}+ e^{2A_2({\tu})}\zeta^{(2)}_{ij}d\tilde{x}^{i}d\tilde{x}^{j}\,,
\label{co2}
\ee
where $\tilde{x}^\a, \a=1,2,\cdots,d$ and $\tilde{x}^i, i=1,2,\cdots,n$  are the internal coordinates of the two  Einstein spaces with $\zeta^{(1)}_{ij}$ and $\zeta^{(2)}_{ij}$ metrics. Here, $\tu$ is the holographic coordinate of the higher-dimensional theory, and $A_1$ and $A_2$ act as scale factors of the two Einstein spaces, determining how the corresponding sub-spaces are scaled along the holographic direction $\tilde{u}$.
In addition, we assume that the $n$-form field and the scalar field take the following background configurations
\be \label{co3}
C_{i_1i_2\cdots i_n} = a(\tu)\sqrt{\zeta^{(2)}} \epsilon_{i_1i_2\cdots i_n} \sp \chi=\chi(\tu)\,,
\ee
where $\zeta^{(2)}$ denotes the determinant of the internal metric $\zeta^{(2)}_{ij}$ and $\epsilon_{i_1 i_2 \cdots i_n}$ is the Levi-Civita symbol on the $n$-dimensional compact space. The corresponding field strength is then given by
\be \label{co4}
(dC)_{\m i_1i_2\cdots i_n }=(n+1) \partial_{[\m}  C_{i_1i_2\cdots i_n]}  \rightarrow  (dC)_{\m 12\cdots n }=
\sqrt{\zeta^{(2)}}\partial_{\m}a(\tu) \epsilon_{12\cdots n}\,.
\ee
We now carry out a dimensional reduction of the $(d+n+1)$-dimensional action \eqref{co1} down to $(d+1)$ dimensions by treating the second manifold as a compact internal space.
To obtain the corresponding $(d+1)$-dimensional metric, we rewrite the higher-dimensional metric \eqref{co2} in the form
\be \label{co5}
ds^2=e^{-{\frac{2n}{d-1}}A_2(\tu)}ds_{d+1}^2+e^{2A_2(\tu)} \zeta^{(2)}_{ij} d\tilde{x}^{i}d\tilde{x}^{j}\,,
\ee
where we have defined
\be\label{co6}
ds_{d+1}^2\equiv e^{\frac{2n}{d-1}A_2(\tu)} d\tu^2+e^{\frac{2n}{d-1}A_2(\tu)+2A_1(\tu)} \zeta^{(1)}_{\a\b}d\tilde{x}^{\a}d\tilde{x}^{\b}\,.
\ee
By a change of variable
\be \label{co7}
\frac{d\tu}{d u}= e^{-{\frac{n}{d-1}}A_2(\tu)}\,,
\ee
we rewrite the $d+1$-dimensional metric of \eqref{co6} as
\be
ds_{d+1}^2 ={g}_{\m\n}d{x}^\m d{x}^\n=d u^2+e^{2 A({u})}\zeta^{(1)}_{\a\b}dx^{\a}dx^{\b} \sp  A\equiv A_1+\frac{n}{d-1}A_2\,.
\label{co8}
\ee
Using the above definitions, the reduction of the action   \eqref{co1} gives rise to
\begin{gather}
	S_{d+1} = M_p^{d-1} \int du d^dx \sqrt{-{g}}\Bigg[{R} +
	e^{-\frac{2  (d+n-1)}{d-1}A_2}R_2 -
	\frac{n \left(d+n-1\right)}{d-1}\pa_{u}A_2 \pa^{u}\!{A_2}
	\nn \\
	-\frac12 \pa_{u} {\chi} \pa^{u} {\chi}- e^{-\frac{2n }{d-1}A_2} V({\chi})-\frac12 e^{-2n A_2} Z(\chi) \pa_{u} {a} \pa^{u} {a}
	\Bigg]\,,\label{co9}
\end{gather}
where $R_2$ is the (constant) Ricci scalar curvature of the $\zeta^{(2)}_{ij}$ metric.
This theory, with metric ${g}_{\m\n}$ in \eqref{co8} contains one more canonically normalized scalar field $\f$ defined by
\be
\f \equiv - \sqrt{\frac{2n(d+n-1)}{d-1}}A_2\,,
\label{co10}
\ee
where the resulting action in \eqref{co9} takes the form
\begin{gather}
	S_{d+1} =  M_p^{d-1}  \int du d^dx \sqrt{-{g}}\Bigg[{R} +
	e^{\sqrt{\frac{2(d+n-1)}{n(d-1)}}\f}R_2-
	\frac12\pa_{{u}}\f \pa^{{u}}\f
	-\frac12 \pa_{{u}} {\chi} \pa^{{u}} {\chi}
	\nn \\
	- e^{\sqrt{\frac{2n}{(d-1)(d+n-1)}}\f} V({\chi})
	-\frac12 e^{ \sqrt{\frac{2n(d-1)}{d+n-1}}\f} Z(\chi) \pa_{u} {a} \pa^{u} {a}
	\Bigg]\,.\label{co11}
\end{gather}
Finally, the complete $(d+1)$-dimensional action can be expressed as
\be\label{co13}
S_{d+1}=  M_p^{d-1}  \int d^{d+1}x\sqrt{-g}\left[ R-{\frac12}(\pa\f)^2-{\frac12}(\pa\chi)^2- V(\f,\chi)
-\frac12 Y(\f,\chi) (\pa {a})^2 \right]\,,
\ee
where the total potential $V(\f,\chi)$ and axion kinetic function $Y(\f)$ are defined by
\be
V(\f,\chi) \equiv e^{\sqrt{\frac{2n}{(d-1)(d+n-1)}}\f} V(\chi)-e^{\sqrt{\frac{2(d+n-1)}{n(d-1)}}\f}R_2\,,\label{co12}
\ee
\be \label{co14}
Y(\f,\chi) \equiv  e^{ \sqrt{\frac{2n(d-1)}{d+n-1}}\f}Z(\chi)\,.
\ee

In the absence of the higher-dimensional scalar $\chi$, the pure gravity and $n$--form theory reduces to the Einstein--axion--dilaton theory with specific exponential potential functions.  These can be mapped to specific leading IR behavior of phenomenologically interesting potentials, as we discuss below.

\subsubsection{Reduction to a confining/deconfining theory}\label{RCDC}

In this subsection, we explore how the dimensional reduction of the higher dimensional theory can lead to effective lower-dimensional models exhibiting confining or deconfining behavior, depending on the geometry of the compact space. Specifically, we begin with a $(d+n+1)$-dimensional gravitational theory in \eqref{co1} characterized by a negative cosmological constant and a constant scalar field
\be {V}(\chi)=-\frac{(d+n)(d+n-1)}{\tilde{\ell}^2} \sp Z(\chi) = 1 \sp \chi=\text{const}\,.\label{co15}
\ee
By considering the compactification of the metric \eqref{co2} on manifolds with different curvature, namely an $S^n$ sphere or a $T^n$ torus, we derive two distinct classes of $(d+1)$-dimensional theories.

Considering the compactification on $S^n$ leads to an effective potential with a confining phase \cite{GK, Ghodsi:2022umc}. We assume that $A_2\to-\infty$, which implies that the scale factor of the internal $S^n$ shrinks to zero size in the $d+n+1$ dimensional theory. According to \eqref{co10}, $\varphi\to +\infty$ and therefore in this limit, the second term in \eqref{co12} is dominant. In this theory, the two functions \eqref{co12} and \eqref{co14} become
\be \label{co16}
V(\f)=-R_2 e^{2b\f} \sp Y(\f)=e^{\g \f}\,,
\ee
where
\be \label{confg}
b = \sqrt{\frac{d+n-1}{2n(d-1)}} \sp \g = \sqrt{\frac{2n(d-1)}{d+n-1}}\,, \quad \rightarrow \quad \g =\frac1b\,.
\ee
If we assume a compactification on a (flat) $T^n$ instead of $S^n$, then the second term in \eqref{co12} vanishes and we find for the parameters in \eqref{co16}
\be \label{dconfg}
b = \sqrt{\frac{n}{2(d-1)(d+n-1)}} \sp \g = \sqrt{\frac{2n(d-1)}{d+n-1}}\,, \quad \rightarrow \quad \g=2b(d-1)\,.
\ee
This corresponds to a deconfined theory,  \cite{GK, Ghodsi:2022umc}.

\subsection{The first order formalism}\label{FO}

To analyze the dynamics of the system in a framework similar to holographic renormalization group (RG) flows, it is often more convenient to reformulate the second-order Einstein equations as a set of first-order differential equations. This approach not only simplifies the structure of the equations but also provides a clearer physical interpretation of the flow of bulk fields along the holographic direction and the associated constants of integration.

To this end, we introduce the following functions of the scalar field $\varphi$ \cite{Ha1}
\be
S(\varphi)\equiv\dot{\varphi} \sp
W(\varphi)\equiv-2(d-1)\dot{A} \,,\nn
\ee
\be
T(\varphi)\equiv R^{(\zeta)}e^{-2A}\sp
X(\f)\equiv Y^2 \dot{a}^2={Q^2}{ e^{-2 d A}}\,.\label{SWTX}
\ee
Then the system of equations \eqref{eom1}--\eqref{eom3} becomes
\be \label{feom1}
S^2-SW'+\frac{2}{d}T+\frac{X}{Y}=0\,,
\ee
\be \label{feom2}
SS'-\frac{d}{2(d-1)}SW-V'-\frac12\frac{XY'}{Y^2}=0\,,
\ee
\be \label{feom3}
\frac{d}{2(d-1)}W^2-S^2-2T+2V-\frac{X}{Y}=0\,.
\ee
Furthermore, from the definitions given in \eqref{SWTX}, we find the following useful relation between the curvature term and the axionic contribution
\be \label{TX}
\left(\frac{T}{R^{(\zeta)}}\right)^d=\frac{X}{Q^2}\,.
\ee

By construction,  $X\geq 0$ and $T$ cannot change sign across a solution. Therefore, only solutions of the first-order system satisfying these constraints are also solutions of the original system of second-order equations.

The Ricci scalar of the bulk geometry can be expressed as
\begin{gather}
	R=-2d\ddot{A}-d(d+1)\dot{A}^2+e^{-2A} R^{(\zeta)}
	\nn \\
	=\frac{1}{2}\left(S^2+\frac{Q^2}{Y e^{2dA}}\right)+\frac{d+1}{d-1}V=\frac{d}{4(d-1)}W^2-T+\frac{2d}{d-1}V\,,\label{Rbulk}
\end{gather}
where we have used the definitions introduced in \eqref{SWTX}.
This expression shows that a curvature singularity at the IR end-point ($e^{A} \to 0$) can be avoided if the function $Y(\varphi)$ diverges sufficiently fast in this limit, effectively regularizing the axionic contribution to the geometry.

By solving $W$ and $T$ from equations \eqref{feom2} and \eqref{feom3}
\be
2T=\frac{2(d-1)}{d}\left(S'-\frac{V'}{S}-\frac{XY'}{2SY^2}\right)^2-S^2+2V-\frac{X}{Y}\,,
\label{1}\ee
\be
W=\frac{2(d-1)}{d}\left(S'-\frac{V'}{S}-\frac{XY'}{ 2SY^2}\right)\,,
\label{2}\ee
and inserting into the \eqref{feom1} one finds the following equation for $S$ and $X$
\begin{gather}
	S^3S''-\frac{1}{d}S^2{S'}^2+\frac{d+2}{d}\left(V'+\frac{XY'}{2Y^2}\right)SS'-\frac12 S^4-\frac{1}{d}\left(V'+\frac{XY'}{2Y^2}\right)^2
	\nn \\
	-\left(V''+\frac{V}{d-1}+
	\frac{1}{2 Y^3}\left( X' Y Y'+X Y (Y''+Y)-2 X{Y'}^2\right)\right)S^2=0\,.\label{SX}
\end{gather}
Moreover, by considering equations \eqref{TX} and \eqref{1} we obtain
\begin{gather}
	2 R^{(\z)} \left(\frac{X}{Q^2}\right)^\frac{1}{d}=\frac{2(d-1)}{d}\left(S'-\frac{V'}{S}-\frac{XY'}{2SY^2}\right)^2-S^2+2V-\frac{X}{Y}\,.\label{SXE}
\end{gather}
Using \eqref{Q} and \eqref{SWTX}, the axion field is given by
\be \label{ap}
{a'}^2=\frac{X}{Y^2 S^2}\rightarrow a(\f)= \pm \int\frac{\sqrt{X}}{|YS|} d\f
=Q\int \frac{1}{|YS|}\left(\frac{T}{R^{(\z)}}\right)^{d/2}d\f\,.
\ee
This relation will be used in the following sections to determine the value of the axion, particularly at the UV boundaries, once a solution is obtained.
Note that for a given value of the charge $Q^2$ that enters the equations, we obtain two possible signs for the axion vev as well as the axion field.

	\section{Asymptotic expansions near the UV boundary and IR end-points}\label{NBE0}
	
	We now analyze the asymptotic behavior of the holographic solutions in the vicinity of the UV boundary and the IR end-points. Such expansions provide crucial insight into the physical interpretation of the bulk fields in terms of the dual field theory quantities.
	
	Near the UV boundary, the geometry approaches an asymptotically AdS region, allowing one to identify the sources and expectation values of the dual operators through the standard holographic dictionary.
	
	In contrast, the IR expansions characterize the deep interior of the geometry, revealing whether the flow terminates smoothly, ends at a curvature singularity, or exhibits confinement-like behavior. By studying these limits, we establish the boundary conditions necessary for constructing regular and physically meaningful holographic solutions.
	
	We shall be interested in holographic theories that, in flat space,  exhibit a UV conformal fixed point and a confining behavior in the IR.
	To this end, we pick a class of scalar potentials that contain a maximum (the UV of the confining theory) and a suitable runaway behavior at $\f\to +\infty$ that will describe the IR asymptotic.

	\subsection{Expansion near the UV boundary}\label{NBE}

	We start by characterizing the UV region. We shall adjust the maximum of the potential $V(\varphi)$ to be at $\varphi=0$. Upon the expansion near this point, the bulk functions $V$ and $Y$ are expected  to behave as
	\begin{equation}\label{VYE}
		V(\f)=-\frac{d(d-1)}{\ell^2}-\frac12{m^2}\varphi^2+\mathcal{O}(\varphi^3) \sp
		Y(\f)=Y_0+Y_1\varphi+\mathcal{O}(\varphi^2) \,,
	\end{equation}
	where we parameterize the mass, $m$, in terms of the UV scaling dimension $\Delta$ of the scalar operator dual to $\f$:
	\begin{equation}\label{m2}
		m^2=\frac{(d-\Delta)\Delta}{\ell^2} \,.
	\end{equation}
	Here, $\ell$ sets the characteristic length scale of the asymptotically AdS geometry, determining the curvature radius of the UV region. The parameter $\Delta$ we restrict to the range $0 < \Delta < d/2$ to ensure that the corresponding mode is normalizable and that the scalar field represents a relevant deformation driving the RG flow away from the conformal fixed point.
	
	The UV expansion of $\f$, $A$ and $a$, can be found by solving the equations \eqref{eom1}--\eqref{eom4} near the maximum. As $u\rightarrow \pm\infty$\footnote{In the case of solutions with one boundary, the boundary is set to be at $u\to -\infty$.
		When we have solutions with two boundaries, then one is at $u\to -\infty$ and the other at $u\to +\infty$.}
	
	\be
	\varphi(u)=\varphi_{-}\ell^{\Delta}e^{\mp\Delta u/\ell}+\frac{dC\left|\varphi_{-}\right|^{(d-\Delta)/\Delta}}{\Delta(d-2\Delta)} \ell^{d-\Delta}e^{\mp(d-\Delta)u/\ell}+\ldots \,, \label{fe}
	\ee
	\begin{gather}
	A(u)=A_{-}\pm\frac{u}{\ell}-\frac{\varphi_{-}^{2} \ell^{2\Delta}}{8(d-1)}e^{\mp 2\Delta u/\ell}-\frac{\mathcal{R}|\varphi_{-}|^{2/\Delta} \ell^{2}}{4d(d-1)}e^{\mp 2u/\ell} \nn \\
	-\frac{(d-\Delta)C|\varphi_-|^{d/\Delta}\ell^d}{d(d-1)(d-2\Delta)}e^{\mp du/\ell}+\ldots \,, \label{Ae}
	\end{gather}
	\be
	a(u)=a_{\textrm{UV}}+\frac{e^{-d A_- }\ell Q}{d Y_0}e^{\mp du/\ell}+\ldots\,,\label{ae}
	\ee
	where the ellipses represent subleading terms of higher order in the expansion parameters, beginning at $\mathcal{O}(C^2, \mathcal{R}^2, Q^2)$. Notably, the axion contribution in the UV in equations \eqref{fe} and \eqref{Ae} enters only at subleading order and is thus parametrically suppressed relative to the leading behavior.
	In expansions above $\f_-$, $A_-$, $C$, $\mathcal{R}$ and $a_{\textrm{UV}}$ are constants of integration. These constants arising in the asymptotic expansions encode the independent UV data of the solution, such as the source and vev of the scalar, the axion charge, and the scale of the geometry, and they are subject to constraints ensuring regularity and consistency of the flow:
	
	\begin{itemize}
		\item $A_-$ is the leading coefficient of the scale factor near the UV boundary, setting the overall normalization of the boundary metric and affecting the holographic identification of curvature and energy scales.
		
		\item $\mathcal{R}$ is the ``dimensionless'' curvature parameter
		\be \label{mR}
		\mathcal{R} \equiv R^{(\xi)}e^{-2A_-}|\varphi_{-}|^{-2/\Delta}\,,
		\ee
		and it encodes the curvature of the boundary metric $R^{UV}$ in units set by the scalar source and the UV scale of the geometry
		\be \label{ruv}
		R^{UV}=\mathcal{R}|\varphi_-|^{{2}/{\Delta}}\,.
		\ee
		
		\item $\varphi_-$ represents the source of the operator dual to the scalar field $\varphi$, while $C = C(\mathcal{R}, a_{\textrm{UV}})$ determines its vacuum expectation value
		\be \label{vevf}
		\langle O_\f\rangle= (M_p\ell)^{d-1} \frac{d\,C}{\Delta}|\varphi_-|^{(d-\Delta)/\Delta}\,.
		\ee
		This establishes the standard holographic relation between the source and vev of a relevant operator in the UV.
		
		\item The expectation value of the operator dual to the axion $a$  is
		\be \label{veva}
		\langle O_a\rangle= \, Q \frac{(M_p\ell)^{d-1}}{N_c}\,,
		\ee
		$N_c$ is the number of colors in the dual gauge theory\footnote{In both (\ref{vevf}) and (\ref{veva}) we have assumed that the dual operator is such that it couples to the relevant bulk field, $\f$ or $a$, with a unit coupling. If the coupling is different, there is an extra multiplicative factor in the vev formulae. }.

		\item Due to the axion shift symmetry of the action \eqref{axact}, the axion source $a_{\textrm{UV}}$ is a free parameter which seems to be independent of the conserved charge $Q$.
		As we have argued in \cite{Ha1, Ha2}, the two are connected by an IR Boundary condition. The  holographic interpretation of $a_{\textrm{UV}}$, when the axion is dual to a four-dimensional instanton density,  relates it to the UV $\theta$-angle of the dual gauge theory through
		\begin{equation} \label{thuv}
			a_{\textrm{UV}}=w\,\frac{\theta_{\textrm{UV}}+2\pi k}{N_c}\,,
		\end{equation}
		where $\theta_{\textrm{UV}}\in[0,2\pi)$ and $k\in\mathbb{Z}$ labels the distinct topological sectors of the gauge theory.
		The constant $w$ in (\ref{thuv}) depends on the holographic theory, and can be computed only if a top-down description exists.
		
	\end{itemize}

	\subsection{Classification of the solutions by their IR end-points}\label{IRAS}

	We now examine the IR region of the bulk solutions, i.e., the region where $e^A\to 0$.  In curved RG-flows, the IR can be characterized by a regular endpoint with a finite value of the dilaton,  $\varphi=\varphi_0$, or by a runaway behavior $\varphi \to +\infty$ \cite{Ghosh:2017big, Jani}.
	
	In fact, finite endpoint geometries require the axion field to be constant:  suppose that the flow terminates at a finite value $\varphi=\varphi_0$.  From \eqref{Rbulk}, unless $Q=0$, regularity of the solution when  $e^A\to 0$ requires   $Y(\f)$ to diverge at that point.  This is incompatible with the fact that $\varphi_0$  appears in these solutions as an IR integration constant, which is related in the UV with the boundary curvature: the value of the dilaton at the end-point should therefore be allowed to take values in a continuous set.  Moreover,  a divergent behavior of   $Y(\f)$ at finite values is typically not found in effective theories derived from string theory.
	Thus, we conclude that a flows terminating at a finite $\varphi$ have necessarily $Q=0$, i.e., a constant axion. Conversely, solutions featuring a non-trivial axion profile with $Q\neq 0$  can only appear when $\varphi$ extends to the boundary of its field space, i.e., $\varphi\to\pm\infty$.
	In what follows, we pick positive infinity.
	
	As $\f\rightarrow +\infty$, we describe the asymptotic behavior of the bulk  potential functions using the following parameterization
	\be \label{VY}
	V(\varphi)= -V_\infty e^{2b\varphi}+\cdots \sp
	Y(\varphi)= Y_\infty e^{\g\varphi}+\cdots\,,
	\ee
	where $V_\infty$ and $Y_\infty$ are two positive numbers and
	\begin{equation}\label{bls}
		\sqrt{\frac1{2(d-1)}}<b<\sqrt{\frac{d}{2(d-1)}} \sp \gamma>0 \,.
	\end{equation}
	The first inequality ensures the existence of confining solutions with a well-behaved singularity as $\f\rightarrow  +\infty$, (see \cite{Jani} for an extended discussion)  while the second inequality guarantees that $Y$ diverges for large $\f$.
	
	Motivated by the dimensional reduction, we choose $\gamma$ according to equation  \eqref{confg}:
	\be \label{gamma-choice}
	\gamma = 1/b,
	\ee
	and this is the value we shall assume from now on. The general case is discussed for completeness in Appendix \ref{IRdetails}.
	
	To determine the IR asymptotics of the various fields, we parameterize  the first-order scalar functions by the following exponential behavior
	\be \label{SXe}
	S(\varphi) = S_\infty e^{s\varphi}+\cdots \sp
	X(\varphi) = X_\infty e^{k\varphi}+\cdots\, ,
	\ee
	and we solve the equations of motion to determine the value of the (leading and subleading) exponents and coefficients.
	
	The definition of $X$ in \eqref{SWTX} leads to the following implication
	\be \label{xpos}
	\x > 0\,.
	\ee
	A detailed analysis of the equations of motion near the IR end-point is provided in appendix \ref{IRdetails}. In the upcoming sections, we briefly outline the possible solutions.
	
	One key requirement in the analysis is that the value of $Q$, if it is non-zero, should {\em not} be completely fixed by the IR \cite{Ha1}: this is because $Q$, via the holographic dictionary, gives the vacuum expectation value of the $Tr \tilde{F}F$ in the dual theory. This should be proportional to the field theory $\theta$-angle, i.e., the axion source term, which can be freely chosen in the UV for any solution. A fixed value of  $Q$ determined from IR considerations only would then be incompatible with the field theory expectation.

	\subsubsection{Type I solution}\label{TY1}

	Within the context of the exponential expansion presented in \eqref{SXe}, a special class of solutions exists where $Q=0$, or equivalently, the axion field is constant. These  solutions are examined in Appendix \ref{apa3}, and their key findings are summarized as follows
	\be \label{t1S}
	S = \pm \sqrt{\frac{2\v}{d-1}} e^{b\f}+\cdots\,,
	\ee
	\be \label{t1W}
	W = \pm b \sqrt{8(d-1)\v}  e^{b\f}+\cdots\,,
	\ee
	\be \label{t1T}
	T = d\left(2b^2-\frac{1}{d-1}\right)\v e^{2b\f}+\cdots\,.
	\ee
	This is a fully determined solution (with no free parameters), previously presented in \cite{Ghodsi:2024jxe}. Within the scope of \eqref{bls}, this solution is exclusively valid for positive slice curvature ($T>0$).
	For our future purposes, we choose the plus sign in \eqref{t1S} or \eqref{t1W}, which as $\f\rightarrow +\infty$ then $S, W\rightarrow \infty$.
	
	Given the first-order scalar functions (\ref{t1S}-\ref{t1T}), we determine the expansions of the scalar field and the scale factor by integrating equations (\ref{SWTX}). By placing the IR end-point at some (arbitrary) position $u_0$, in the limit $u \rightarrow u_0^-$, we obtain
	\be \label{t1fi}
	\f(u)=-\frac1b \log \left(b  \sqrt{\frac{2\v}{d-1}}  (u_0-u)\right)+\cdots\,,
	\ee
	\be \label{t1A}
	A(u)=A_{IR} + \log \left(b  \sqrt{\frac{2\v}{d-1}}  (u_0-u)\right)+\cdots\,.
	\ee
	The regularity at end-point fixes the value of $A_{IR}$ by using equation \eqref{Rbulk}
	\be \label{RTI}
	e^{2A_{IR}} = \frac{R^{(\z)}}{2 b^2 d V_\infty}\,.
	\ee
	On the other hand, in this case, the axion is identically vanishing because,
	
	(a) Since in the higher-dimensional cases the internal sphere volume shrinks, one should require that $a_{IR}=0$.
	
	(b) The Type I solution requires $Q=0$, therefore $a=0$ everywhere.

	\subsubsection{Type II solutions}\label{SFQ}

	A class of regular solutions exists in which the value of $Q$ is not determined by the asymptotic values of the potentials (see appendix \ref{apa3} for further details). For these solutions, the near-IR end-point expansions of the functions $S$, $W$, $T$, and $X$  defined in (\ref{SWTX}) are given by
	\be
	S= \si e^{b\f} + \ss e^{(\frac{1}{b(d-1)}-b)\f}+\cdots\,,\label{SR}
	\ee
	\be
	W=\frac{\si}{b} e^{b\f} \frac{1}{b d \si \y}\left( (d-2) \si \y \ss
	+(d-1) \x\right) e^{(\frac{1}{b (d-1)}-b)\f}		+\cdots\,,\label{WR}
	\ee
	\be
	T=\ti e^{\frac{1}{b (d-1)}\f}+
	\frac{\Rz}{d} Q^{-\frac2d} \x^{\frac1d -1} \xx e^{-2 \left(b-\frac{1}{b(d-1)}\right)\f}+\cdots\sp
	\label{TR}
	\ee
	\be
	X = \x e^{\frac{d}{b(d-1)}\f}+
	\xx e^{\frac{-2 b^2 (d-1)+d+1}{b (d-1)}\f}+\cdots\,,\label{XR}
	\ee
	and\footnote{It should be emphasized that there are two signs for $S_\infty$, the plus or minus sign represents the two possible ways to parameterize the flow by taking $u \to -u$.}
	\be \label{SiR}
	\si =  2b \sqrt{\frac{(d-1)\v}{d-2b^2(d-1)}}\sp
	T_{\infty} = \Rz ~Q^{-\frac2d} \x^{\frac1d}\,.
	\ee
	This solution is characterized by two free parameters, $\x$ and $\ss$, with all other constants expressible in terms of these parameters. For $Q$, we obtain
	\be \label{Si1R}
	Q^{\frac{2}{d}}=-\frac{2 b^2  \y \Rz \x^{\frac1d}\si \left(d-2 b^2 (d-1)\right)}{ 4 b^2 \left(2 b^2 (d-1)+d-2\right)\v\y \ss
		+\left(d-2 b^2 (d-1)\right) \left(b^2+1\right)\si \x }\,.
	\ee
	We also find that
	\begin{gather}
		\xx =
		\frac{\x\left(\left( d-2 b^2 (d-1)\right) \si \x + 8 b^2 (d-1) \ss \v \y\right)}{4 b^2 \left(2 b^2 (d-1)-1\right) \si \v \y}
		\,.
		\label{Xi1R}
	\end{gather}
	From the definition of $\ti$ in \eqref{SiR} and by applying equation \eqref{Si1R}, we obtain the following result
	\be \label{Si1Rz}
	T_\infty = - \frac{2\left(2 b^2 (d-1)+d-2\right) }{d-2 b^2 (d-1) }\frac{\v\ss}{\si}-\frac{b^2+1}{2b^2}\frac{\x}{\y}\,.
	\ee
	By defining
	\be \label{sstar}
	S_\infty^{*} \equiv
	\frac{\si \left(2 b^2 (d-1)-d\right) \left(b^2+1\right)\x}{4 b^2 \left(2 b^2 (d-1)+d-2\right)\v\y }\,,
	\ee
	and noting that we have assumed $\x, \y, \v>0$, by choosing either sign of $S_\infty$ together with the condition $d - 2(d - 1)b^2 > 0$ from \eqref{bls}, there are two bounds that arise due to the sign of curvature
	\begin{equation} \label{smax}
		\begin{cases}
			\ss > S_\infty^{*}\sp & T_\infty < 0\,, \\[12pt]
			\ss < S_\infty^{*}\sp & T_\infty > 0\,.
		\end{cases}
	\end{equation}
	
	As $u\rightarrow u_{IR}^-$, the expansions for the scalar field, scale factor, and axion field near the IR endpoint, expressed in terms of the holographic coordinate $u$, are given by
	\begin{gather}
		\f(u) = -\frac{1}{b}\log \left(b \si (u_{IR}-u)\right)
		\nn \\
		-\frac{(d-1)b^2}{3 b^2 (d-1)-1}   \left(b\si\right)^{1-\frac{1}{b^2 (d-1)}} \ss \left(u_{IR}-u\right)^{2-\frac{1}{b^2 (d-1)}}+\cdots\,,\label{fiir}
	\end{gather}
	\begin{gather}
		A(u)=A_{IR}+\frac{1}{2 b^2 (d-1)}\log\left(u_{IR}-u\right)
		\nn \\
		+\frac{\left(b \si\right)^{2-\frac{1}{b^2 (d-1)}}}{2 d \si^2 \y \left(b^2 (d-1) \left(6 b^2 (d-1)-5\right)+1\right)}\Big(\si \y \ss \big(2 b^2 (d-1) (2 d-3)-d+2\big)
		\nn \\
		+(d-1) \x \left(3 b^2 (d-1)-1\right)\Big)( u_{IR}-u)^{2-\frac{1}{b^2 (d-1)}}+\cdots\,,\label{Air}
	\end{gather}
	\begin{gather}
		a(u) = a_{IR}\pm \frac{2b^2 (d-1)Q e^{-d A_{IR}}}{\left(2 b^2 (d-1)+d-2\right)\y}\left(b\si\right)^\frac{1}{b^2}\left(u_{IR}-u\right)^{\frac{d-2}{2 b^2 (d-1)}+1}+\cdots\,.\label{fAa}
	\end{gather}
	Using the relation \eqref{Rbulk}, one finds that the scalar curvature of the bulk geometry admits the following expansion near the IR end-point,
	\be \label{RT2}
	R = \frac{d \left(4 b^2 (d-1)- (d+1)\,\right)}{4 b^4 (d-1)^2 \left( u_{\rm IR}-u\right)^2} + \cdots \, .
	\ee
	This expression shows that the curvature diverges when approaching the IR end-point, indicating the presence of a curvature singularity in the effective lower dimensional description. However, irrespective of the curvature of the slice geometry, this singular behavior disappears upon uplifting the solution to higher dimensions:  the geometry becomes regular at $u=u_{\rm IR}$ where the internal $S^n$ smoothly shrinks to zero size (provided that the d-dimensional manifold remains finite size), resolving the singularity in the lower-dimensional theory.

	Using the last equation in \eqref{SWTX}, one finds
	\be \label{XiR}
	\x = Q^2 e^{-2d A_{IR}}\,,
	\ee
	where $A_{IR}$ represents the value of $A(u)$ at the IR endpoint in \eqref{Air}  and remains unconstrained.
	
	The expansions obtained in this section are analyzed for two specific cases: the zero axion limit and the flat limit. Detailed results can be found in appendix \ref{zl} and \ref{FL}. As shown there, there are,  in particular, two limits in which Type II solutions reduce to previously known cases:
	
	\begin{itemize}
		\item Zero axion limit: We find the Type II regular solution described in \cite{Ghodsi:2024jxe}.
		
		\item Flat limit:  One recovers the `` sub-leading solution" previously found in \cite{Ha1}.
	\end{itemize}

	\subsubsection{Type III solution}\label{TY3}

	When the slice curvature is positive, there is another type of IR endpoint which, as in the case of Type I solutions, requires a vanishing axion vev ($Q=0$).  Following the classification in \cite{Jani}, we call these solutions Type III.  They are characterized by a polynomial expansion for $S(\f)$ near an IR end-point at a (generic) final value of $\f_0$.
	
	To see how this solution arises, we expand the potential near $\f_0$ as
	\be \label{t3V}
	V = V_0 + V_1 (\f_0-\f) + V_2 (\f_0-\f)^2+\cdots\,,
	\ee
	where $V_0<0$.  The solution will exist for $\f<\f_0$ when $V_1>0$ and for $\f>\f_0$ for $V_1<0$.
	Assuming the vanishing of $S(\f)$ at  $\f=\f_0$,  one finds two possible expansions, which correspond respectively to a true IR endpoint, or to a $\f$-bounce.  The latter case is not an endpoint of the geometry, but rather a point where the parameterization in terms of scalar functions of $\f$ breaks down because $\f(u)$ fails to be monotonic: in this case one needs to glue two different solutions for $S(\f)$ and $W(\f)$ across $\f_0$.  We discuss each case separately below:
	
	\begin{itemize}
		\item IR end-point:
	\end{itemize}
	\be \label{t3S}
	S= S_0 \left(\f_0-\f\right)^\frac12 + \frac{3 (d-1) V_2+V_0}{(d-1) (d+3) S_0} \left(\f_0-\f\right)^\frac32 + \cdots\,,
	\ee
	\be \label{t3W}
	W = \frac{2V_1 (d-1)}{S_0(d+1)} \left(\f_0-\f\right)^{-\frac12}+
	\frac{ (d+4) {V_0}-(d-1) d {V_2}}{d (d+3) {S_0}} \left(\f_0-\f\right)^{\frac12}+\cdots\,,
	\ee
	\be \label{t3T}
	T =\frac{(d-1) d {V_1}}{2 (d+1) } \left(\f_0-\f\right)^{-1}+ \frac{(d+2) {V_0}+(d-1) d {V_2}}{2 (d+3)}+\cdots\,,
	\ee
	where
	\be \label{S0IR}
	S_0=  \sqrt{\frac{2V_1}{d+1}}\,.
	\ee
	Since $V_1>0$, then according to \eqref{t3T} the slice curvature should be positive to have a Type III solution. Considering the IR end-point at $\f=\f_0$ to be reached as $u\rightarrow u_0^-$, the expansions of the scalar field and scale factor are given by
	\be \label{t3fi}
	\f(u) = \f_0 - \frac{1}{2(d+1)} V_1 (u-u_0)^2+\cdots\,,
	\ee
	\be \label{t3A}
	A(u) = A_{IR} +\log \left|u-u_0\right|+\cdots\,.
	\ee
	The regularity of this solution at the IR end-point fixes the value of $A_{IR}$. Using equation \eqref{Rbulk} we find
	\be \label{RT3}
	e^{2A_{IR}} = \frac{R^{(\z)}}{d(d-1)}\,.
	\ee
	
	\begin{itemize}
		\item $\f$-bounce:
	\end{itemize}
	A $\f$-bounce refers to a situation in which the bulk scalar field $\f(u)$ reaches a turning point and reverses its direction of flow, namely
	\[
	\dot{\phi}(u_b)=0 \sp \ddot{\phi}(u_b)\neq 0 \,,
	\]
	at some finite radial position $u=u_b$. Importantly, this turning point does not correspond to a singularity of the bulk geometry, i.e., the metric functions and curvature invariants remain finite at $u_b$. A  $\f$-bounce indicates that the holographic RG flow cannot be described globally using $\f$ as a monotonic coordinate, even though the radial evolution itself remains smooth. The superpotential description becomes multi-valued across the bounce, reflecting the fact that the same value of $\f$ is encountered at two distinct radial positions. Physically, $\f$-bounces typically arise due to the competition between the scalar potential and the curvature of the slices, and are especially common for flat or negatively curved geometries. They signal a breakdown of monotonicity of the bulk solution.
	
	Near a $\f$-bounce  at $\f=\f_0$ the expansions are given by
	\be
	\label{t3bS}
	S= S_0 \left(\f_0-\f\right)^\frac12 + S_1 \left(\f_0-\f\right)+
	\frac{1}{S_0}\left(\frac{(d+9) S_1^2}{4d}+ \left(\frac{{V_0}}{d-1}+{V_2}\right)\!\right)\!
	\left(\f_0-\f\right)^\frac32 + \cdots,
	\ee
	\be \label{t3bW}
	W = \frac{3 (d-1) {S_1}}{d}+ \frac{ 9 (d-1) S_1^2+4 d {V_0}}{d^2 {S_0}} \left(\f_0-\f\right)^{\frac12}+\cdots\,,
	\ee
	\be \label{t3bT}
	T =\left(\frac{9 (d-1) S_1^2}{4 d}+{V_0}\right)+ \frac{3 {S_1}  \left(9 (d-1) S_1^2+4 d {V_0}\right)}{2 d^2 {S_0}} \left(\f_0-\f\right)^{\frac12}+\cdots\,,
	\ee
	where
	\be \label{S0b}
	S_0= \pm \sqrt{2V_1}\,,
	\ee
	and $S_1$ is a free parameter.

	\subsection{Classification from the perspective of the uplifted theory}\label{class}

	In the Euclidean framework, in the confining regime, the two Einstein spaces in \eqref{co2}, when $R_{\zeta_1}>0$, can be regarded as two spheres, $S^d$ and $S^n$. Here, $S^n$ represents the internal sphere on which the $n$-form field is defined, as specified in \eqref{co3}. Near the point where the scale factor
	$A$ shrinks to zero or $\f\to\infty$, the solutions map to the higher-dimensional solutions, and the flow is interpretable in the higher-dimensional theory. In particular, its regularity can be interpreted in the higher-dimensional theory.
	
	There are three possible ways in which the higher-dimensional spacetime can terminate at the IR endpoint: either $A_1$ diverges to $-\infty$, or $A_2$ diverges to $-\infty$, or both do so.
	
	\begin{itemize}
		\item If the internal $S^n$ shrinks to zero size while $S^d$ remains finite, or equivalently if $A_2 \rightarrow -\infty$, then according to \eqref{co10}, we have $\f \rightarrow +\infty$. Furthermore, from \eqref{co14}, it follows that $Y \rightarrow +\infty$ as well. This behavior corresponds to what we previously identified as {\bf{Type II}} solutions. Regularity of the higher-dimensional solution implies that the value of the axion at the IR endpoint is $a_{IR} = 0$.
		
		\item If the $S^d$ shrinks to zero size while $S^n$ remains finite, or equivalently, if $A_1 \rightarrow -\infty$, then according to \eqref{co8} ($A \rightarrow -\infty$), the scalar field $\f$ can take any finite value. This implies that at the shrinking point, both $V$ and $Y$ remain finite. To ensure that the curvature in the compactified theory stays finite, we must require $Q \rightarrow 0$ according to \eqref{Rbulk}. This is consistent with the {\bf{Type III}} solutions discussed above. Equation \eqref{Q} then implies that the axion remains constant everywhere and equals its IR value, i.e., $a(u) = a_{IR}$.
		
		\item When both spheres shrink simultaneously, one expects the two previous behaviors to occur together, namely, $\f \rightarrow +\infty$ and the axion remaining constant everywhere. This corresponds to what we refer to as {\bf{Type I}} solutions. Since both Type II and Type III flow toward this configuration, the constant value of the axion is expected to be zero, i.e., $a(u) = a_{IR} = 0$.
		Although the higher-dimensional solution is not singular in this limit, it is acceptable in the Gubser spirit, \cite{good}, as a continuous limit of acceptable solutions.

		\item When $R_{\zeta_1}\leq 0$, the d-dimensional Einstein space cannot shrink to zero size regularly, and therefore Type III solutions do not exist.  On the other hand,  the internal $n$-sphere can shrink smoothly to zero size while the $d$-dimensional geometry remains of finite size. From the perspective of the lower-dimensional effective theory, this behavior again corresponds to the {\bf Type II} solutions. As before, the requirement of regularity of the uplifted higher-dimensional geometry imposes a non-trivial constraint on the IR data, fixing the integration constant to $a_{\rm IR}=0$.
		\end{itemize}

	\section{A holographic Vafa-Witten-like ``theorem"\label{VW}}

	In vector-like gauge QFTs\footnote{The definition of such QFTs is that they are gauge theories where all charged degrees of freedom can acquire masses without breaking the gauge symmetry and that their Euclidean action is real. In particular, such theories must have $\theta=0$ as such parity-breaking terms are always imaginary in Euclidean space. } several properties were proven some time ago by Vafa and Witten, \cite{vw2,vw}.
	Of particular interest to us is the proof that in vector-like gauge theories, parity cannot be broken spontaneously.
	
	The proof goes as follows: If there is a parity-odd operator that can acquire an expectation value, then the ground-state energy of the theory perturbed by such an operator would have its minimum away from the parity-invariant vacuum. It then proceeds to show that vector-like theories have their minimum at a zero value of the perturbation coupling, \cite{vw}.
	
	This phenomenon has been observed in \cite{Ha1}. Indeed, it was shown for the particular operator associated to the instanton density $F\wedge \tilde F$ that couples to a massless axion, like the one we consider in this paper, that the free energy, as a function of $a_{UV}$, behaved as
	\be
	{\cal F}={\cal F}_0+\chi a_{UV}^2+{\cal O}(a_{UV}^4)
	\ee
	where  $\chi$ is known as the topological susceptibility.
	
	The same statement can be proven as follows.  All non-trivial axion solutions (i.e., not constant)
	have a non-zero vev $Q$ for the dual operator (topological density). For solutions that are IR singular (but acceptable), there is an IR regularity condition that
	sets the value of the axion to zero in the IR. This condition corresponds to an IR regularity condition for form components in higher dimensions.
	Moreover, as is obvious from the first order axion equation in (\ref{Q}), the evolution of the axion, as it evolves towards the boundary, is monotonic, and therefore if $Q\not=0$, then $a_{UV}\not=0$.
	Therefore, if $a_{UV}=0$ then $Q=0$ and the analogous holographic theorem is proven for the axion.
	
	This result can be generalized to other relevant string theory fields that belong to the ``massless" sector of string theory.
	The first candidates for these are RR forms of type II string theory. Such fields transform non-trivially under the $(-1)^{F_L}$ symmetry of type II string theory, which acts as left-moving space-time fermion number.
	This acts as a minus on R states on the left side of the string and  plus on NS states on the left side of the string.
	In particular, this implies that fields that transform non-trivially under such a symmetry appear in the effective action in pairs (like the massless RR forms).
	This goes together with the fact that no perturbative string state is minimally coupled to the RR forms.
	
	Any such RR $n$-form gauge field, on a compact $n$-dimensional internal manifold, gives rise to an axion in lower dimensions. This axion must vanish for regularity at the tip of the holographic geometry.
	Moreover, its radial evolution is always monotonic due to the positivity of the associated kinetic term. There are two facts that imply the result: if the source is near the boundary, then the vev is always zero.
	
	One can also provide a proof that is closer to the QFT proof in \cite{vw}.  Because RR forms do not minimally couple to other fields of supergravity, their equations are ``massless equations," and their kinetic terms are quadratic in the first derivatives and therefore the holographic vevs. This argument can be extended to higher $\alpha'$ corrections, as such fields always appear via their field strengths and are also even in the field strengths. Then, the on-shell action is an even function of the vevs. Now the vev is an odd function of the source, and this implies that the on-shell action is an even function of the source, which proves the theorem.
	
	It is important to stress that the only terms where such forms can appear with a field strength are the string theory (and supergravity) Chern-Simons terms. These are also even under $(-1)^F$ as they always contain an NSNS antisymmetric term. This can be immediately seen by obtaining the action by dimensional reduction of the eleven-dimensional Chern-Simons term of the three-form. Under such a reduction, one of the forms must contain an 11-th index, and this automatically makes it the NSNS two-form.
	
	There is a final issue that one would need to take into account in order to consider all possible competing solutions. This will be non-perturbative from the point of view of the effective supergravity theory. It involved the inclusion of appropriate D-branes in the bulk. A given $n$-form field couples naturally to a $D_{n-1}$ brane. Wrapping a number of such branes on the appropriate compactification manifold will provide a gas of D-instantons in the lower dimension. Such D-istantons can affect the bulk solution. We expect that when the axion values are positive in the IR, positive D-instanton charge will materialize in the bulk, while when the axion is negative, it will trigger positive D-instanton charge. The homogeneity of the ansatz implies that this charge will distribute uniformly at a single radial position and will affect the bulk solution via Israel matching conditions. Because of the correlation mentioned above, the presence of the charge will retain the monotonicity of the axion evolution that was instrumental for the Vafa-Witten theorem.
	
	We have not considered such instanton effects in this paper, but it is an interesting complement to the present analysis.

	\section{Axionic solutions at positive curvature}

	Having given a general classification of the UV and IR regimes, we now pick a specific theory and conduct a numerical investigation of the space of solutions. We start in this section with the case of positive curvature slices.
	
	A significant feature of holographic confining quantum field theories on positively curved space-times is the emergence of discrete scaling behavior, known as Efimov oscillations \cite{Aharony, Edwan, Jani, Raymond}. These oscillations arise in the bulk solutions to Einstein-scalar equations when the scalar potential follows an exponential asymptotic form, leading to distinct infrared behaviors.
	In the presence of an appropriately tuned scalar potential, the dual gravitational theory exhibits oscillatory solutions that influence the phase transition structure of the QFT. The oscillatory regime is found to be present when $b$ is what we call the Efimov range
	\be \label{Ef}
	\sqrt{\frac{1}{2(d-1)}} <b<  b_E \sp b_E \equiv \frac{2}{\sqrt{(d-1)(9-d)}}\,.
	\ee
	We call $b_E$ the Efimov bound.
	In this regime, variations in space-time curvature produce a sequence of oscillatory transitions, affecting the free energy and structure of the quantum phase transition \cite{Jani}.
	
	In this section, we are going to examine the same transition in the presence of the axion field. The non-constant axion enters only in Type II solutions. For Type I and Type III, the axion is constant, as discussed in Section  3.
	
	To establish a numerical framework for analyzing the regular solutions, we consider a concrete theory defined by a specific confining potential. We start by examining the action
	\begin{equation}\label{actN}
		S=M_p^{d-1}\int d^{d+1}x\sqrt{-g}\left[R-\frac{1}{2}g^{AB}\partial_A\varphi\partial_B\varphi-\frac{1}{2}Y(\varphi)g^{AB}\partial_A a\partial_B a-V(\varphi)\right]\,,
	\end{equation}
	where the scalar potential is defined as
	\be\label{SVN}
	V(\f)=-\frac{d(d-1)}{\ell^2}
	\left(k \f^2+\cosh^2(b\f)\right) \sp   k=\frac{\Delta(d-\Delta)}{2d(d-1)}-b^2 \,,
	\ee
	so that, when $\f\rightarrow \pm \infty$ the potential diverges as
	\be \label{ASVN}
	V(\varphi)\to-\frac{d(d-1)}{4\ell^2}e^{\pm2b\varphi}\equiv - V_\infty e^{\pm2b\varphi}\,.
	\ee
	
	We assume the kinetic term of the axion is controlled by the following function
	\be \label{YN}
	Y(\f) = 2\y \cosh{\frac{\f}{b}}\,,
	\ee
	which is motivated by our uplift to higher dimensions.
	
	To solve the equations numerically, we have assigned fixed numerical values to the parameters as follows
	\be \label{Nums}
	(\ell = 1, d=4) \rightarrow \v = 3 \sp \y = 1 \sp
	\Delta = \frac32\,.
	\ee
	
	To analyze the solutions of the theory, we shall consider two distinct cases, one above and one below the Efimov bound in (\ref{Ef}).
	For an exponent \eqref{confg} that emerges from dimensional reduction, the condition $b>b_E$ in order to be in the Efimov regime translates into $d + n < 9$ in the higher-dimensional theory.
	In $d=4$ that we shall be numerically working in this paper,  the bound \eqref{Ef} is $b_E=\frac{2}{\sqrt{15}}\approx 0.51$. In the following sections, we shall consider several values for $b$, one above and several below the Efimov bound, i.e.
	\be \label{bv}
	b(n=4) = \sqrt{\frac{7}{24}}\approx 0.54 \sp b(n=5)=\frac{2}{\sqrt{15}}\simeq 0.51\;,
	\ee
	\be
	b(n=6) = 0.5\sp b(n=9)=\frac{\sqrt{2}}{3}\simeq 0.47\,.
	\ee
	
\begin{figure}[htbp]
		\centering
		\includegraphics[width=0.5\linewidth]{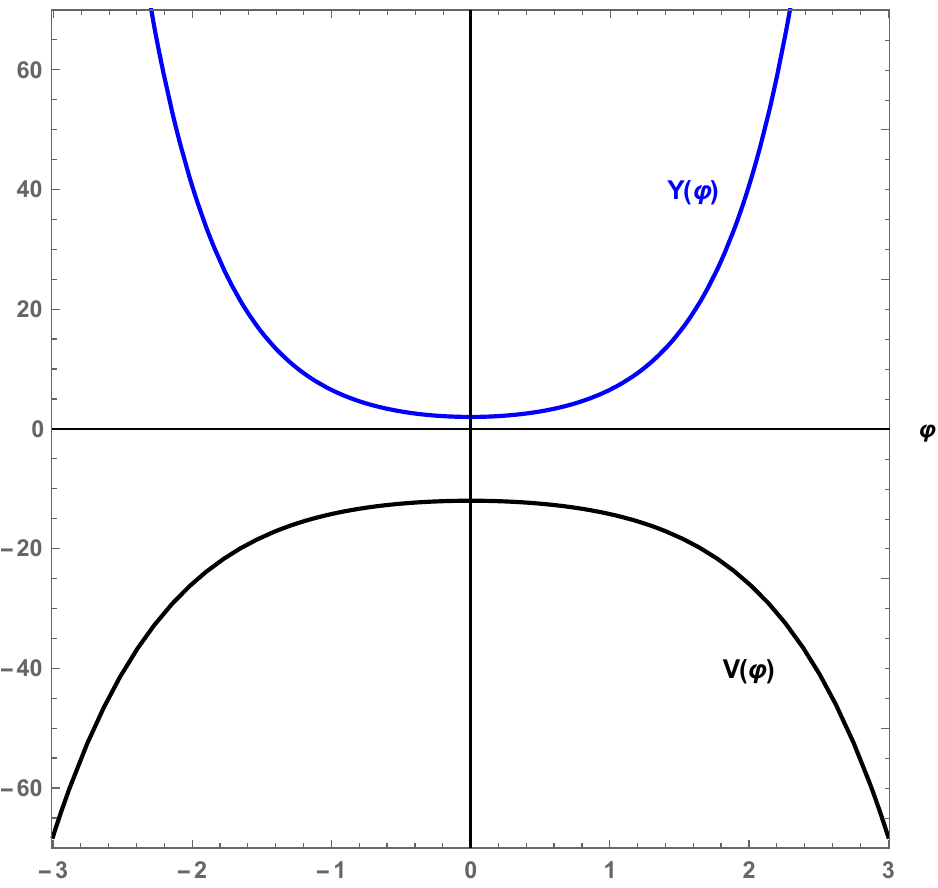}
		\caption{\footnotesize{The scalar potential \eqref{SVN} and axion function \eqref{YN} are evaluated using the numerical values specified in \eqref{Nums}.}}
		\label{pots}
	\end{figure}

	The scalar potential $V(\f)$ and axion function $Y(\f)$ are illustrated in figure \ref{pots} for $b = \sqrt{\frac{7}{24}}$ and the values given in \eqref{Nums}.
	
	\subsection{The structure of the solutions}\label{pcs}
	
	We assume a metric with positive constant curvature slices, and a non-constant scalar and axion field
	\be \label{metN}
	ds^2=du^2+e^{2A(u)}\zeta_{\mu\nu}dx^\mu dx^\nu\sp\varphi=\varphi(u)\sp a=a(u)\,.
	\ee
	As discussed in section \ref{IRAS}, there exist three types of solutions that originate from an IR end-point and terminate at a single boundary in the UV at $\f=0$. The value of the axion field in the IR is further constrained if we require the solution to uplift to higher dimensions in the way described in Section \ref{class}.
	
	\begin{itemize}
		\item Type I: These solutions feature a constant axion, with the IR end-point positioned at $\f \to \infty$  and $a(u)=a_{IR}=0$.
		
		\item Type II: These solutions have their IR endpoint at infinity in field space.  They may feature a non-constant axion field $a(u)$,  which is constrained to vanish at the IR endpoint.
		
		\item Type III: These solutions terminate at a finite  $\f=\f_0$. They also feature a constant axion field, but unlike for Type I, its value is not required to vanish at the IR endpoint and is therefore unconstrained.
		
	\end{itemize}
	
	Figure \ref{all} shows the different solutions (in terms of $W(\f)$) when we fixed the scalar potential as in (\ref{SVN}),  with $b=\sqrt{\frac{7}{24}}$. The properties of the solutions in figure \ref{all} are as follows:
	
	\begin{figure}[htbp]
		\centering
		\includegraphics[width=0.5\textwidth, height=0.5\textwidth]{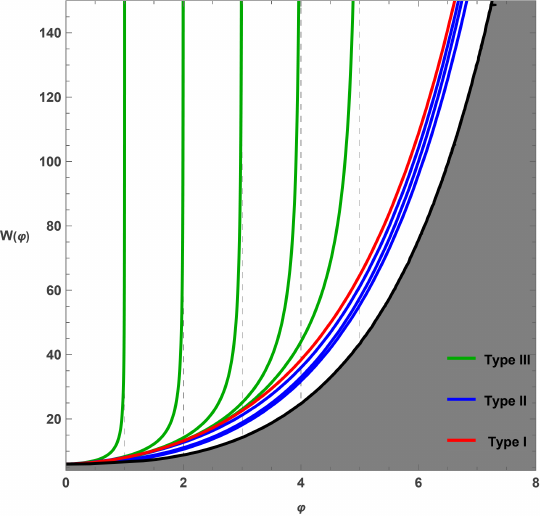}
		\caption{\footnotesize{The UV-IR Type I, II, and III solutions correspond to positive curvature slices, with $b = \sqrt{7/24}\simeq 0.54$ chosen for numerical analysis. The gray region is defined by \eqref{greg}. Among these, the Type I solution emerges as the limiting case of both Type II and Type III solutions.
		}}
		\label{all}
	\end{figure}

	\begin{enumerate}
		
		\item According to \eqref{feom3}, for positive curvature slices where $T>0$, there exists a bounded region that the value of $W$ can attain
		\be \label{greg}
		W \geq \left(-\frac{4(d-1)}{d}V\right)^\frac12\,.
		\ee
		This is illustrated by the gray-colored region in figure \ref{all}.
		
		\item Type II solutions are labeled by two IR parameters, $X_\infty$ and $S_\infty^{(1)}$, which appear in equations \eqref{SR}--\eqref{XR}. These solutions extend to infinity at the IR end-point on one side, while on the other side, they approach the UV at $\f=0$. Examples of such solutions are shown in blue curves in figure \ref{all}. There is always an upper bound, $S_\infty^{*}$, given in \eqref{sstar} for $\ss$ according to \eqref{smax}.
		
		\item Type III solutions are characterized by their IR end-point occurring at a finite value of $\f$, see the expansion in \eqref{t3S}--\eqref{t3T}. Some examples of these solutions are represented in green curves. The dashed lines indicate the IR end-points at $\f=\f_0$.

		\item The Type I solution is distinguished by its IR end-point at infinity. It is a unique solution. It serves as the limiting case of Type II as $\x \rightarrow 0$, while simultaneously acting as a limiting case of Type III as the IR end-point extends to infinity ($\f_0\rightarrow +\infty$). This solution is represented by the red curve in figure \ref{all}.
		
	\end{enumerate}

	\subsection{The QFT parameters}\label{qftp}

	Given the three types of numerical solutions, the UV parameters, the dimensionless curvature $\mathcal{R}$, and the dimensionless vacuum expectation value $C$, as defined in \eqref{mR} and \eqref{vevf}, can be determined as functions of the IR parameters. Additionally, in Type II solutions with a non-zero axion, a further dimensionless parameter emerges: the axion's value at the UV, denoted by $a_{\textrm{UV}}$.
	It is worth emphasizing that for Type II, the relevant IR parameters are $X_\infty$ and $\ss$, whereas for Type III, the IR behavior is governed by the end-point $\f_0$. Type I, on the other hand, corresponds to a unique solution with no free parameters.

	\subsubsection{ Above the Efimov bound}\label{aeb}

	The behavior of the QFT parameters can be understood from  figures \ref{rall} to \ref{as1}, which can be summarized  as follows:
	
	\begin{enumerate}
		
		\item For Type II solutions, we have portrayed four different cases with fixed $X_\infty = 0, 0.01, 0.1, 1$. As mentioned in \eqref{smax}, there exists an upper bound for $\ss$ or a lower bound for $-\ss$.
		Smaller values of $X_\infty$ give rise to smaller lower bounds, as shown by the blue, black, and red curves in figures \ref{rall} and \ref{call}.
		The value of curvature has a finite maximum, and it reaches its minimum as $\ss\rightarrow 0$. However, the vev $C$ is a monotonic
		function with a maximum at the point where  $\ss\rightarrow 0$. The dashed orange curve corresponds to the case where the value
		of the axion is constant, or equivalently, $X_\infty = 0$.
		
		\item The green curve  in figures \ref{rall} and \ref{call} represents the Type III solution. As the IR end-point moves towards the UV fixed point at $\f_0=0$, the dimensionless curvature and the vev monotonically decrease and increase, respectively.
		We should emphasize that we have put in the same diagram, both Type II and Type III solutions, by using as the horizontal axis either $\f_0$ or $-\ss$.
		\begin{figure}[htbp]
			\begin{center}
				\begin{subfigure}{0.49\textwidth}
					\includegraphics[width=\textwidth]{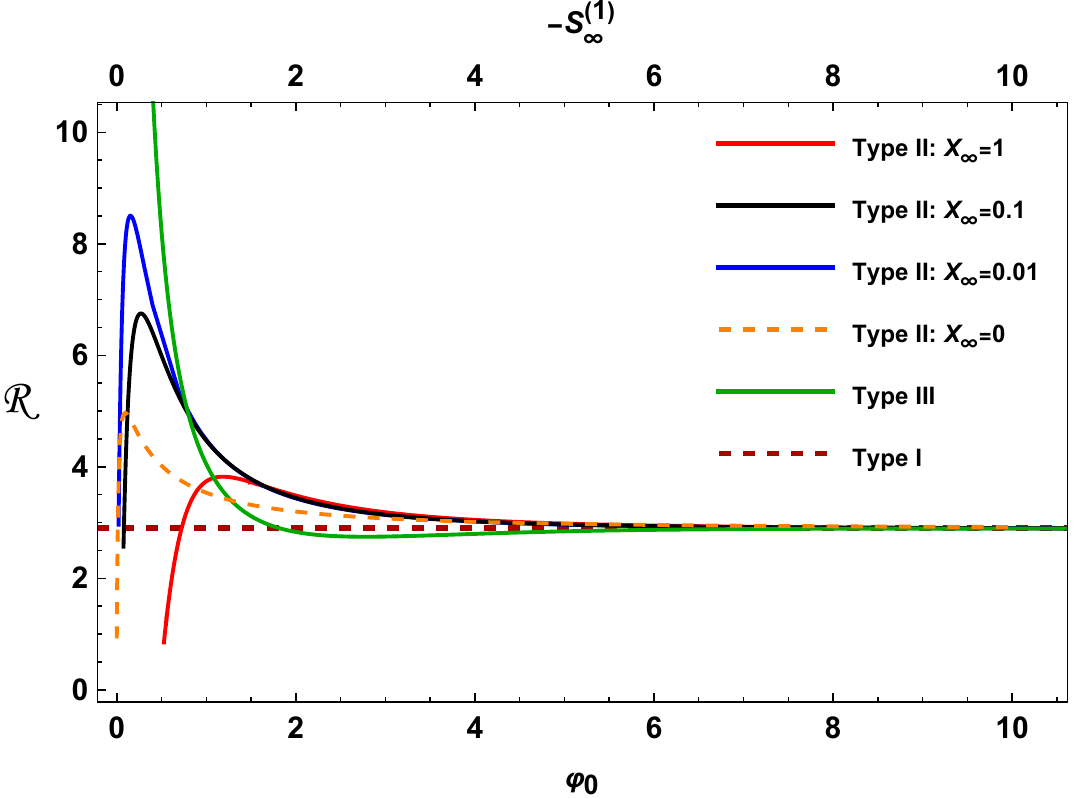}
					\caption{}\label{rall}
				\end{subfigure}\hspace{0.1cm}
				\begin{subfigure}{0.49\textwidth}
					\includegraphics[width=\textwidth]{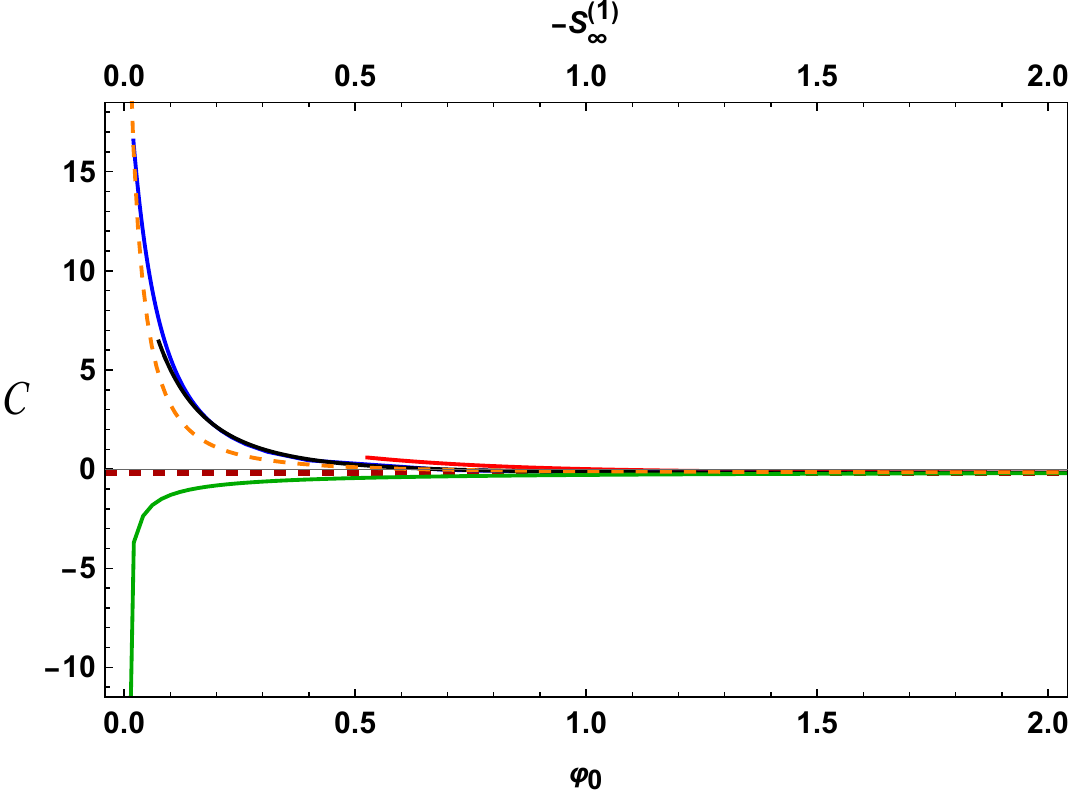}	
					\caption{}\label{call}
				\end{subfigure}
			\end{center}
			\caption{\footnotesize{Above the Efimov bound and for $b=0.54$, (a) and (b) illustrate the dimensionless curvature and $C$ vev
					as functions of the IR parameters. The $\f_0$ axis is at the bottom and the $S^{(1)}_{\infty}$ axis is at the top.
					For the Type II solution, we have depicted four examples with fixed $X_\infty$, represented by the blue, black,
					and red curves and a dashed orange curve for $X_\infty=0$. The green curve shows the data for Type III solutions
					(that do not depend on $X_{\infty}$ as they have a constant axion). All solutions ultimately converge to a unique Type I solution, which is indicated by a horizontal red dashed line.
			}}
		\end{figure}
		\begin{figure}[htbp]
			\begin{center}
				\begin{subfigure}{0.48\textwidth}
					\includegraphics[width=\textwidth]{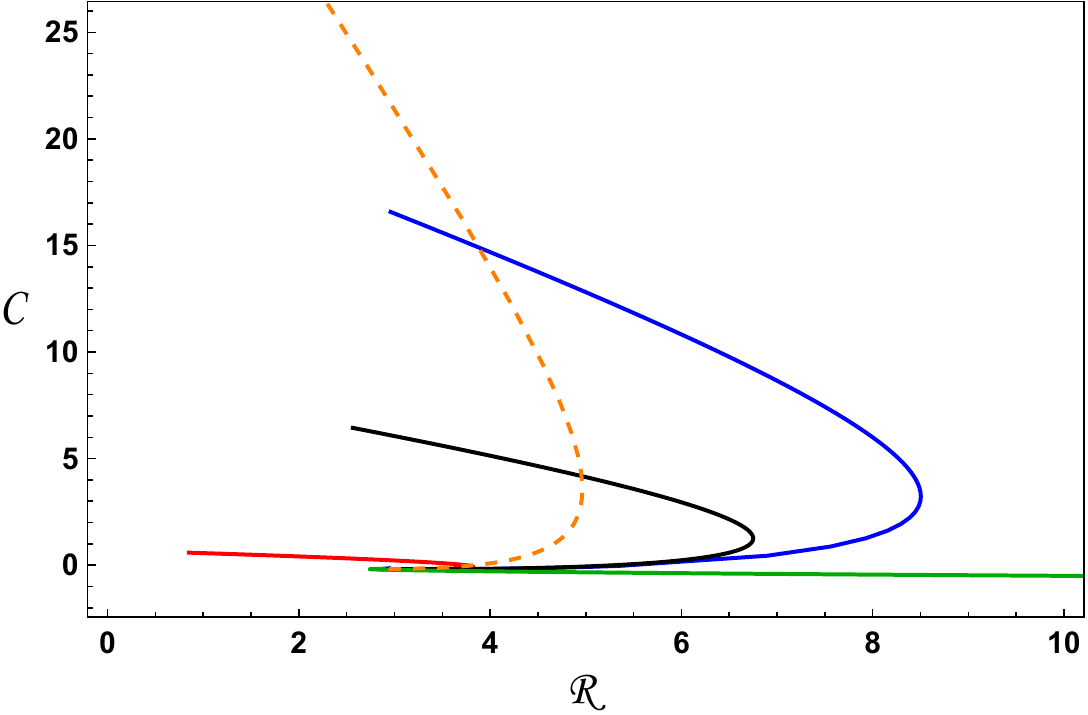}	
					\caption{}\label{crall}
				\end{subfigure}\hspace{0.2cm}
				\begin{subfigure}{0.495\textwidth}
					\includegraphics[width=\textwidth]{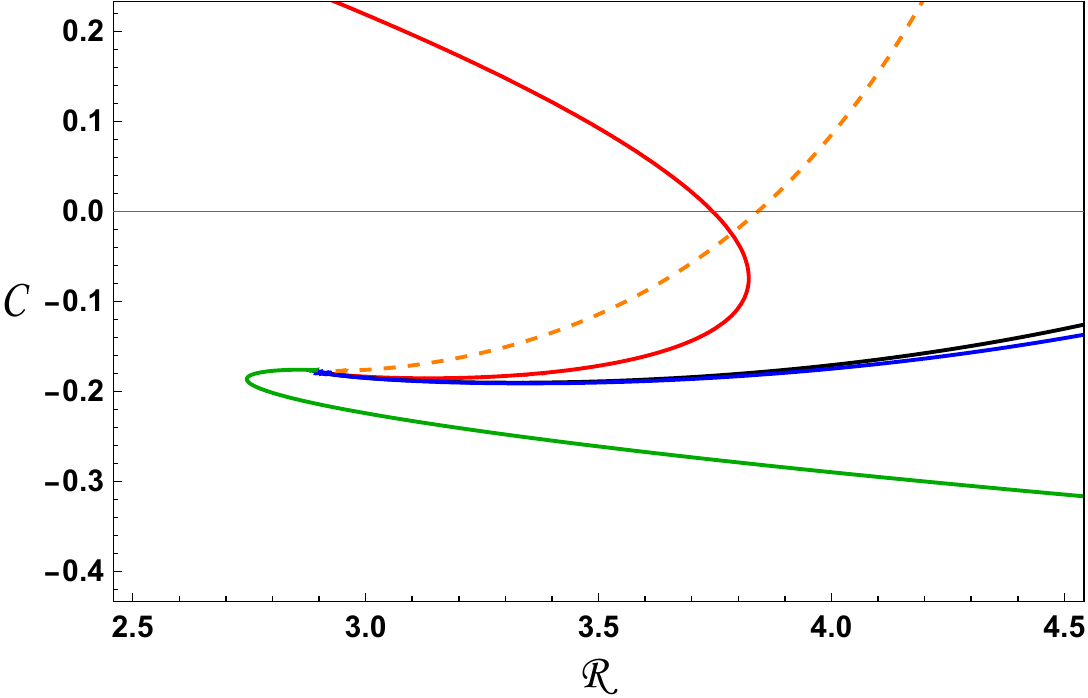}	
					\caption{}\label{crallz}
				\end{subfigure}
			\end{center}
			\caption{\footnotesize{Figures (a) and its zoomed-in counterpart (b) illustrate the emergence of the Efimov spiral in the behavior of $C(\mathcal{R})$.}}
		\end{figure}
		\begin{figure}[htbp]
			\centering
			\includegraphics[width=0.5\linewidth]{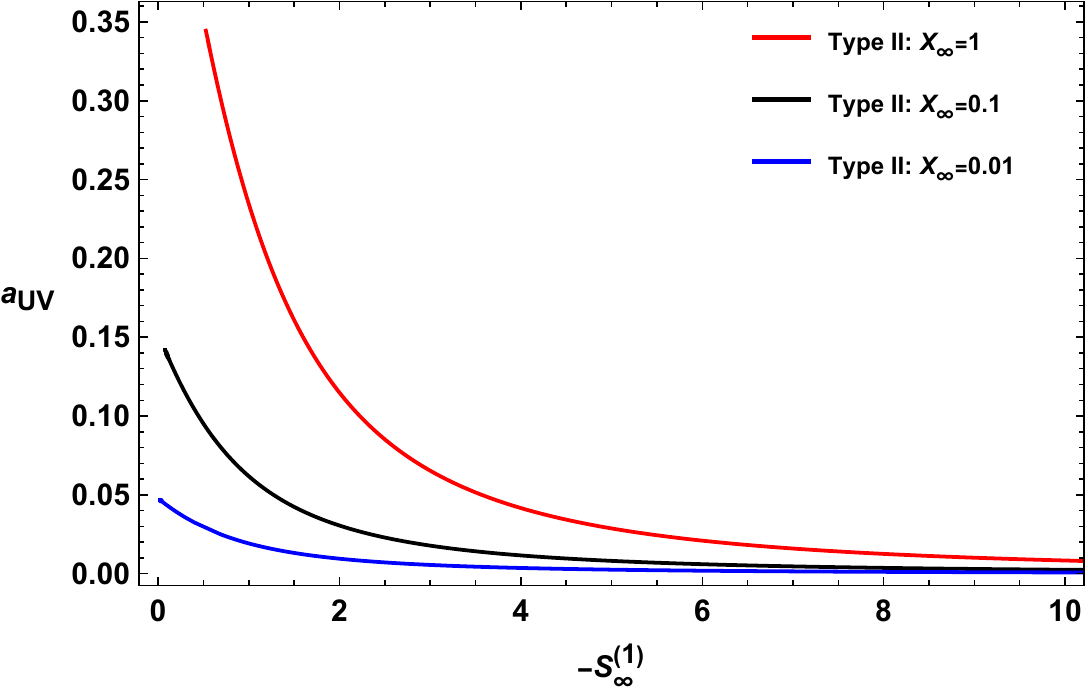}
			\caption{\footnotesize{Above the Efimov bound, the UV axion value is plotted as a function of $\ss$
					for three representative choices of $X_\infty$. For the allowed ranges of $\ss$, all values extend from a
					finite value down to zero. In all solutions UV is located at $-\infty$ and $Q<0$.}}
			\label{as1}
		\end{figure}

		\item When either of the values of $\f_0$ or $-\ss$ increases, all solutions (Type II and III) tend to
		approach the Type I solution. The QFT data for this unique solution is represented by a horizontal red dashed line.
		
		\item
		Figures \ref{crall} and its zoomed-in version in \ref{crallz} illustrate the dependence of the vev
		from the source, $C(\mathcal{R})$, for the same solutions in figures \ref{rall} and \ref{call}.
		All Type II and Type III curves converge to a point where the Type I solution exists.
		Because of the oscillatory
		behavior of $\mathcal{R}$ and $C$ as functions of the IR parameters, $C(\mathcal{R})$ exhibits
		an Efimov spiral which is visible in Figure \ref{crallz}. Because of the spiral, $C(\mathcal{R})$ is
		a multi-valued function of $\mathcal{R}$ in an intermediate regime, suggesting the presence of a quantum
		phase transition from the Type III branch to the Type II as $\mathcal{R}$ becomes smaller.
		
		\item
		The axion source $a_{\textrm{UV}}$  at the UV boundary is illustrated in figure \ref{as1}. In all cases, $a_{\textrm{UV}}$
		reaches a maximum when $-\ss$ is at its minimum, and gradually approaches zero as $-\ss$ becomes large.
		By decreasing the values of $X_\infty$, the values of the axion become smaller. Note that here, the solutions
		under consideration have a UV at $u \rightarrow -\infty$ with $Q<0$. Consequently, imposing the boundary
		condition $a_{IR}=0$ ensures that all values of $a_{\textrm{UV}}$ are positive. One should obtain opposite values of $a_{\textrm{UV}}$ by changing the sign of $Q$.
		
	\end{enumerate}

	\subsubsection{ Below the Efimov bound}\label{beb}

\begin{figure}[htbp]
		\begin{center}
			\begin{subfigure}{0.47\textwidth}
				\includegraphics[width=\textwidth]{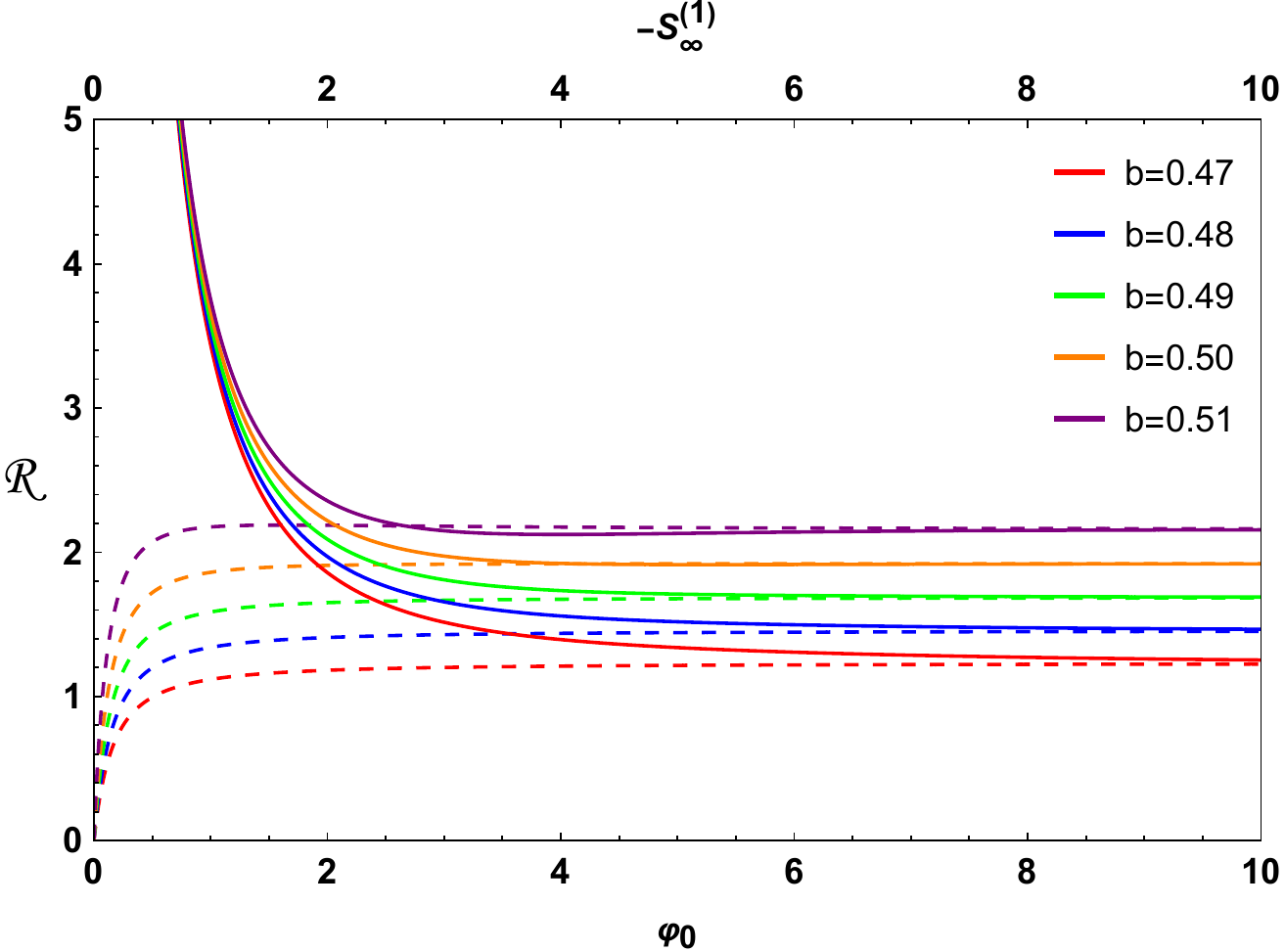}
				\caption{}\label{mub1}
			\end{subfigure}\hspace{0.25cm}
			\begin{subfigure}{0.495\textwidth}
				\includegraphics[width=\textwidth]{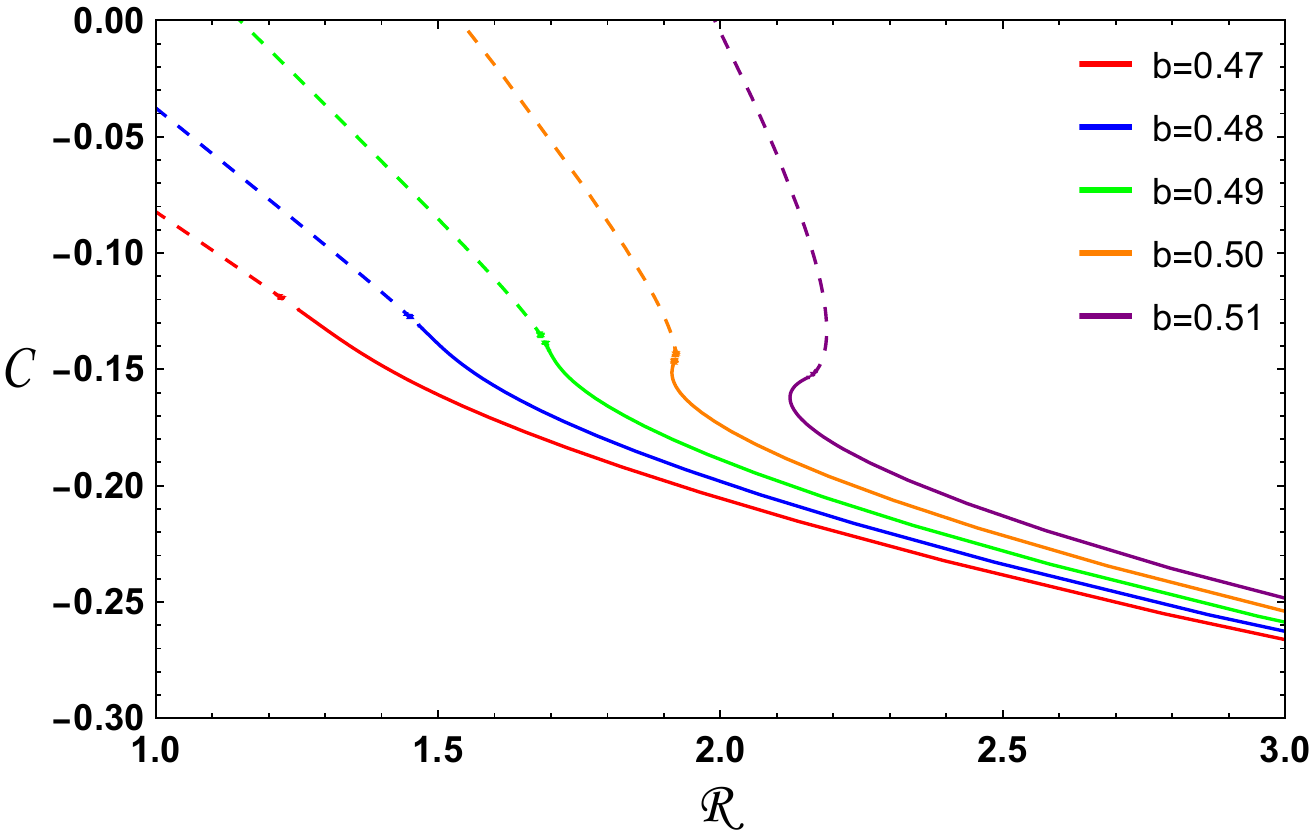}
				\caption{}\label{mub2}
			\end{subfigure}
		\end{center}
		\caption{\footnotesize{ (a): Scalar curvature profiles for Type II (dashed) and Type III (solid) solutions across decreasing values of $b$, illustrating the gradual disappearance of Efimov spirals. (b): Plot of vev $C$ versus $\mathcal{R}$ showing the spiral structure opening and ultimately vanishing as $b$ approaches a critical value.
		}}
	\end{figure}

	Before discussing solutions with a nontrivial axion field, we recall the properties of  Type II and Type III solutions across a range of $b$ values situated below the Efimov bound, \cite{Jani}. This is displayed in the two figures \ref{mub1} and \ref{mub2}.
	
	Figure \ref{mub1} illustrates the scalar curvature behavior for both Type II (dashed curves) and Type III (solid curves) corresponding to five distinct values of the parameter $b$. As $b$ decreases, the intersection point between Type II and Type III solutions shifts to the right. This trend indicates that reducing $b$ causes the Efimov spirals to gradually fade. For a clearer visualization, we refer to figure \ref{mub2}, where the vacuum expectation value $C$ is plotted against $\mathcal{R}$. Once again, we observe that decreasing $b$ causes the spiral structure to broaden, and beyond a certain critical value, the spiral pattern disappears entirely.

	We now incorporate the axion field into the analysis. For values below the Efimov bound, we re-evaluate the solutions corresponding to the same IR values of $X_\infty$ as previously considered. The examination of figures \ref{brall} through \ref{as2} reveals:

	\begin{enumerate}
		
		\item In Figure \ref{brall}, the dimensionless curvature for the Type III solution (green curve) begins at a higher value when $\ss\rightarrow 0$ and decreases smoothly. In contrast, Type II curves (red, blue, black) show initial variation influenced by $X_\infty$, with their curvatures $\mathcal{R}$ increasing as $-\ss$ increases. This trend also holds for the Type II solution with zero axion (orange dashed curve). However, the behavior of the vev $C$ exhibits the opposite trend.
		\begin{figure}[htbp]
			\begin{center}
				\begin{subfigure}{0.48\textwidth}
					\includegraphics[width=\textwidth]{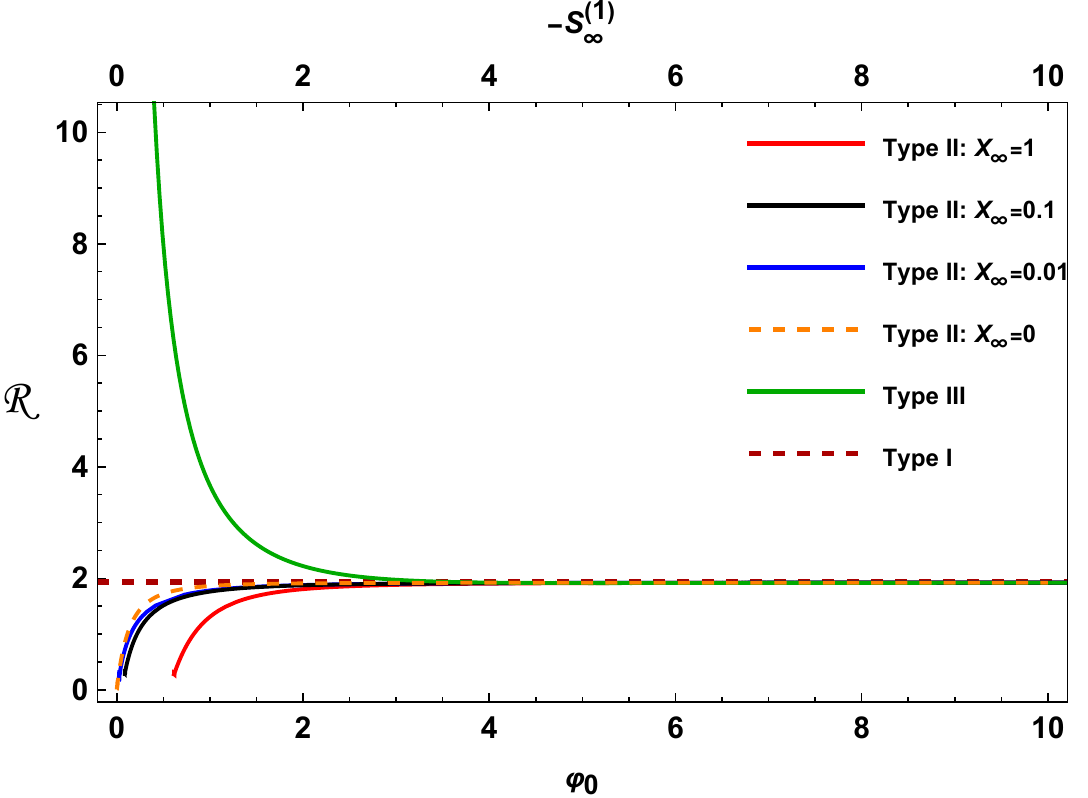}
					\caption{}\label{brall}
				\end{subfigure}\hspace{0.2cm}
				\begin{subfigure}{0.49\textwidth}
					\includegraphics[width=\textwidth]{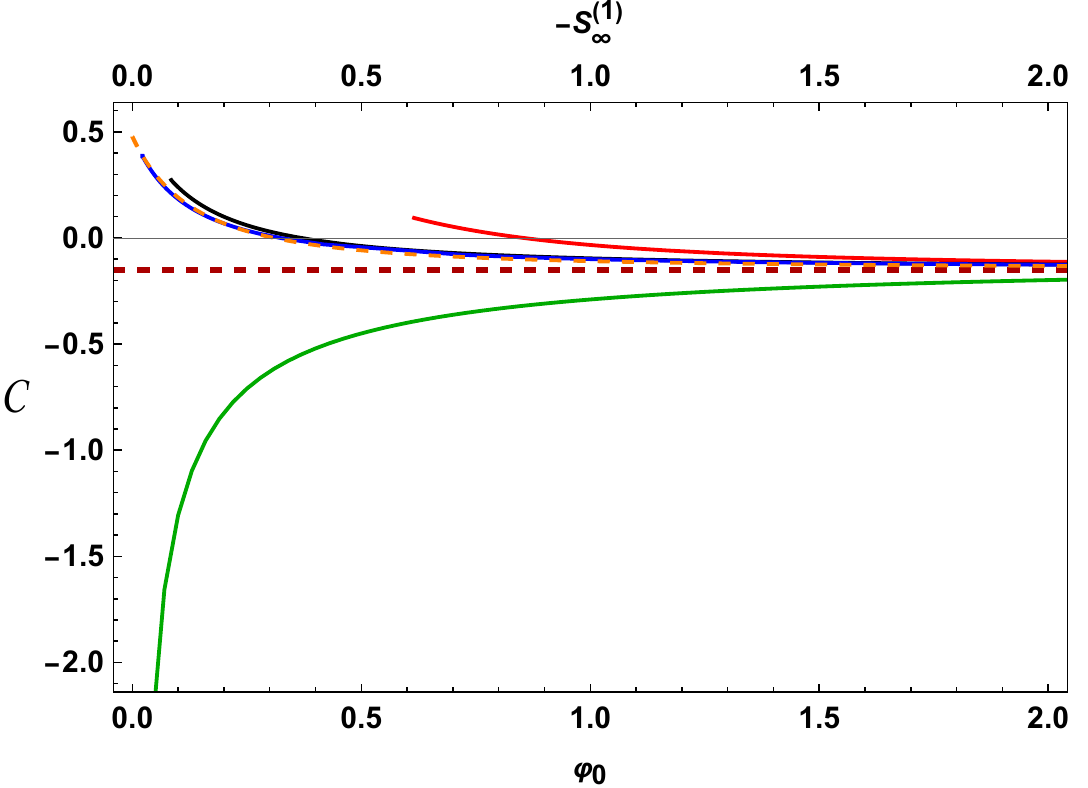}
					\caption{}\label{bcall}
				\end{subfigure}
			\end{center}
			\caption{\footnotesize{Below the Efimov bound and for $b=0.5$, (a) and (b) illustrate the dimensionless curvature and $C$ vev as functions of the IR parameters.
			}}
		\end{figure}
		\item Figures \ref{bcrallz} and  \ref{bcrall} illustrate the correlation between $C$ and $\mathcal{R}$ for $b=0.5$ and $b=0.47$ respectively. By decreasing $b$, the Efimov spiral disappears gradually.
		\begin{figure}[htbp]
			\begin{center}
				\begin{subfigure}{0.48\textwidth}
					\includegraphics[width=\textwidth]{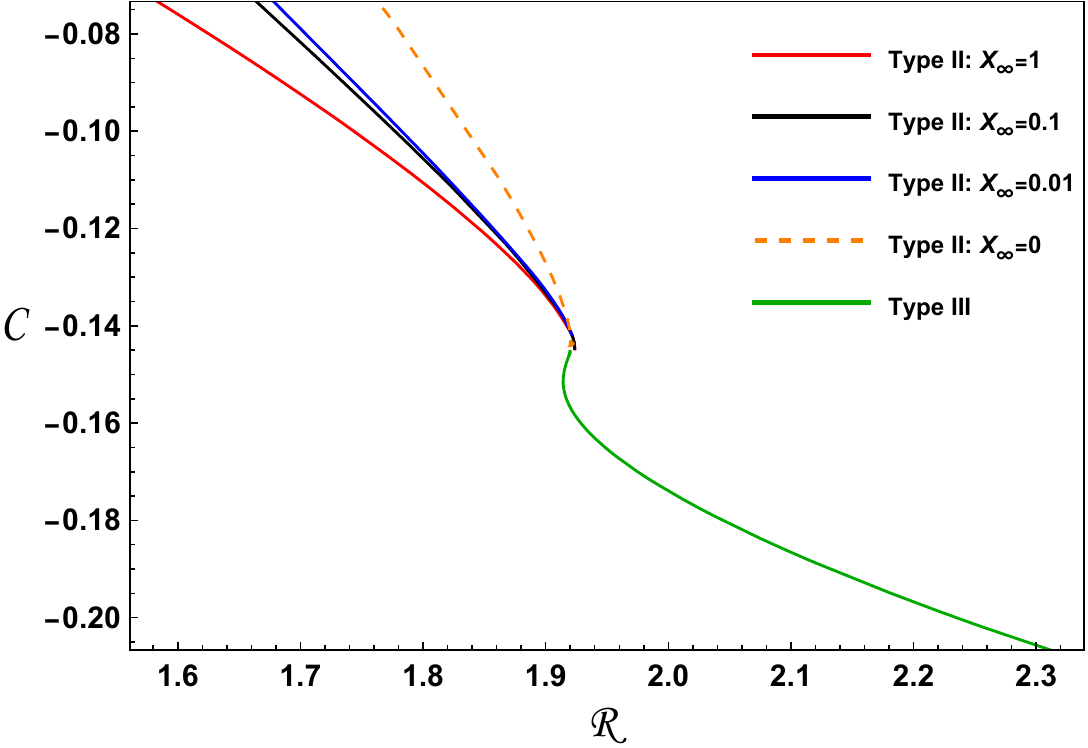}	
					\caption{$b=0.5$}\label{bcrallz}
				\end{subfigure}\hspace{0.2cm}
				\begin{subfigure}{0.48\textwidth}
					\includegraphics[width=\textwidth]{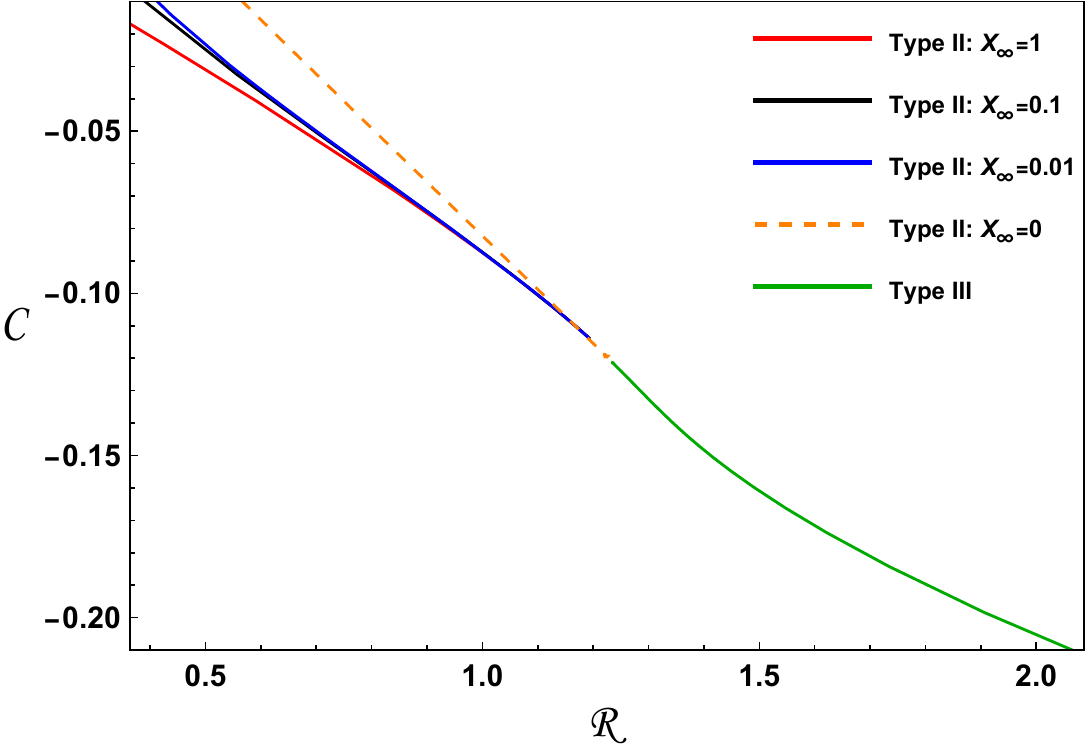}	
					\caption{$b=0.47$}\label{bcrall}
				\end{subfigure}
			\end{center}
			\caption{\footnotesize{The behavior of $C(\mathcal{R})$ for two different values of $b$.}}
		\end{figure}
		\item As Figure \ref{as2} indicates, the variation of $a_{\textrm{UV}}$ with respect to the IR parameters appears to be independent of the Efimov bound. For comparison, refer to Figure \ref{as1}.
		\begin{figure}[htbp]
			\centering
			\includegraphics[width=0.5\linewidth]{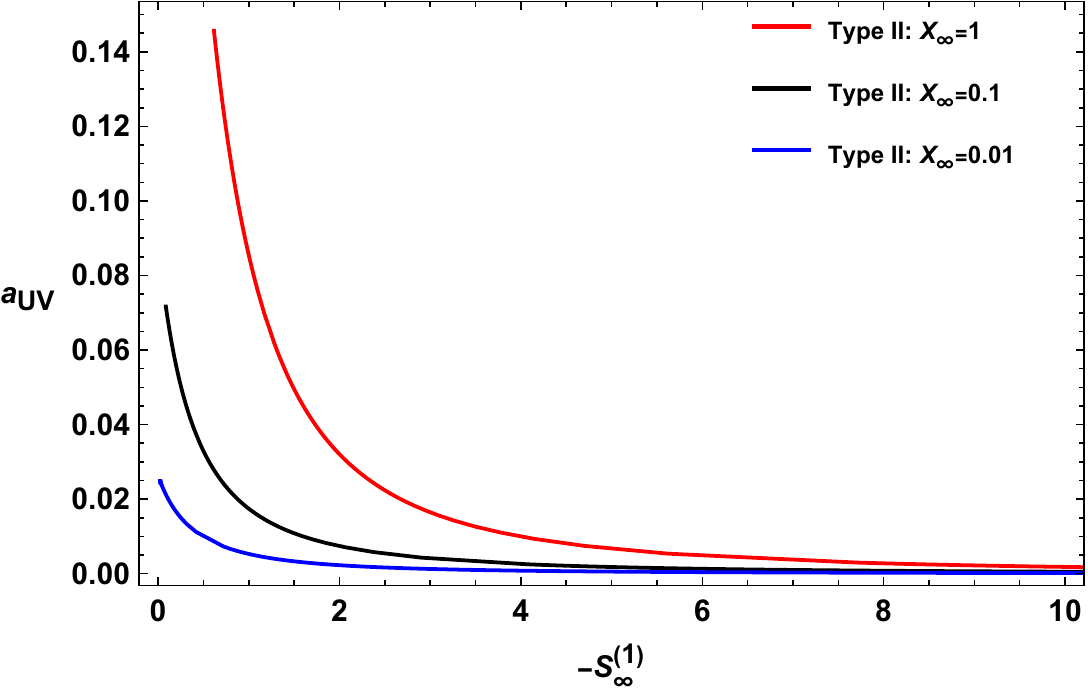}
			\caption{\footnotesize{Below the Efimov bound, for three selected values of $X_\infty$, the UV axion value is shown as a function of $\ss$. The data belong to $b=0.5$.	}}
			\label{as2}
		\end{figure}
		
	\end{enumerate}

	\subsection{The free energy and the phase structure} \label{FEP}
	
	Since we have found that there may be multiple solutions corresponding to the same values of the UV parameters, we now turn to determining which of the competing solutions dominates.  To this end, in this section, we compute the free energy and thermal entropy of one-boundary solutions with positively curved spatial slices, $S^d$ or $dS_d$. These geometries possess a single UV boundary, while their IR end-points depend on the type of regular behavior discussed in section \ref{IRAS}. More details of the computations are addressed in Appendix \ref{FE}.
	
	The free energy of the dual field theory is computed holographically from the Euclidean on-shell action, including the Gibbons--Hawking--York term and suitable counter-terms. For the metric ansatz
	\be  \label{FEP1}
	ds^2 = du^2 + e^{2A(u)} \a^2 ds_{S^d}^2\,,
	\ee
	the renormalized free-energy density in $d=4$ takes the form \cite{Ghodsi:2024jxe}
	\be
	\mathcal{F}(\mathcal{R},a_{\textrm{UV}}) = \frac{3}{2} \left(1 - 96\,\frac{C(\mathcal{R},a_{\textrm{UV}})}{\mathcal{R}^2}
	- 48\,\frac{\mathcal{B}(\mathcal{R},a_{\textrm{UV}})}{\mathcal{R}}\right)\!,
	\label{FEP2}
	\ee
	where $\mathcal{R}$ and $a_{\textrm{UV}}$ are the UV sources associated with curvature and the axion, respectively, and $C$, $\mathcal{B}$ are the corresponding vacuum expectation values (vevs) all defined from the near-boundary expansions, (\ref{fe})--(\ref{mR}), and (\ref{vevf}).
	
	In \cite{Ghosh:2018qtg}, an interesting structure emerged when computing the free energy of holographic solutions for positive curvature slices (de Sitter).
	The free energy could be written as in thermodynamics
	with temperature the Hawking temperature of the de Sitter slice, and entropy the thermal entropy of the static patch.
	This same entropy was equal to the entanglement entropy in global coordinates between the two hemispheres of the spatial
	sphere. Even more unexpectedly, this picture, with some modifications, persisted in the case of negative slice curvature, \cite{Ghodsi:2022umc}.
	
	This thermal (or entanglement) entropy, discussed above, can be derived either from the area of the cosmological
	horizon, or equivalently, from the first law of thermodynamics. The two approaches yield consistent results
	\be  \label{FEP3}
	\mathcal{S}_{th}(\mathcal{R},a_{\textrm{UV}})
	= -\!\left(\frac{1}{2}\mathcal{R}\frac{\partial}{\partial \mathcal{R}}+1\right)
	\mathcal{F}(\mathcal{R},a_{\textrm{UV}})\,,
	\ee
	and, using the UV expansions of the bulk fields,
	\be  \label{FEP4}
	S_{th} = 72\,(M_p\ell)^3 V_{S^4}\,\frac{\mathcal{B}(\mathcal{R},a_{\textrm{UV}})}{\mathcal{R}}\,,
	\ee
	where $V_{S^4}$ is the volume of the $S^4$ slice with a radius equal to one.
	The consistency between these expressions implies a differential relation among the vevs,
	\be  \label{FEP5}
	\frac{\pa}{\pa \mathcal{R}}C(\mathcal{R},a_{\textrm{UV}}) = \frac12 \mathcal{B}(\mathcal{R},a_{\textrm{UV}}) - \frac{\mathcal{R}}{2}\frac{\pa}{\pa \mathcal{R}}\mathcal{B}(\mathcal{R},a_{\textrm{UV}}) + \frac{\mathcal{R}}{48}\,,
	\ee
	where the last term is due to the conformal anomaly.
	(\ref{FEP5}) in turn, determines the $\mathcal{R}$-dependence of the free energy as
	\be  \label{FEP6}
	\frac{\partial \mathcal{F}}{\partial \mathcal{R}}
	= 144\!\left(\frac{2C(\mathcal{R},a_{\textrm{UV}})}{\mathcal{R}^3}-\frac{1}{48\mathcal{R}}\right)\!.
	\ee
	
	In summary, the one-boundary solutions define a consistent thermodynamic system in which both $\mathcal{F}$ and $\mathcal{S}_{th}$ are fully characterized by the UV data $(\mathcal{R}, a_{\textrm{UV}})$ and their associated vevs $(C,\mathcal{B})$.
	It is shown in Appendix \ref{FE} that the trace of the energy-momentum tensor at the CFT point  satisfies the standard anomaly formula  with
	\be
	a=c=2\pi^2 (M_p\ell)^3\,.
	\ee
	It is also shown at the end of subsection \ref{TES} that the trace in the general case is given by the field-theoretic formula
	\be
	\langle {T^{\m}}_{\m}\rangle =\beta\cdot \langle O\rangle +R^2~~{\rm anomaly}\,.
	\ee
	
	In the subsequent sections, to analyze and compare the free energy among different solutions, we use the relation \eqref{FEP2}.  We show that in the two-dimensional phase space spanned by $(a_{\textrm{UV}}, \mathcal{R})$, a phase transition takes place between Type II and Type III solutions.

	\subsubsection{Above the Efimov bound}\label{pta}

\begin{figure}[htbp]
		\begin{center}
			\begin{subfigure}{0.49\textwidth}
				\includegraphics[width=\textwidth]{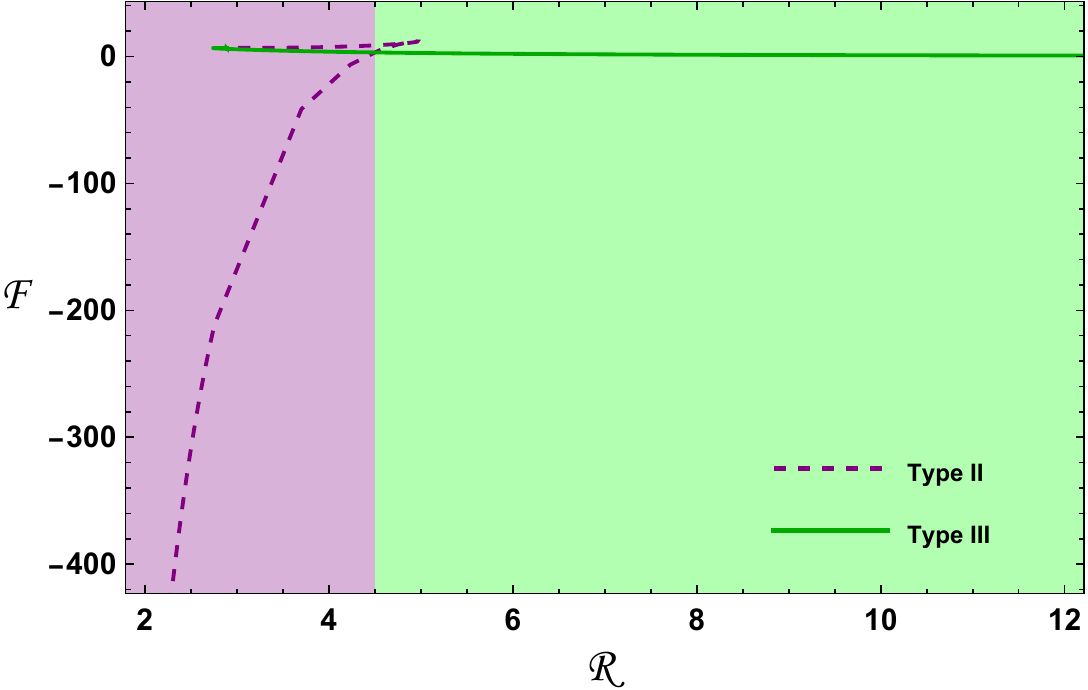}	
				\caption{}\label{zax}
			\end{subfigure}
			\begin{subfigure}{0.485\textwidth}
				\includegraphics[width=\textwidth]{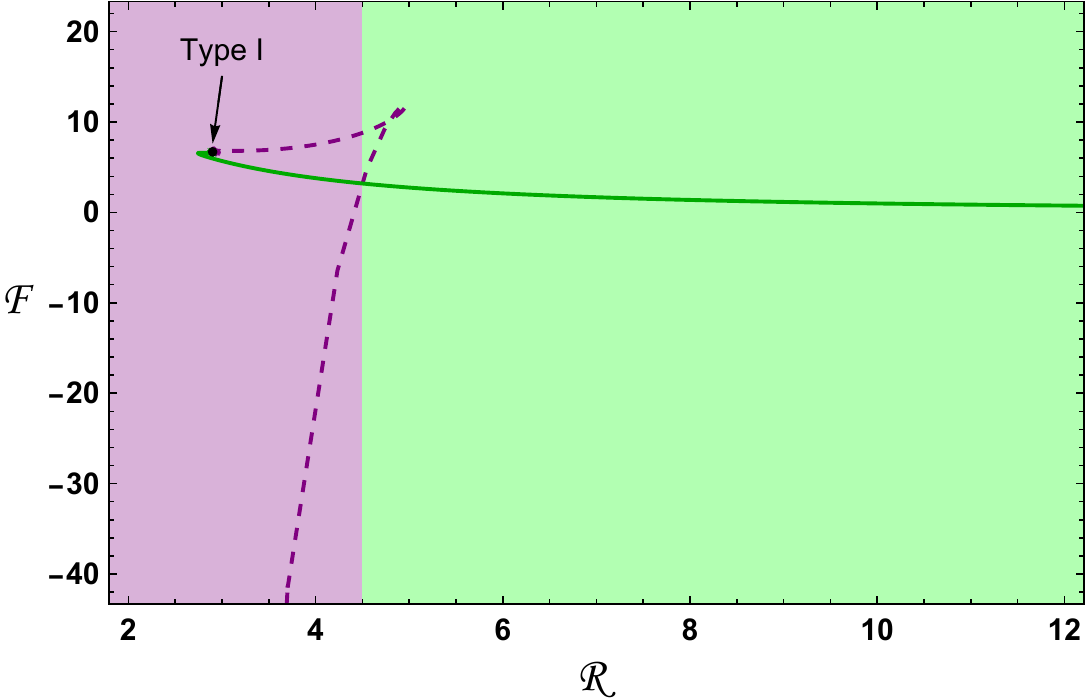}
				\caption{}\label{zaxz}
			\end{subfigure}
		\end{center}
		\caption{\footnotesize{Above the Efimov bound for $b=0.54$ and in the zero axion limit, a phase transition occurs at a critical curvature value $\mathcal{R} \approx 4.5$. The figure on the right is a zoomed-in view of the figure on the left.
		}}
	\end{figure}

	Before considering the axions, it is worth reviewing the results obtained by comparing the free energy of Type II and Type III when $a_{IR}=0$. This is illustrated in figure \ref{zax} and its zoomed-in version \ref{zaxz}, where the free energy is plotted as a function of the dimensionless curvature.
	
	Below a critical value of curvature, the free energy of Type II is lower (purple region). Above that, Type III is always the favored solution (green region). This indicates a phase transition, with the order parameter $\mathcal{R}$, from Type II to Type III solutions. As illustrated in Figure \ref{zaxz}, this corresponds to a first-order phase transition, as the free energy remains continuous at the transition point, while its first derivative exhibits a discontinuity.
	\begin{figure}[htbp]
		\begin{center}
			\begin{subfigure}{0.49\textwidth}
				\includegraphics[width=\textwidth]{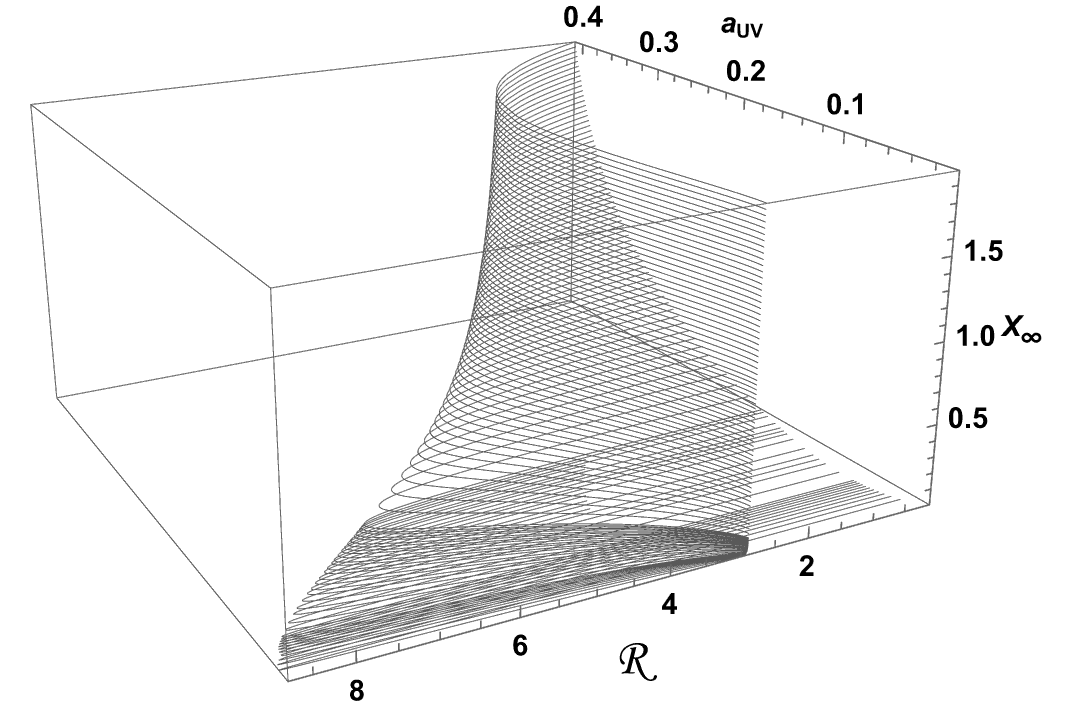}	
				\caption{}\label{xar}
			\end{subfigure}
			\begin{subfigure}{0.49\textwidth}
				\includegraphics[width=\textwidth]{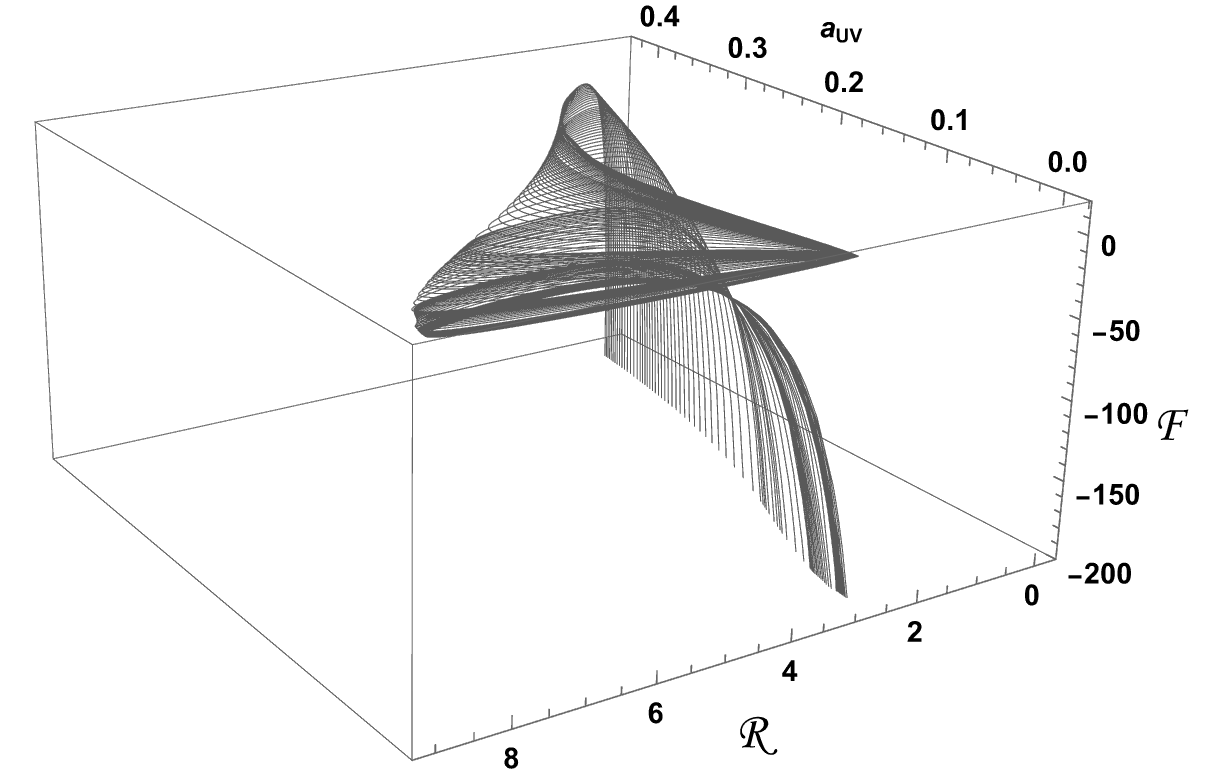}	
				\caption{}\label{far}
			\end{subfigure}
		\end{center}
		\caption{\footnotesize{Above the Efimov bound at $b = 0.54$:
				(a): The values of the sources $\mathcal{R}$ and $a_{\textrm{UV}}$ are shown for various Type II solutions. At each fixed $X_\infty$, the parameter $Q$ varies along the corresponding curve.
				(b): For each point depicted in panel (a), the corresponding free energy has been plotted.
		}}
	\end{figure}
	
	In the presence of the axion, one should consider that for each solution there are two sources in the dual QFT: the value of the axion in the UV, denoted by $a_{\textrm{UV}}$, and the dimensionless curvature $\mathcal{R}$.
	
	Figure \ref{xar} illustrates, for Type II solutions, how the values of the sources on the UV boundary ($\mathcal{R}$, $a_{\textrm{UV}}$) change for different values of the IR parameter $X_\infty$. The other IR parameter is $\ss$, or equivalently $Q$ (see \eqref{Si1R}), which varies along each curve at fixed $X_\infty$. For small $|Q|$, all curves converge to $a_{\textrm{UV}} = 0$ with the same non-zero finite $\mathcal{R}$. As $|Q|$ increases, each curve eventually approaches a non-zero finite value for  $a_{\textrm{UV}}$ and $\mathcal{R} \rightarrow 0$. Moreover, as $X_\infty$ grows, the boundary value of the axion, $a_{\textrm{UV}}$, also increases.
	
	Figure \ref{far}, in contrast, presents the free energy density \eqref{FEP2}  corresponding to each point in Figure \ref{xar}. As $|Q|$ decreases, all curves once again converge to a single point. This marks the location where the Type I solution emerges.
	
	To determine which solution is preferred from the free energy perspective, there are two distinct cases to consider. When, at a given point ($\mathcal{R}, a_{\textrm{UV}}$), there is a unique solution, this is the only available choice. However, it is also necessary to compare solutions that share the same sources on the QFT dual side. This means identifying, among the various solutions in Figure \ref{xar}, the points where two solutions have identical values of $\mathcal{R}$ and $a_{\textrm{UV}}$.
	
\begin{figure}[htbp]
		\begin{center}
			\begin{subfigure}{0.49\textwidth}
				\includegraphics[width=\textwidth]{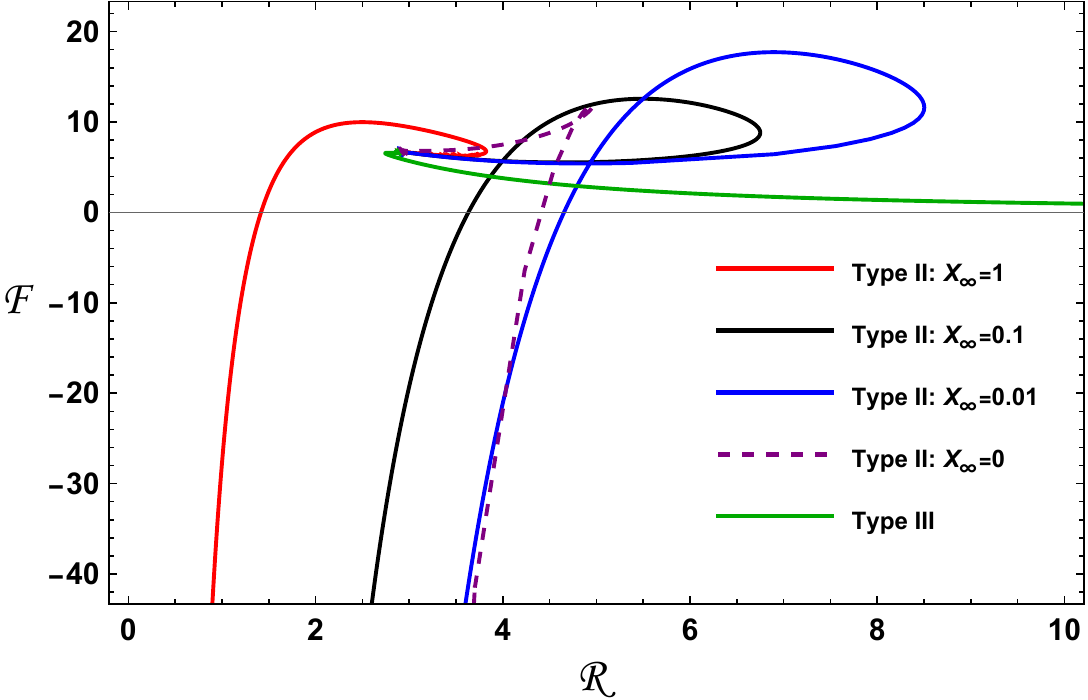}	
				\caption{}\label{fall}
			\end{subfigure}
			\begin{subfigure}{0.49\textwidth}
				\includegraphics[width=\textwidth]{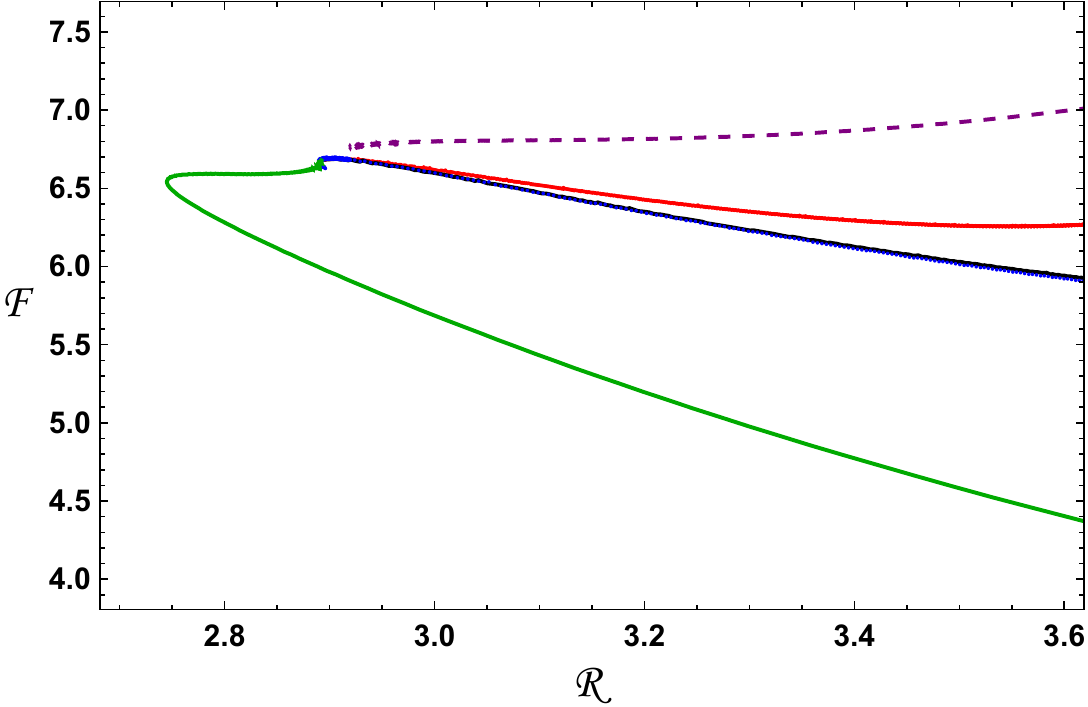}
				\caption{}\label{fallz}
			\end{subfigure}
		\end{center}
		\caption{\footnotesize{(a) Free energy profiles of four sample Type II solutions at fixed $X_\infty$, alongside the Type III solution, plotted as functions of the dimensionless curvature. Each point along the Type II curves (except $X_\infty=0$)  corresponds to a distinct value of $a_{\textrm{UV}}$, whereas all points on the Type III curve share the same arbitrary constant value of $a_{\textrm{UV}}$.
				(b) A magnified view focusing on the region near the Type I solution in panel (a).
		}}
	\end{figure}

	To illustrate the latter case, we show Figure \ref{fall} (and its zoomed-in version, Figure \ref{fallz}) as a representative example. In this figure, the free energy of three samples (labeled by $X_\infty$) of Type II solutions is shown as a function of $\mathcal{R}$. It is important to note that different points on these curves correspond to different values of $a_{\textrm{UV}}$. Also included for comparison are the Type II solution with zero axion ($X_\infty=0$) and the Type III solutions. For the latter, note that the axion is constant and may take any value. In such a case, the free energy is independent of $a_{\textrm{UV}}$.
	
\begin{figure}[htbp]
		\begin{center}
			\begin{subfigure}{0.5\textwidth}
				\includegraphics[width=\textwidth]{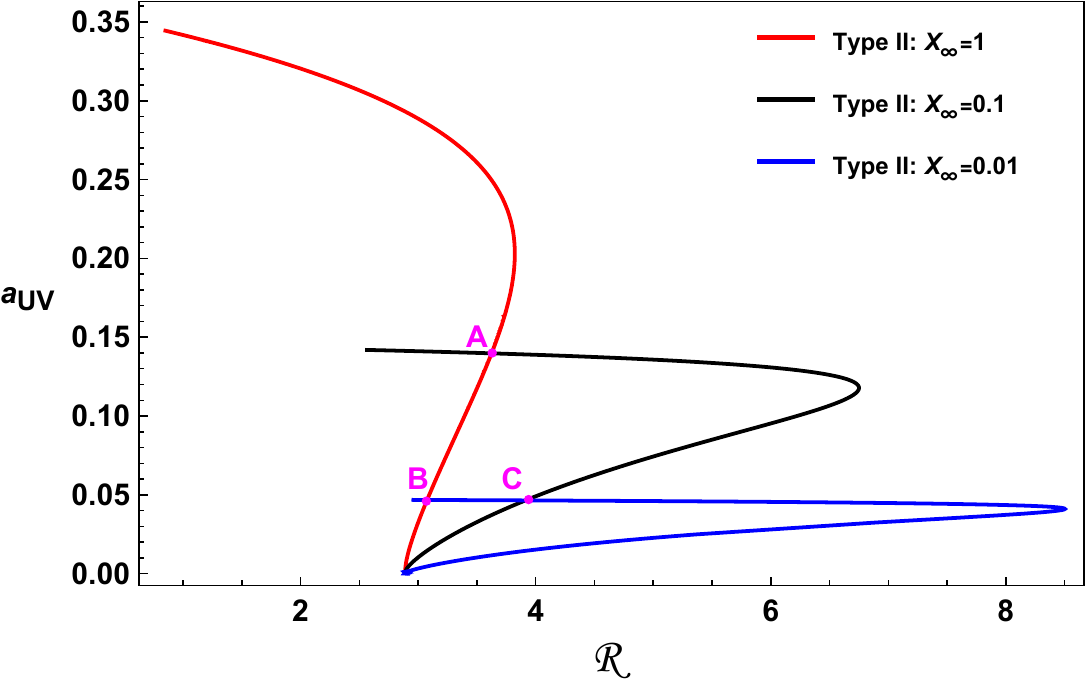}	
				\caption{}\label{ar1}
			\end{subfigure}
			\begin{subfigure}{0.47\textwidth}
				\includegraphics[width=\textwidth]{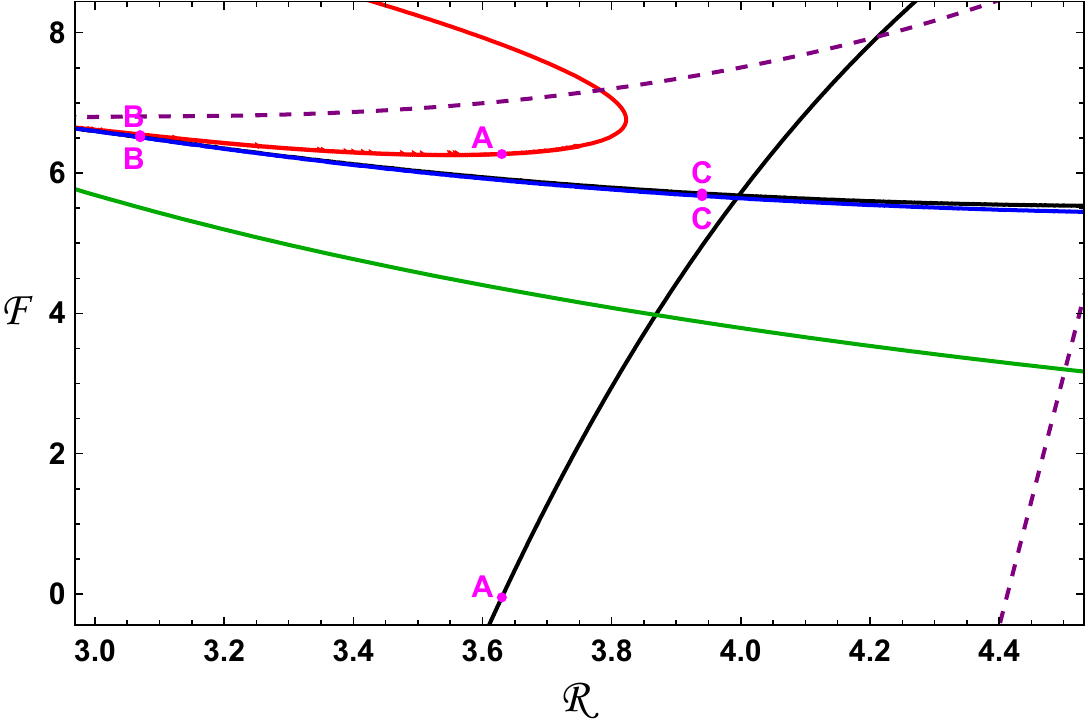}	
				\caption{}\label{ar2}
			\end{subfigure}
		\end{center}
		\caption{\footnotesize{(a) For the three particular cases where $X_\infty = 0.01, 0.1, 1$, the axion value is plotted against the dimensionless curvature. Intersection points A, B, and C indicate where two distinct solutions share the same sources. (b) Illustrates the free energy as a function of curvature, with the points from figure (a) now displayed separately. Among the points A, B, and C, only point A on the black curve has a smaller free energy than the Type III (green) curve.
		}}
	\end{figure}

	To perform a meaningful comparison of free energies, we must focus on solutions with matching $\mathcal{R}$ and $a_{\textrm{UV}}$. By plotting data from curves corresponding to different $X_\infty$ values, the intersection points of these curves indicate the relevant points to be compared as shown in  Figure \ref{ar1}.
	
	The final step involves determining the free energy values at these specific points, as shown in Figure \ref{ar2}. This allows us to identify, at any given $\mathcal{R}$, the solution with the lowest free energy.
	
\begin{figure}[htbp]
		\begin{center}
			\begin{subfigure}{0.445\textwidth}
				\includegraphics[width=\textwidth]{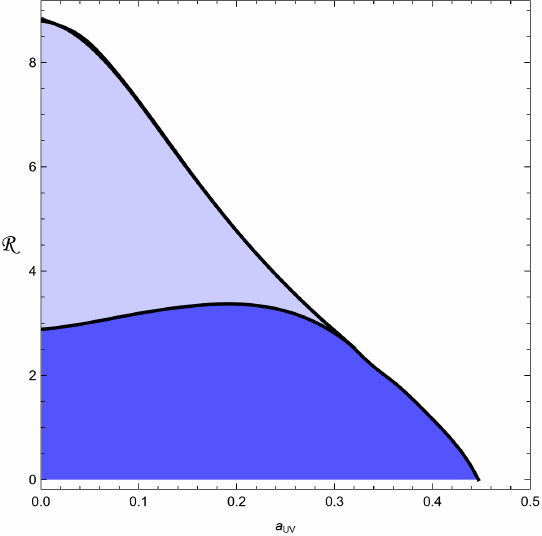}	
				\caption{}\label{ov1}
			\end{subfigure}
			\hspace{0.5cm}
			\begin{subfigure}{0.45\textwidth}
				\includegraphics[width=\textwidth]{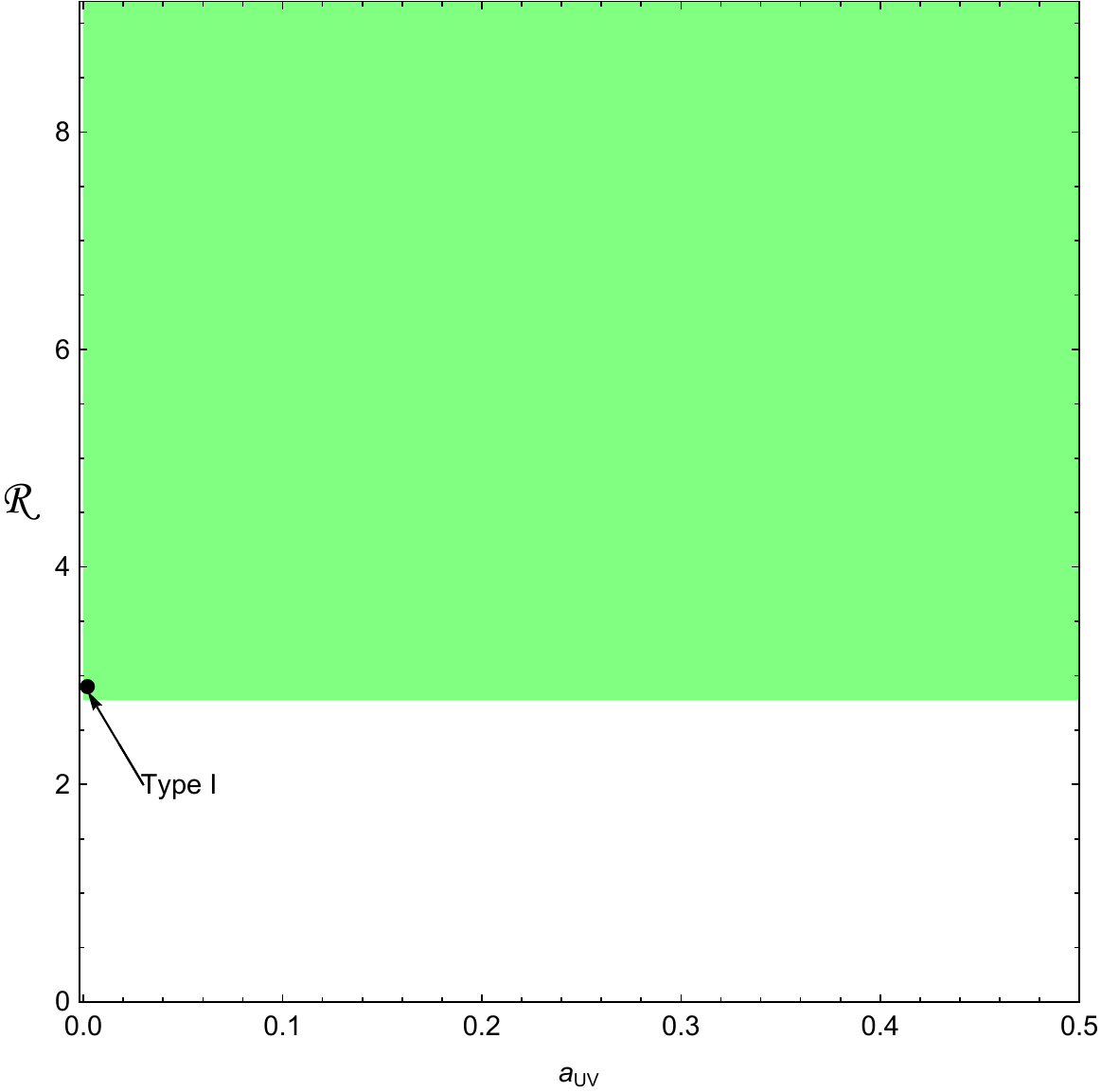}	
				\caption{}\label{ov2}
			\end{subfigure}
		\end{center}
		\caption{\footnotesize{(a): All points in the light-blue region correspond to two Type II solutions with the same source values but with different free energies. In the dark-blue region, only one unique Type II solution exists. (b): The region allowed by the Type III solution. The black dot shows that unique Type I solution at $a_{\textrm{UV}}=0$ and $\mathcal{R}\approx 2.9$. The lower bound of the green region is at $\mathcal{R}\approx 2.75$.
		}}
	\end{figure}

	To identify the phase transition, we examine the points across all Type II solutions that correspond to the same source values. Figure \ref{ov1} shows this region. In this figure, the light-blue region indicates the points in $\mathcal{R}-a_{\textrm{UV}}$ space that belong to two different solutions, while the dark-blue region shows the points that are associated with only one solution. Our numerical analysis shows that there is no region with more than two Type II solutions with the same UV sources. We emphasize that the solutions considered here correspond to $Q<0$. For $Q>0$, the result is the mirror image of figure \ref{ov1} with respect to the vertical axis.
	
	For the Type III solution, there is a lower bound on $\mathcal{R}$ (this can also be recognized from figure \ref{zax}); however, the value of $a_{\textrm{UV}}$ can be any constant. This region is shown in green in figure \eqref{ov2}.
	
\begin{figure}[htbp]
		\centering
		\includegraphics[width=0.45\linewidth]{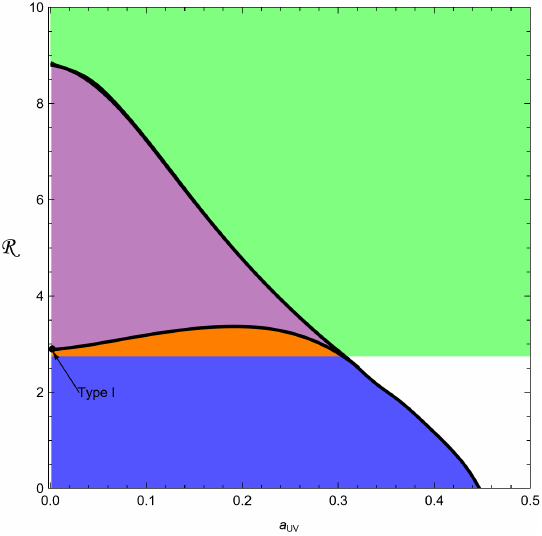}
		\caption{\footnotesize{Regions in the $(\mathcal{R}, a_{\textrm{UV}})$ plane showing the possible solutions according to the numerical analysis.
				Green areas indicate the regions that correspond to points where Type III is the unique solution.
				Blue regions contain only one Type II solution.
				Orange areas mark where there is competition between one Type II solution and Type III.
				The purple region shows where two Type II solutions coexist and compete with the Type III solution.
		}}
		\label{ov3}
	\end{figure}

	By identifying figures \ref{ov1} and \ref{ov2} we find figure \ref{ov3}, which summarizes as follows:
	
	\begin{itemize}
		
		\item At each point in the green region of figure \ref{ov3}, there is only one Type III solution.
		
		\item For each light-blue point in figure \ref{ov1}, we should find the solution with the lowest free energy and then compare it with the Type III solution that has the same values of $\mathcal{R}$ and $a_{\textrm{UV}}$. This should be done in the purple region in figure \ref{ov3}.
		
		\item Similarly, we should compare the free energy of the dark-blue points in figure \ref{ov1} with the corresponding Type III solutions on the overlap regions. These are the points in the orange region in figure \ref{ov3}.
		
		\item In the blue regions of figure \ref{ov3}, there is only one Type II solution, and no Type III exists here. This region extends to zero values of the curvature.
		
		\end{itemize}
	\begin{figure}[htbp]
		\begin{center}
			\begin{subfigure}{0.45\textwidth}
				\includegraphics[width=\textwidth]{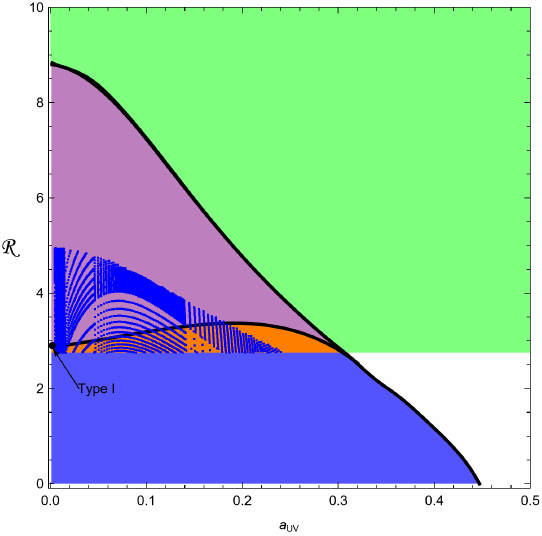}	
				\caption{}\label{ov4}
			\end{subfigure}
			\hspace{0.5cm}
			\begin{subfigure}{0.45\textwidth}
				\includegraphics[width=\textwidth]{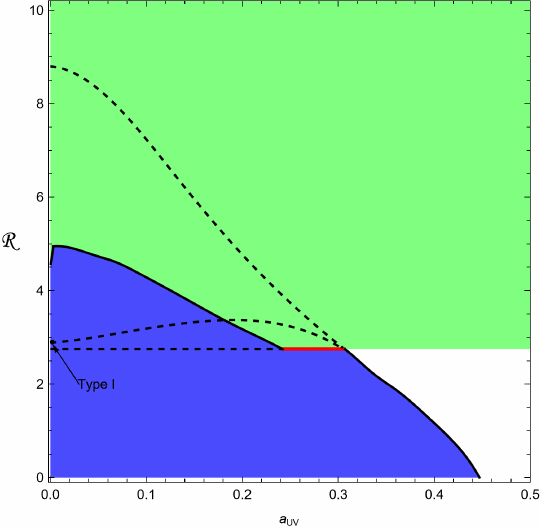}	
				\caption{}\label{ov5}
			\end{subfigure}
		\end{center}
		\caption{\footnotesize{(a) Numerical data (blue points) have been incorporated into Figure \ref{ov3} to highlight the overlap regions.
				(b) These points from panel (a) have been extrapolated, indicating the blue region where Type II solutions dominate and the green region corresponding to Type III. The transition boundary consists of two segments: along the black curve, the free energy remains continuous, whereas it becomes discontinuous along the red line.
		}}
	\end{figure}

	To better understand the overlap regions between different solution types, we have incorporated numerical data
	into Figure \ref{ov3}, represented by blue points in figure \ref{ov4}. These data points allow us to pinpoint the regions
	in which Type II and Type III solutions either coexist or undergo a transition. Figure \ref{ov5} builds upon this
	by extrapolating the numerical data, allowing us to visualize the domains in which each solution type dominates:
	The blue-shaded region primarily hosts Type II solutions, while the green-shaded area corresponds to Type III.

\begin{figure}[htbp]
		\centering
		\includegraphics[width=0.45\linewidth]{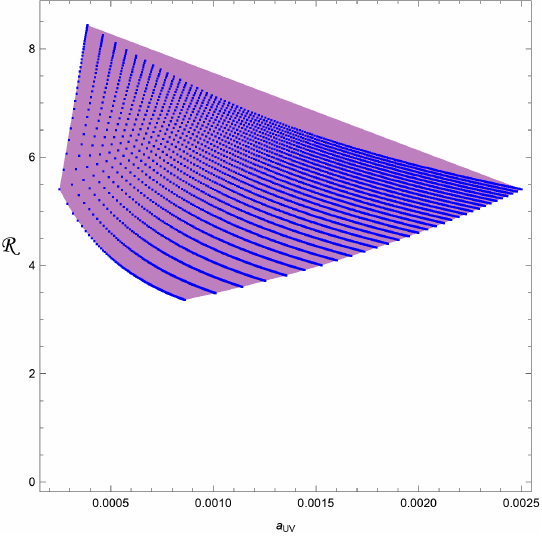}
		\caption{\footnotesize{Zoomed region of figure \ref{ov4} when $a_{\textrm{UV}}\rightarrow 0$. The data belong to the values in the range $10^{-6}<X_\infty<10^{-4}$ and $10^{-4}<Q<10^{-3}$
		}}
		\label{ov6}
	\end{figure}

	It is important to note that in the limit $a_{\textrm{UV}} \rightarrow 0$, the results should match those of figures \ref{zax} or \ref{zaxz}. Specifically, in this limit, Type III solutions should appear above $\mathcal{R} \approx 4.5$, while Type II solutions should exist below that value. To show this,  we have zoomed figure \ref{ov4} very near to the vertical axis. As figure \ref{ov6} shows, near $a_{\textrm{UV}} = 0$, the blue region (dots) is expected to drop and intersect the vertical axis at $\mathcal{R} \approx 4.5$.
	\begin{figure}[htbp]
		\centering
		\includegraphics[width=0.5\linewidth]{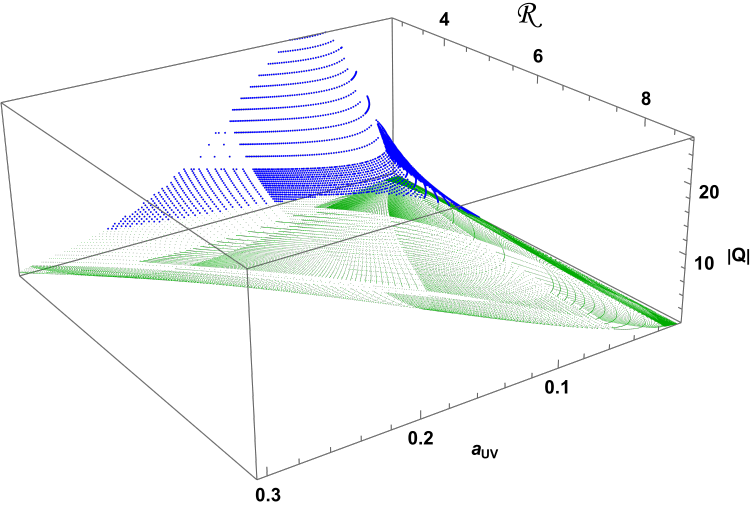}
		\caption{\footnotesize{ Visualization of the competing Type II solutions. The blue dots correspond to the
				energetically favorable Type II configurations with lower free energy and exhibit larger vacuum expectation values
				$|Q|$ compared to the green dots, which represent the less preferred Type II alternatives.
		}}
		\label{Qc}
	\end{figure}
	
	The transition boundary separating Type II and Type III domains comprises two distinct segments. Along the black curve in figure \ref{ov5}, the free energy varies continuously, indicating a smooth crossover between solution types (first-order phase transition as we show below). In contrast, the red line marks a sharp boundary where the free energy exhibits a discontinuity, signaling a problem we shall discuss later on.
	
	An interesting observation arises in the region of competition between two Type II solutions and one Type III solution, corresponding to the purple area in Figure \ref{ov3}. In cases where the preferred solution, i.e., the one with lower free energy, is a Type II solution (indicated by blue points in figure \ref{ov4}), its associated vev $|Q|$ is larger compared to the other competing Type II solution. This behavior is illustrated in the Figure \ref{Qc}, where the blue dots represent the energetically favored Type II solutions, exhibiting higher values of $|Q|$ than the alternative Type II solutions shown as green dots.

	\subsubsection{A first order phase transition}\label{fot}

\begin{figure}[htbp]
		\begin{center}
			\begin{subfigure}{0.475\textwidth}
				\includegraphics[width=\textwidth]{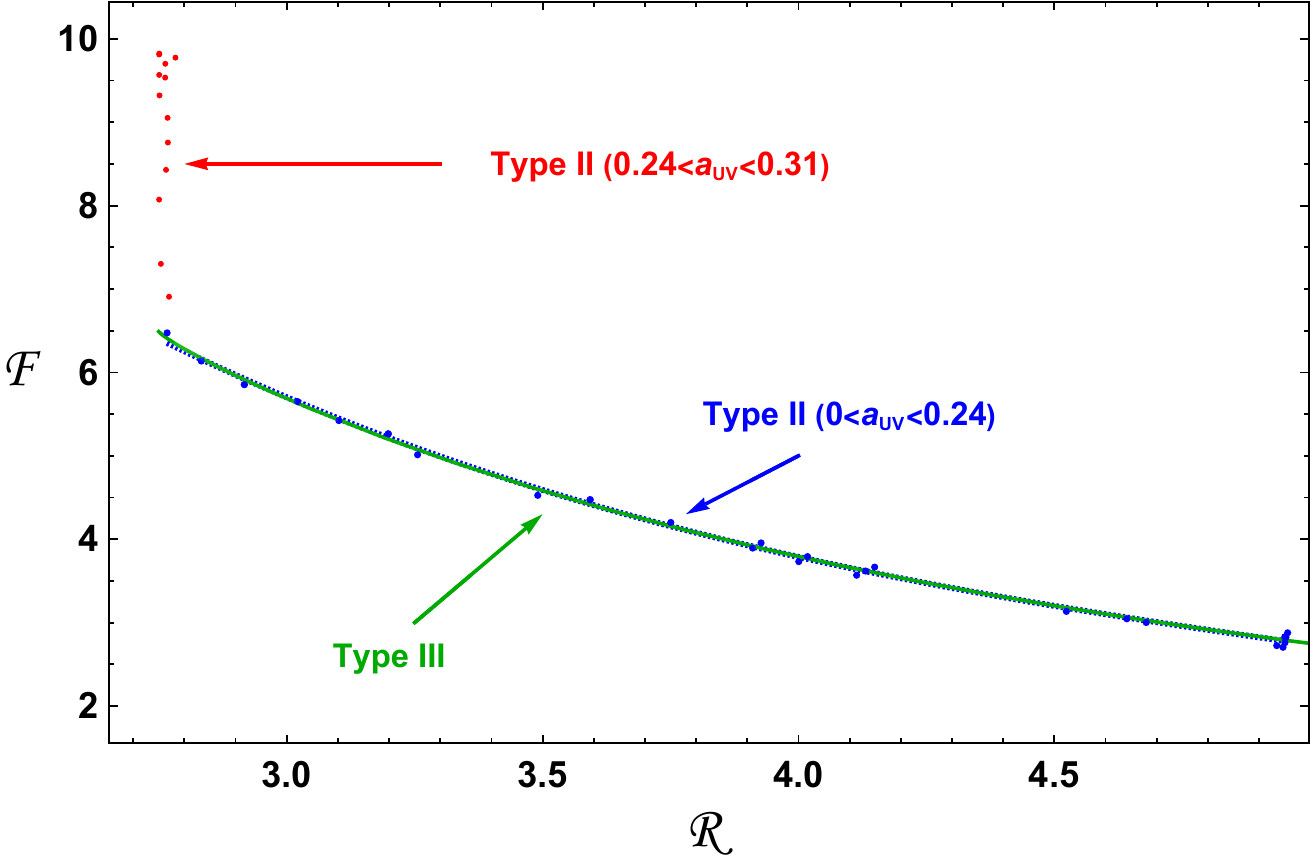}	
				\caption{}\label{tr1}
			\end{subfigure}
			\hspace{0.2cm}
			\begin{subfigure}{0.495\textwidth}
				\includegraphics[width=\textwidth]{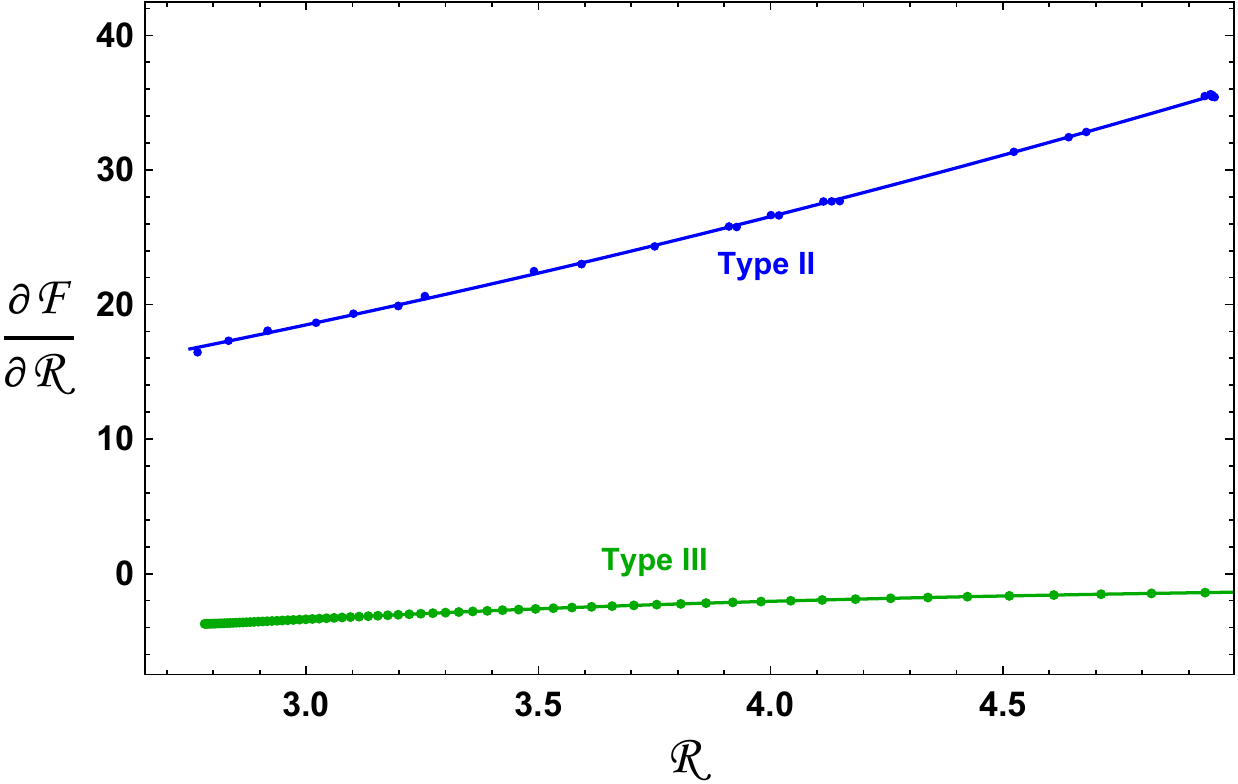}	
				\caption{}\label{tr2}
			\end{subfigure}
		\end{center}
		\caption{\footnotesize{
				Free energy $\mathcal{F}$ (a) and its derivative with respect to the dimensionless curvature $\mathcal{R}$ (b) evaluated along the black and red boundary curve separating Type II and Type III regions in Fig.~\ref{ov5}. In panel (a), the blue branch corresponds to Type II solutions in the range $0<a_{\textrm{UV}}<0.24$, where the free energies of Type II and Type III coincide, while the green branch represents Type III solutions. The red points show a disconnected Type II branch appearing for $0.24<a_{\textrm{UV}}<0.31$, as indicated by a red line in Fig.~\ref{ov5}. Panel (b) shows the corresponding behavior of $\partial\mathcal{F}/\partial\mathcal{R}$, exhibiting a finite jump across the transition. This discontinuity confirms that the black curve separating the blue and green regions in Fig.~\ref{ov5} corresponds to a first-order phase transition between Type II and Type III solutions.
		}}
	\end{figure}

\begin{figure}[htbp]
		\centering
		\includegraphics[width=0.49\linewidth]{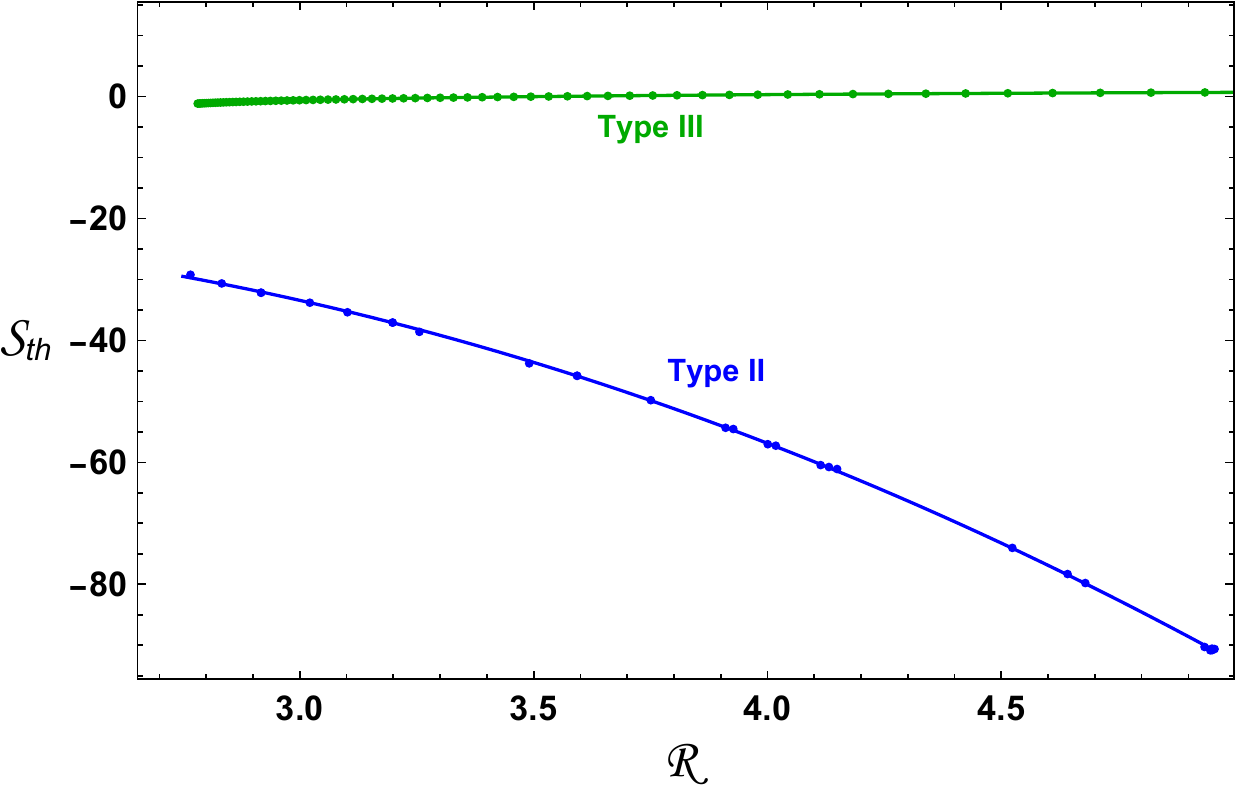}
		\caption{\footnotesize{Renormalized thermal (Entanglement) entropy at the phase transition points.
		}}
		\label{sth}
	\end{figure}

	To illustrate the nature of the transition, Figures \ref{tr1} and \ref{tr2} show the behavior of the free energy and its first partial derivative with respect to the dimensionless curvature at the transition points
	\begin{itemize}
		\item As shown in Figure \ref{tr1}, for the range $0 < a_{\textrm{UV}} < 0.24$, corresponding to the points on the black curve in Figure \ref{ov5}, the free energy values of Type II (blue points) and Type III (green points) solutions coincide.
		
		\item In the interval $0.24 < a_{\textrm{UV}} < 0.31$, represented by points on the red line in Figure \ref{ov5}, the free energy of Type II solutions exhibits a discontinuity, visible as red dots in Figure \ref{tr1}.
		
		\item By examining the first derivative of the free energy with respect to $\mathcal{R}$ (by applying relation \eqref{FEP6}), we arrive at Figure \ref{tr2}. This clearly indicates that the phase transition occurring along the black curve boundary in Figure \ref{ov5} is of first order.
	\end{itemize}
	Finally, we compare the renormalized thermal entropy at the phase transition point using the expression in equation \eqref{FEP3}. This comparison is illustrated in Figure \ref{sth}, where thermal entropy $\mathcal{S}_{th}$ is plotted as a function of $\mathcal{R}$. The entropy difference at each fixed value of $\mathcal{R}$ corresponds to the latent heat.

	\subsubsection{Below the Efimov bound}\label{ptb}

\begin{figure}[htbp]
		\centering
		\includegraphics[width=0.55\linewidth]{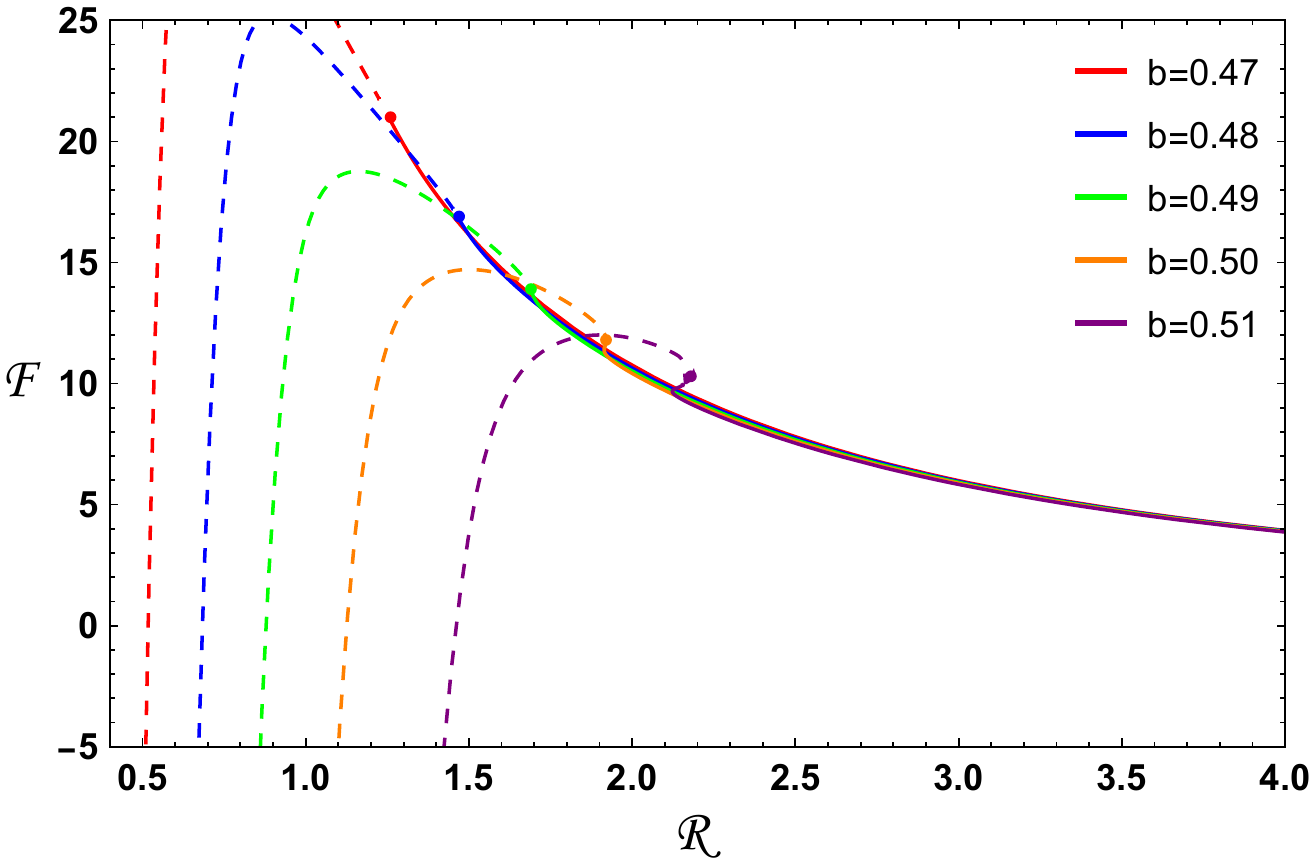}
		\caption{\footnotesize{Comparison of free energies for Type II and Type III solutions as functions
				of the UV parameter $\mathcal{R}$, illustrating the influence of $b$ on spiral behavior below the Efimov
				bound. The colored dots are Type I solutions.}}
		\label{bfr}
	\end{figure}

	Assuming the zero axion limit and working below the Efimov bound, one can compare the free energies of the Type II and Type III solutions.
	As we have already shown in section \ref{beb}, the spiral behavior is sensitive to the value of $b$.
	To explore the free energy profiles of the Type II and Type III solutions as functions of the UV parameter
	$\mathcal{R}$, we have presented figure \ref{bfr}. This figure illustrates that as the value of $b$ decreases,
	the spiral behavior characteristic of the Efimov effect gradually vanishes.
	\begin{figure}[!htpb]
		\begin{center}
			\begin{subfigure}{0.48\textwidth}
				\includegraphics[width=\textwidth]{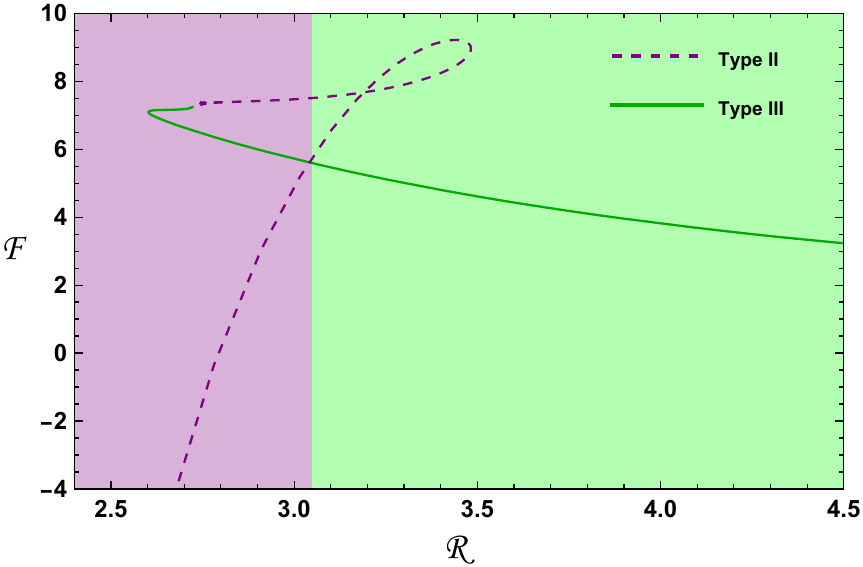}	
				\caption{$b(n=4.25)=\sqrt{\frac{29}{102}}\simeq 0.533$}\label{n425}
			\end{subfigure}
			\begin{subfigure}{0.48\textwidth}
				\includegraphics[width=\textwidth]{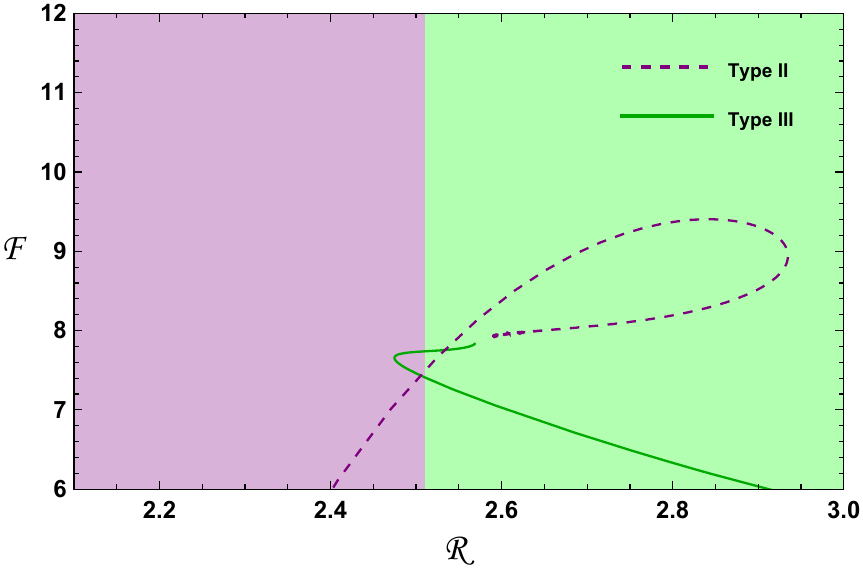}	
				\caption{$b(n=4.5)=\frac13\sqrt{\frac52}\simeq 0.527$}\label{n92}
			\end{subfigure}
			\begin{subfigure}{0.48\textwidth}
				\includegraphics[width=\textwidth]{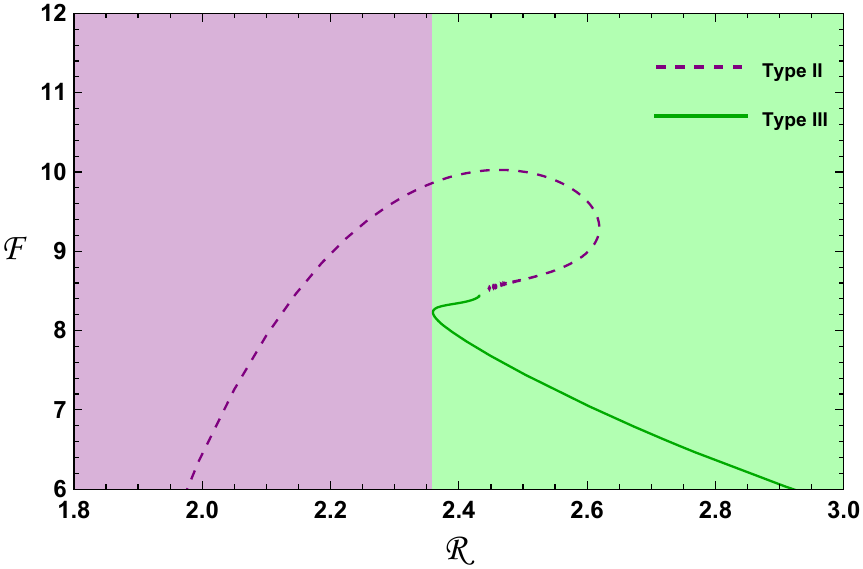}	
				\caption{$b(n=4.75)={\sqrt{\frac{31}{114}}}\simeq 0.521$}\label{n475}
			\end{subfigure}
			\begin{subfigure}{0.49\textwidth}
				\includegraphics[width=\textwidth]{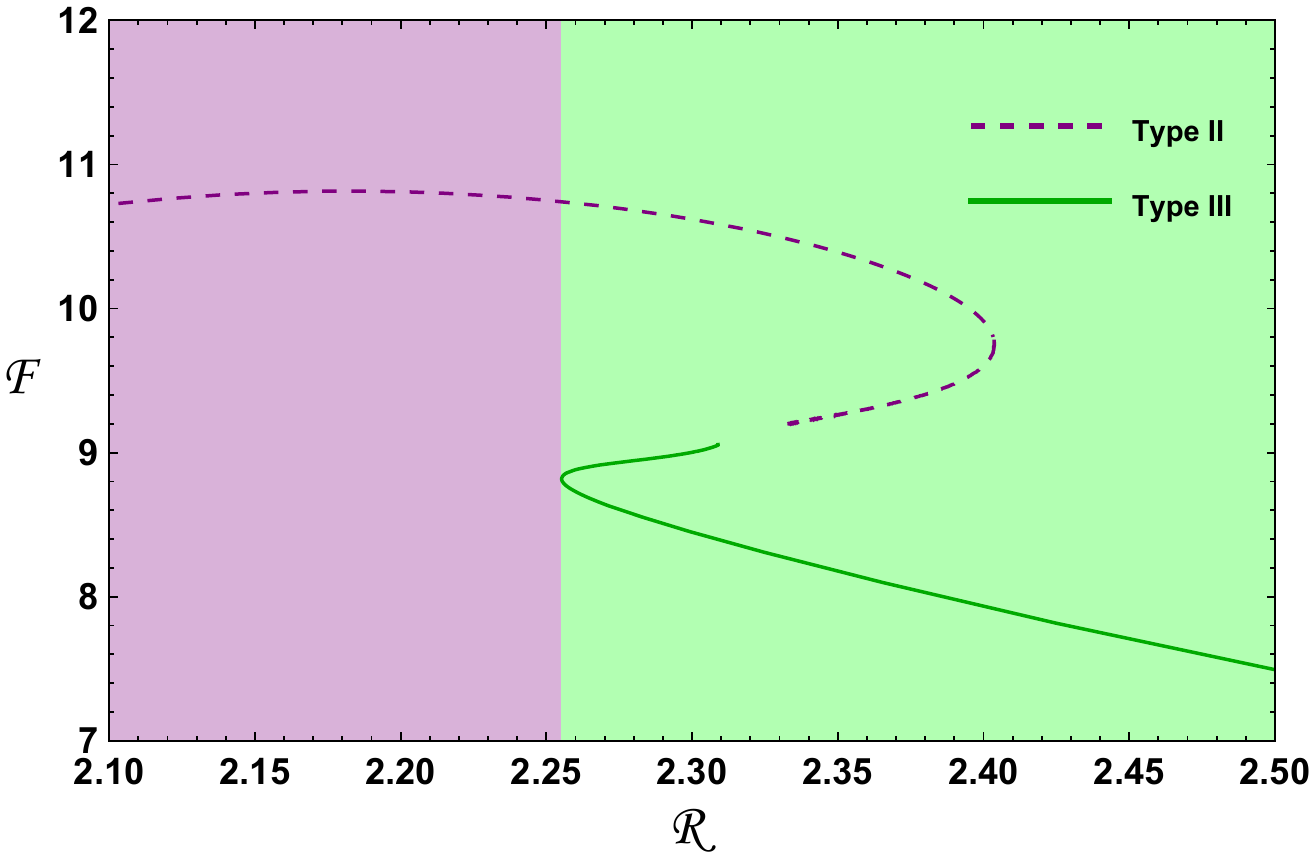}	
				\caption{$b(n=5)=b_E=\frac{2}{\sqrt{15}}\simeq 0.516$}\label{bzaxc}
			\end{subfigure}
			\begin{subfigure}{0.48\textwidth}
				\includegraphics[width=\textwidth]{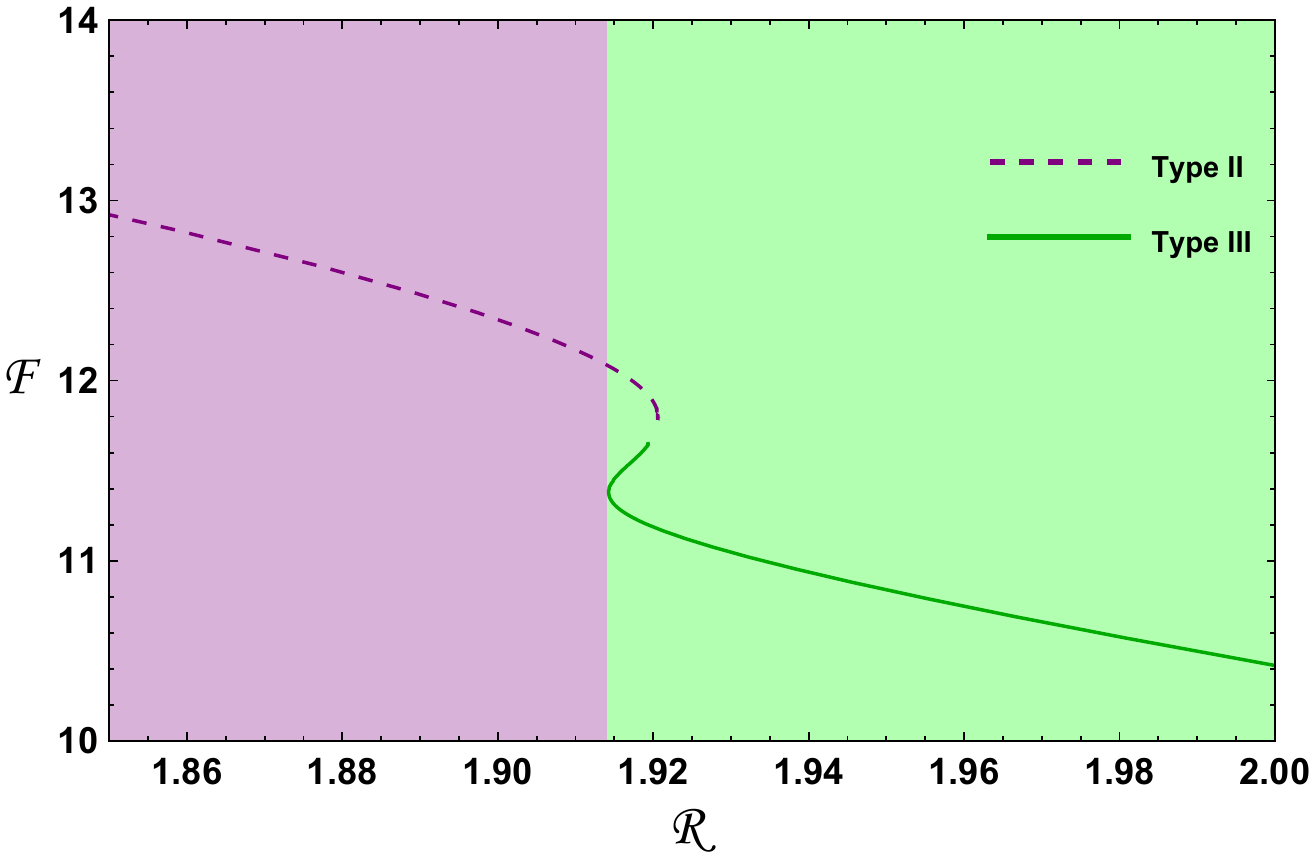}	
				\caption{$b(n=6)=0.5$}\label{bzax}
			\end{subfigure}
			\begin{subfigure}{0.48\textwidth}
				\includegraphics[width=\textwidth]{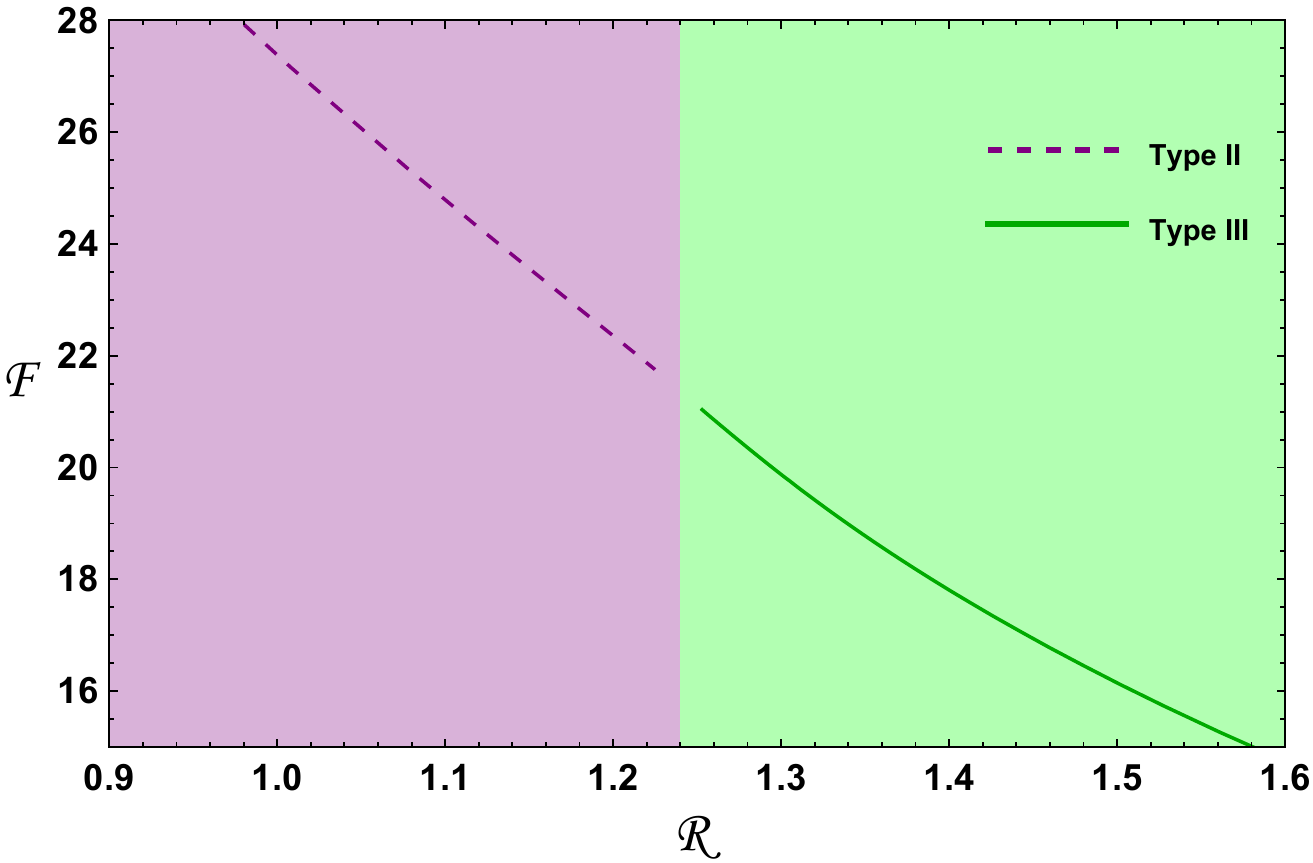}
				\caption{$b(n=9)=\frac{\sqrt{2}}{3}\simeq 0.471$}\label{bzaxb}
			\end{subfigure}
		\end{center}
		\caption{\footnotesize{
				Phase transition between Type II and Type III solutions slightly above, at, and below the Efimov bound, with a trivial axion profile.
				In figures (a) and (b) $b$ is just above the Efimov bound. The theory exhibits a first-order phase transition with a continuous free energy. In figure (c), $b$ is above the Efimov bound, but the theory exhibits a phase transition with a discontinuous free energy.
				In figure (d), we are at exactly the Efimov bound, $b=2/\sqrt{15}$. The theory still exhibits a phase transition with a discontinuous free energy.
				In figure (e), we are just below the Efimov bound. Similarly, here there is a transition with a discontinuous free energy.
				In figure (f), we are well below the Efimov bound.  Here we have a second-order transition with a continuous free energy.
		}}
		\label{below}
	\end{figure}
	
	To be more precise, the effect of the different values of $b$ is shown in figures \ref{n425} to \ref{bzaxb}, where the free energy is plotted as a function of the dimensionless curvature.

	Figures \ref{n425} to \ref{bzaxb} display the phase structure between Type II (dashed) and Type III (solid) solutions for different values of $b(n)$ around the Efimov bound. The shaded regions indicate the domains where each branch exists.
	
	In figures \ref{n425} and \ref{n92}, which lie above the Efimov bound, the system exhibits a characteristic spiral structure. For curvatures below a critical value (purple region), the Type II solution has lower free energy, while above this critical value (green region), the Type III solution becomes thermodynamically preferred. The transition is accompanied by a continuity in the free energy, indicating a first-order phase transition.
	
	In figure \ref{n475}, still above the Efimov bound, the spiral becomes less pronounced, and the free energy is discontinuous. In figure \ref{bzaxc}, corresponding to the Efimov bound, the structure simplifies further; nevertheless, the transition continues to exhibit a discontinuity in the free energy. In figure \ref{bzax}, slightly below the Efimov bound, the spiral structure has essentially disappeared, and the branches become smoother, yet the transition remains discontinuous.
	
	This suggests that we may be missing other possible solutions in this case.   One obvious possibility is that one of the many fields that do not appear in simplified bottom-up actions to obtain a vev and provide the missing dominant solution. We shall comment on such possibilities in the last section.
	
	Finally, in figure \ref{bzaxb}, well below the Efimov bound, the behavior changes qualitatively. The two branches merge smoothly without a coexistence region, and the free energy becomes continuous across the transition, signaling a second-order phase transition.

\begin{figure}[!htbp]
		\begin{center}
			\begin{subfigure}{0.49\textwidth}
				\includegraphics[width=\textwidth]{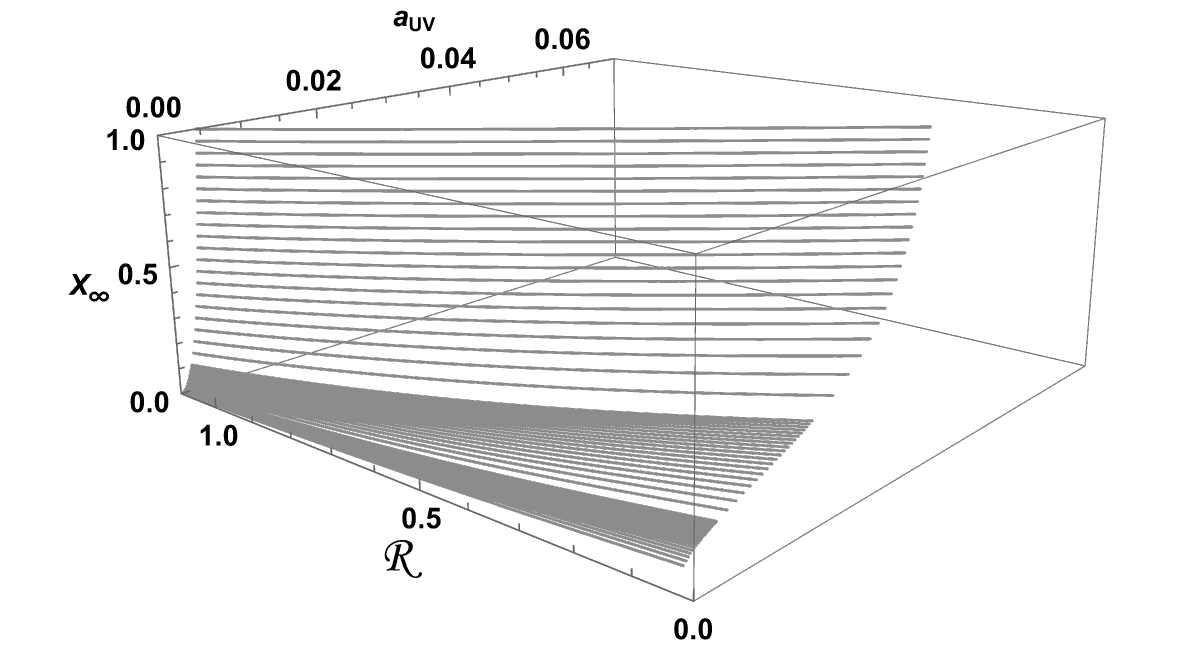}	
				\caption{}\label{bxar}
			\end{subfigure}
			\begin{subfigure}{0.45\textwidth}
				\includegraphics[width=\textwidth]{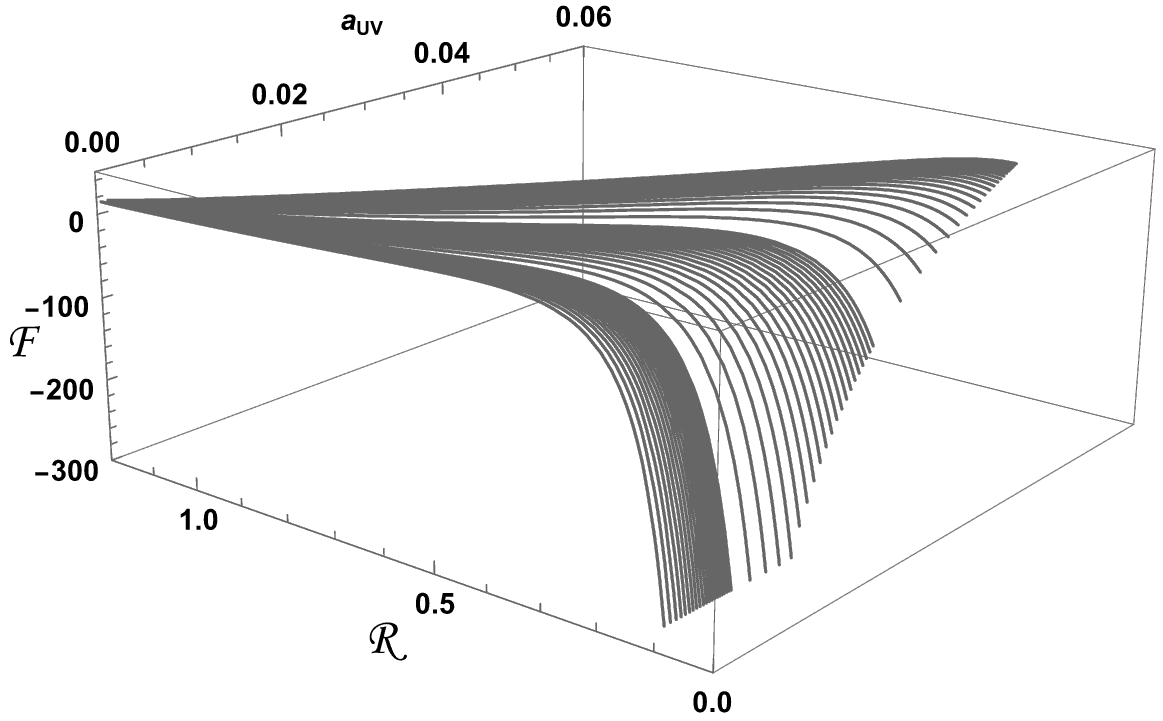}
				\caption{}\label{bfar}
			\end{subfigure}
		\end{center}
		\caption{\footnotesize{(a): Variation of the source values (${\cal R},a_{\textrm{UV}}$) with the IR parameter $X_\infty$. Each curve is traced by changing $\ss$ (or equivalently $Q$) at fixed $X_\infty$.
				(b): Free energy density for the solutions in (a). In contrast to the case above the Efimov bound, the free energy here is single-valued, with a unique Type II solution for each set of sources.
		}}
	\end{figure}

	In the presence of the axion, the behavior of the parameters of the solutions is illustrated in figures \ref{bxar} and \ref{bfar}. Figure \ref{bxar} shows how the UV sources, $(\mathcal{R}, a_{\textrm{UV}})$, behave as functions of the IR parameters $X_\infty$ and $|\ss|$ . Along each curve, the other IR parameter $\ss$ is varied while keeping $X_\infty$ fixed. For small $|\ss|$, all curves converge to $a_{\textrm{UV}} = 0$ at a common finite value of $\mathcal{R}$. As $|\ss|$ increases, each curve shows that $\mathcal{R}$ eventually decreases to zero, while $a_{\textrm{UV}}$ increases and approaches a finite maximum.
	\begin{figure}[htbp]
		\begin{center}
			\begin{subfigure}{0.48\textwidth}
				\includegraphics[width=\textwidth]{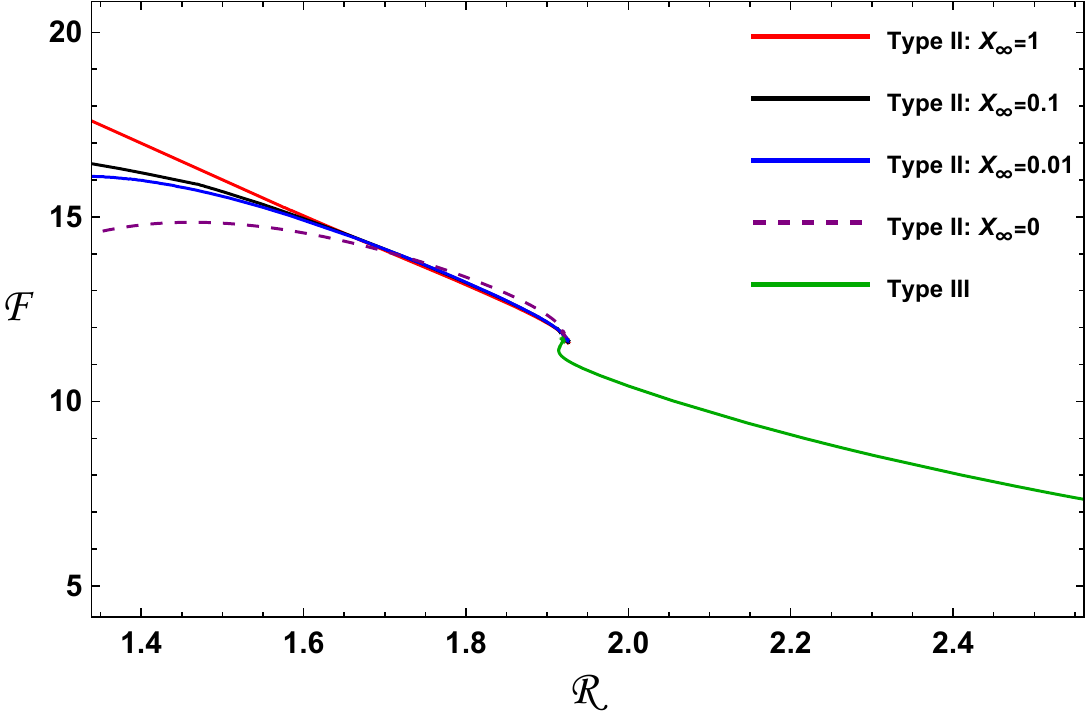}	
				\caption{b=0.5}\label{bfallz}
			\end{subfigure}\hspace{0.2cm}
			\begin{subfigure}{0.48\textwidth}
				\includegraphics[width=\textwidth]{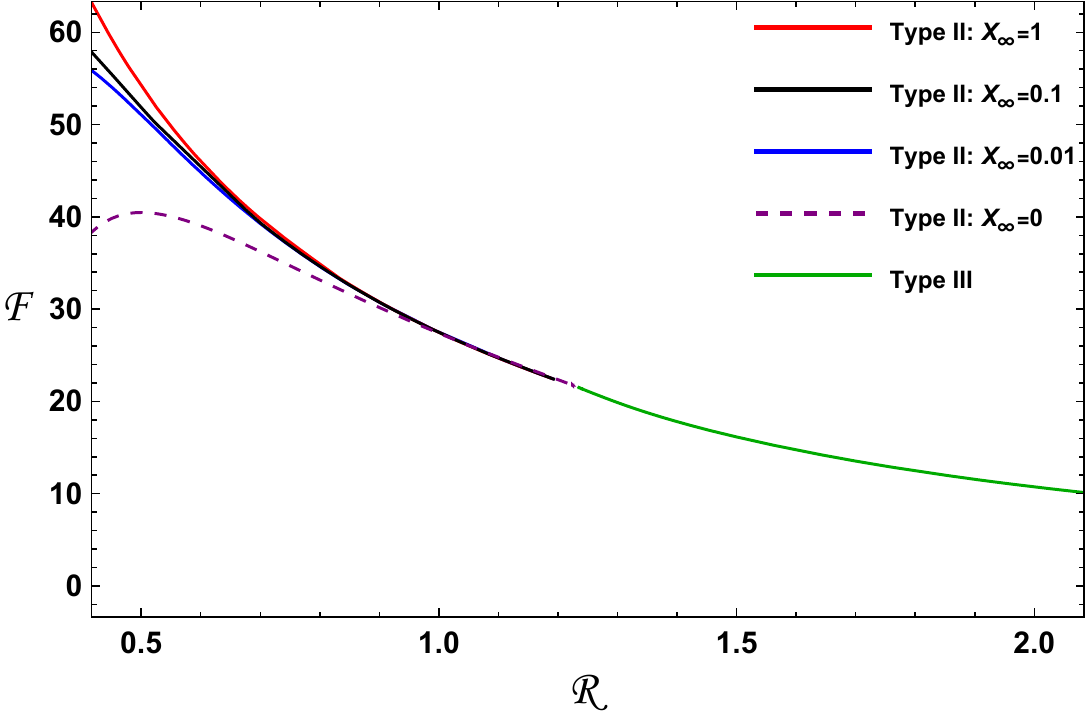}	
				\caption{b=0.47}\label{bfall}
			\end{subfigure}
		\end{center}
		\caption{\footnotesize{(a): Free energy as a function of $\mathcal{R}$ for three representative values of $X_\infty = 0.01$, $0.1$, and $1$.
				To compare, we have included Type II at zero axion limit and Type III as well.}}
	\end{figure}
	\begin{figure}[htbp]
		\centering
		\includegraphics[width=0.5\linewidth]{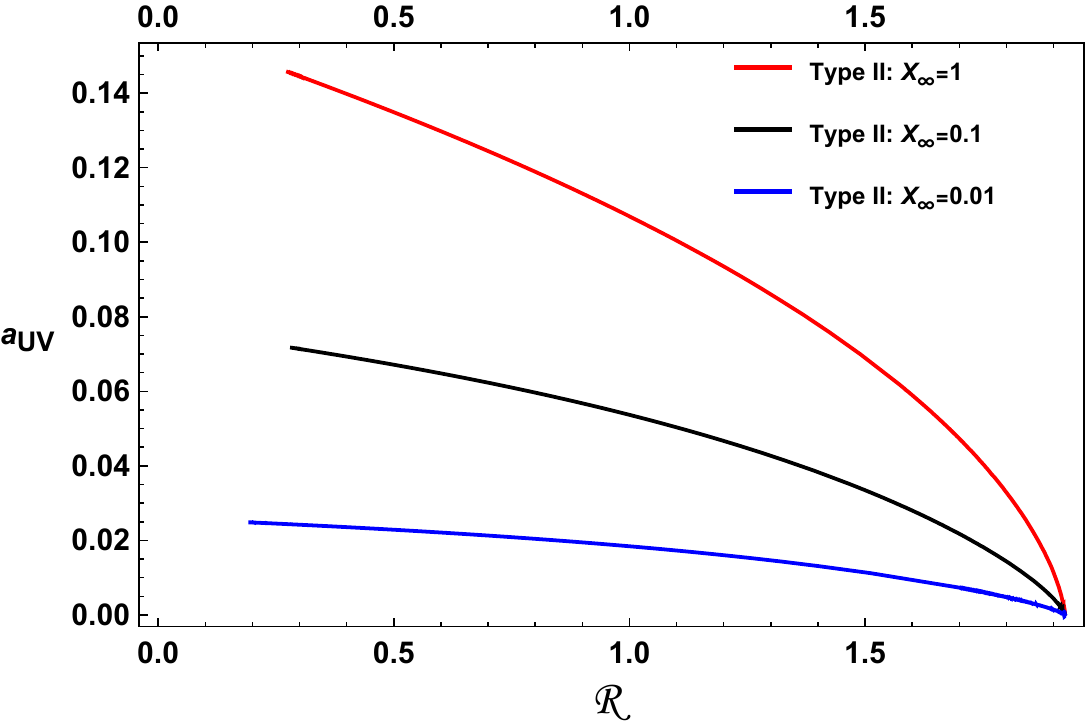}
		\caption{\footnotesize{The source space for three representative sets of solutions with $X_\infty = 0.01$, $0.1$, and $1$ is shown.
				Except at the Type I joining point, there is no overlap in the source space; each solution is uniquely determined.}}
		\label{bar}
	\end{figure}
	
	Figure \ref{bfar} displays the corresponding free energy density for the same set of solutions shown in Figure \ref{bxar}. As $|\ss|$ decreases, all curves converge to a single point, corresponding to the Type I solution. For increasing $|\ss|$, the free energy consistently decreases, regardless of the value of $X_\infty$.
	\begin{figure}[htbp]
		\begin{center}
			\begin{subfigure}{0.48\textwidth}
				\includegraphics[width=\textwidth]{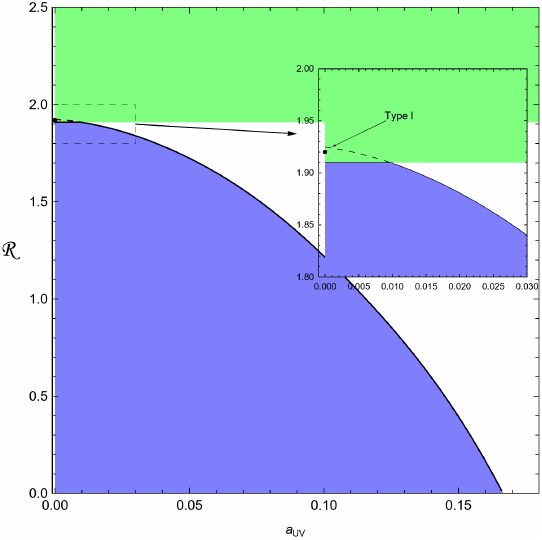}	
				\caption{b=0.5}\label{bov3}
			\end{subfigure}\hspace{0.2cm}
			\begin{subfigure}{0.495\textwidth}
				\includegraphics[width=\textwidth]{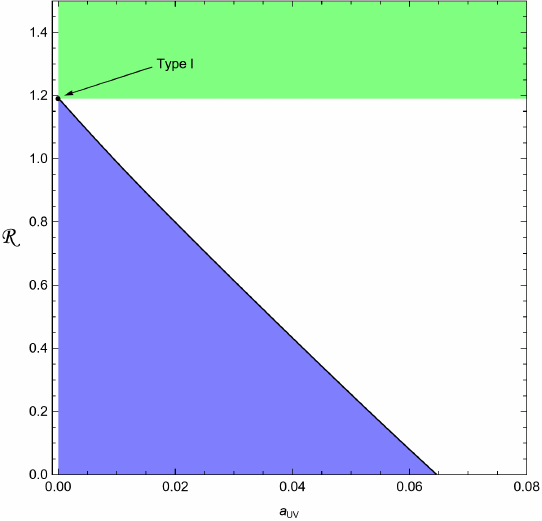}
				\caption{b=0.47}\label{bov4}
			\end{subfigure}
		\end{center}
		\caption{\footnotesize{
				Phase diagram illustrating regions characterized by unique Type II (blue) and Type III (green) solutions. In the narrow overlap region, such as in case (a), both solutions coexist, with Type III being energetically preferred. In case (b), no overlap region is present, except at the junction point, where a Type I solution emerges.
		}}
	\end{figure}
	
	A key distinction from the previous case (above the Efimov bound in figures \ref{xar} and \ref{far}) is that here the free energy is a single-valued function of the sources. That is, for each point in the space of couplings,  $(\mathcal{R}, a_{\textrm{UV}})$, of Type II solutions, there exists a unique solution.

	To further illustrate this behavior, we consider four specific sets of solutions with $X_\infty = 0, 0.01, 0.1$, and $1$. Figures \ref{bfallz} and \ref{bfall} show how the free energy varies as a function of $\mathcal{R}$ for two different values of $b$. Each point on these curves corresponds to a different value of $a_{\textrm{UV}}$, as shown in figure \ref{bar}. There is no overlap in the space of sources, indicating that each solution is distinct.

	Similar to the previous case, here we should analyze the free energies of different solutions in the phase space of sources. The summary of the results are depicted in figure \ref{bov3} and \ref{bov4} and explained as follows:
	
	\begin{itemize}
		\item For $b$ above a critical threshold, there consistently exists an overlap region between Type II and Type III solutions. Within this region, Type III exhibits a lower free energy, refer to figure \ref{bov3}. The free energy displays a discontinuity at $a_{\textrm{UV}}=0$ as we discussed above.

		\item For $b$ below the critical threshold, as Figure \ref{bov4} shows, no overlap region exists between Type II and III solutions, except at $a_{\textrm{UV}}=0$. At this point, a Type I solution emerges, and the free energy remains continuous.
		
	\end{itemize}

	\section{Axionic solutions at negative curvature}\label{SRS}

	In this section, we investigate axionic solutions in backgrounds with negative slice curvature, focusing on the possible
	configurations that connect IR and UV regions.
	
	Unlike the positively curved cases discussed previously,
	these geometries exhibit distinct boundary structures. Depending on the boundary conditions and the
	presence of an $A$-bounce, the regular solutions can be classified into two main categories: those connecting
	two UV boundaries (UV-UV) and those extending from a regular IR end-point to a UV fixed point (IR-UV).
	The UV-UV solutions have the holographic interpretation as holographic interfaces. The IR-UV solutions
	correspond to single holographic theories on a constant negative curvature manifold.
	Below, we summarize the general features and parameter spaces associated with each class:
	
	\begin{itemize}
		\item {\bf{UV-UV:}} The UV-UV solution refers to solutions extending between two UV fixed points at the same location,
		$\f = 0 $\footnote{In general, such solutions can connect different maxima of the bulk scalar potential as was
			shown in \cite{Ghodsi:2022umc}. However, here, our potential has a single maximum, and therefore such solutions
			return to the same maximum.}. Since an A-bounce is required for such solutions, the parameter space of
		the UV-UV solution forms a subset of the space of possible A-bounces. As discussed in appendix \ref{EB}, an A-bounce
		belongs to a three-parameter family characterized by $\f_0$, $S_0 $, and $Q$, subject to the inequality specified in \eqref{Abc1}.
		The expansion of fields near the A-bounce is described by equations \eqref{AbS}--\eqref{AbT}.
		
		\item {\bf{IR-UV:}} These solutions originate from infinity as a regular IR end-point and extend to a UV fixed point at $\f=0$. The set of such solutions is determined by the IR asymptotic expansions outlined in section \ref{IRAS}. This space is two-dimensional, characterized by the IR parameters
		$\x$ and $\ss$.
	\end{itemize}

	\subsection{UV-UV solutions}\label{UVUV}

	UV-UV solutions define holographic interfaces when the slice geometry is non-compact, while they represent wormholes when it is compact. These solutions extend between two asymptotically AdS boundaries, both situated at the maximum of the potential where $\f = 0$ (this is true for our potential in \eqref{SVN}, see also figure \ref{pots}), but they differ in general in their boundary (UV) sources.
	
\begin{figure}[!t]
		\begin{center}
			\begin{subfigure}{0.32\textwidth}
				\includegraphics[width=\textwidth]{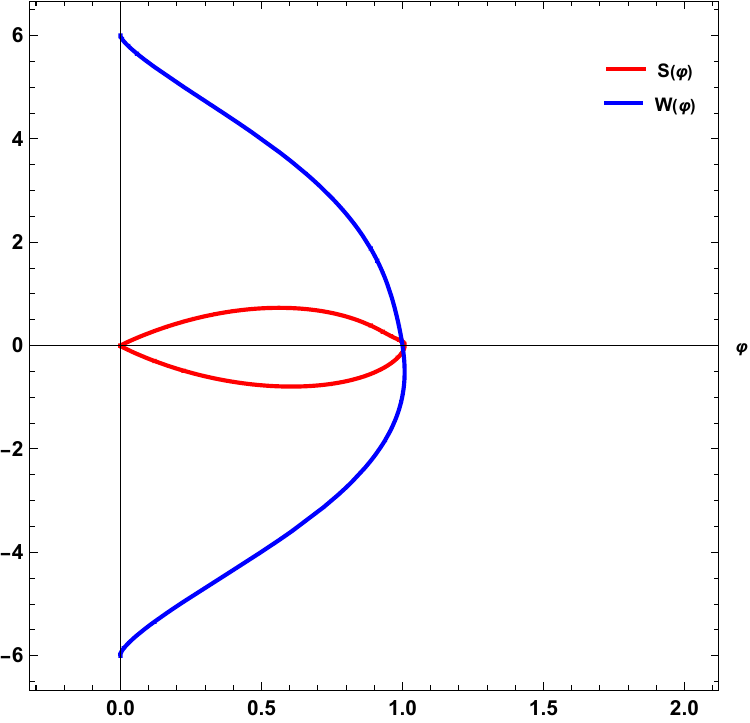}
				\caption{\footnotesize{$ S_0=0.080, Q=2.08$}}
			\end{subfigure}
			\begin{subfigure}{0.32\textwidth}
				\includegraphics[width=\textwidth]{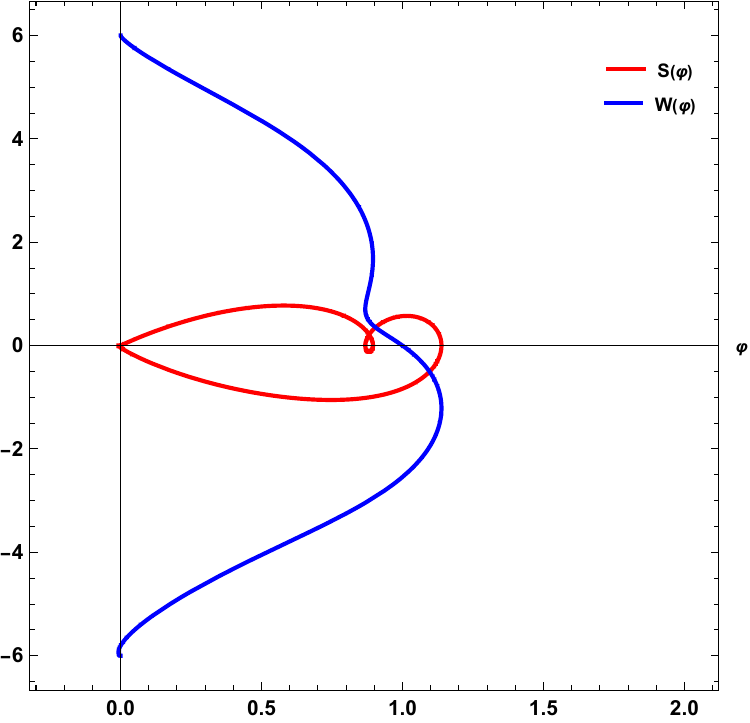}	
				\caption{\footnotesize{$ S_0=0.571, Q=2.08$}}
			\end{subfigure}
			\begin{subfigure}{0.32\textwidth}
				\includegraphics[width=\textwidth]{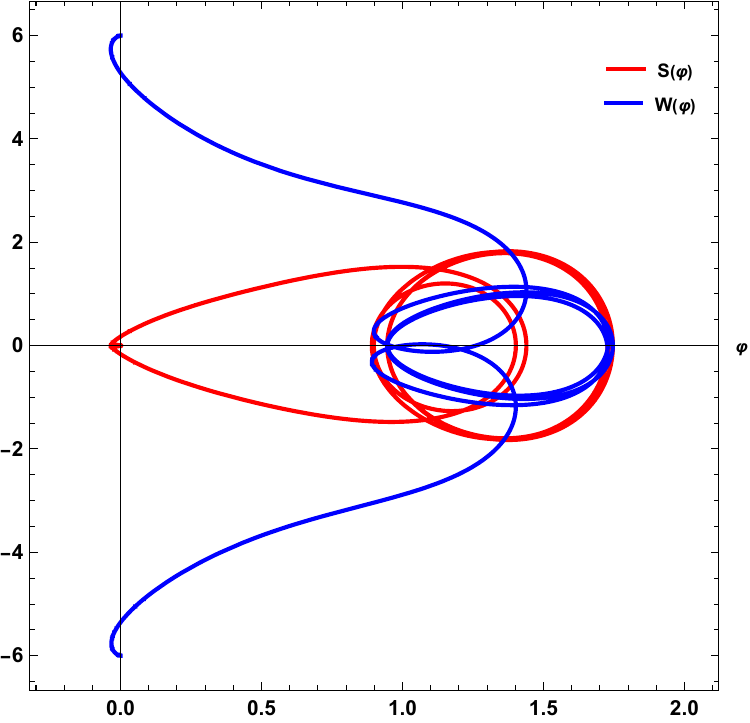}	
				\caption{\footnotesize{$ S_0=0.978, Q=2.08$}}
			\end{subfigure}
		\end{center}
		\begin{center}
			\begin{subfigure}{0.32\textwidth}
				\includegraphics[width=\textwidth]{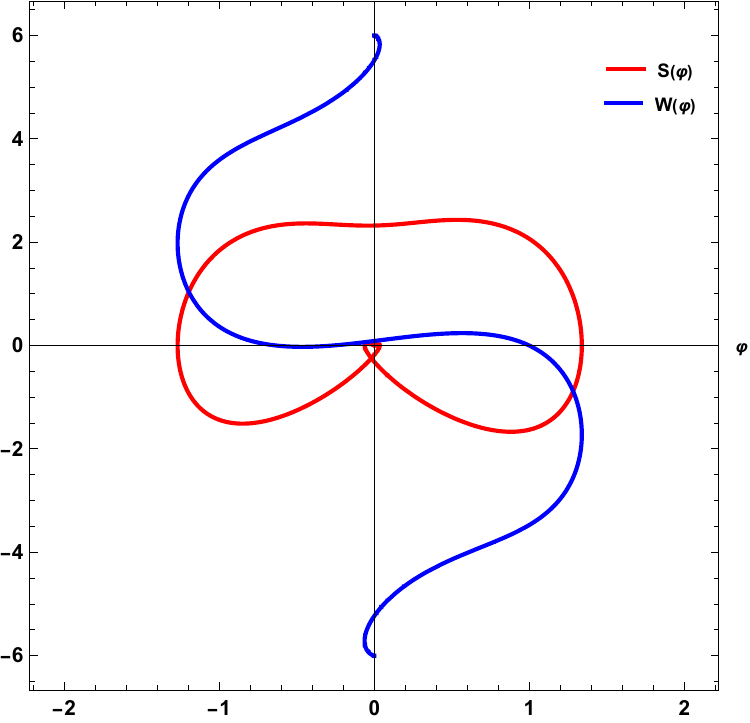}
				\caption{\footnotesize{$ S_0=2.064, Q=1$}}
			\end{subfigure}
			\begin{subfigure}{0.32\textwidth}
				\includegraphics[width=\textwidth]{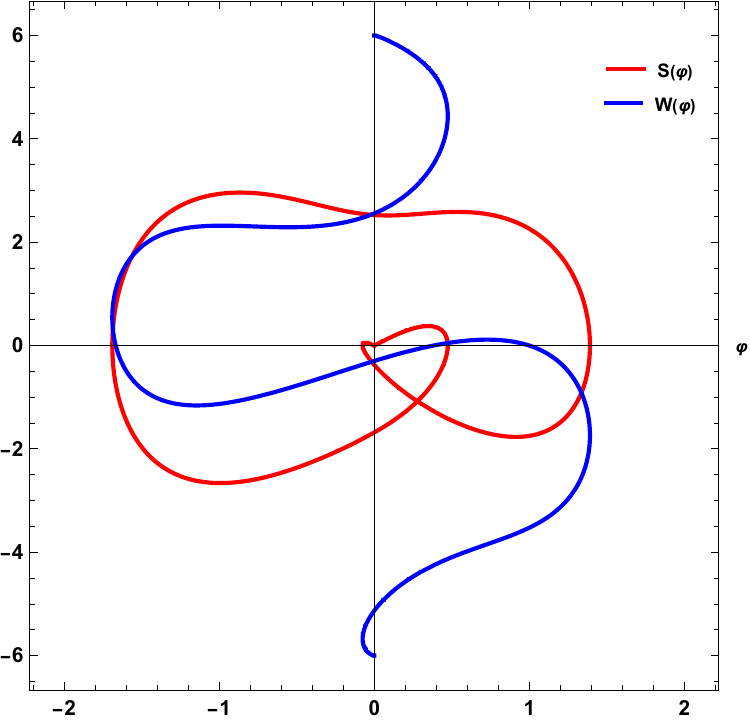}	
				\caption{\footnotesize{$S_0=2.260, Q=1$}}
			\end{subfigure}
			\begin{subfigure}{0.32\textwidth}
				\includegraphics[width=\textwidth]{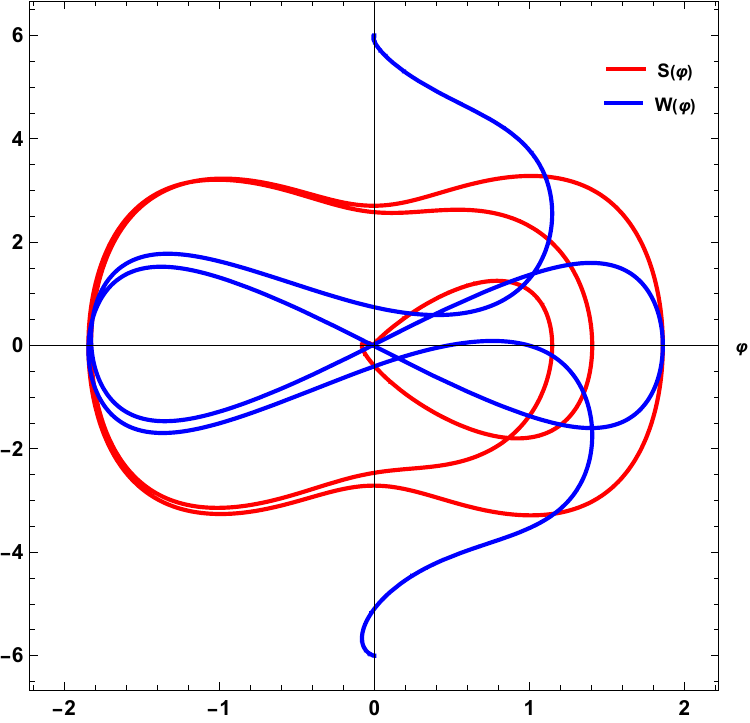}
				\caption{\footnotesize{$ S_0=2.312, Q=1$}}	
			\end{subfigure}
		\end{center}
		\caption{\footnotesize{(a)-(f): The functions $W(\f)$ and $S(\f)$ are examined for six distinct UV-UV solutions for  $b=\sqrt{\frac{7}{24}}$, all sharing an A-bounce at $\f=1$, with varying values of $S_0$ and $Q$.
		}}\label{6fig}
	\end{figure}

	Figure \ref{6fig} illustrates six distinct solutions within this category (in all figures we have fixed $b=\sqrt{7/24}$). Each solution shares the same A-bounce point at $\f_0=1$, with their differences arising from the values of $S_0$ and $Q$. These parameters affect the number of $\f$-bounces, along with other A-bounces present in the solutions. Since UV-UV solutions inherently contain at least one A-bounce, the solution space is defined by three parameters: $\f_0$, $Q$, and $S_0$.
	\begin{figure}[!htbp]
		\centering
		\includegraphics[width=0.45\textwidth]{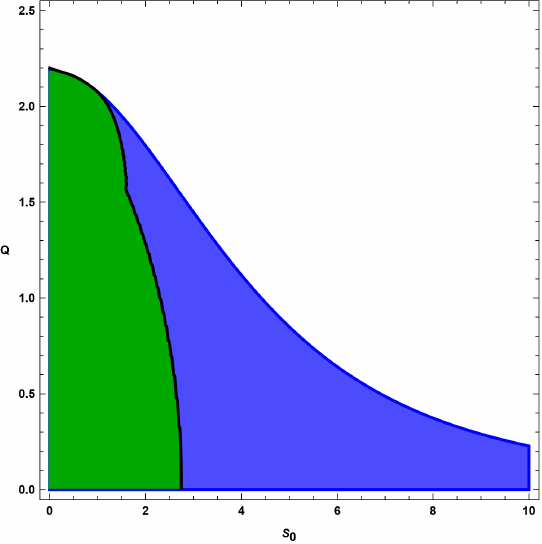}
		\caption{\footnotesize{For the fixed value of $\f_0=1$, the permissible space for UV-UV solutions is limited to the green region ( $b=\sqrt{\frac{7}{24}}$). As one moves toward the black boundary, the number of $\f$-bounces increases indefinitely. Beyond this region, in the blue area, one has either IR-UV or infinity to infinity solutions.
		}}\label{sreg}
	\end{figure}
	
	A-bounces are constrained within the region specified by the inequality in \eqref{Abc1}. In the particular case of $\f_0=1$, this region appears as the union of the green and blue areas in figure \ref{sreg}. However, not all solutions in this area qualify as UV-UV solutions. The green-marked section represents the actual UV-UV solution space. Outside this subset, the blue region consists of IR-UV solutions, which are mostly singular. However, certain points within this area correspond to solutions that remain regular in the IR. These will be examined further in the next section. In the blue region, some solutions lack a UV boundary entirely. This is also the structure found earlier in the absence of an axion field, \cite{Ghodsi:2024jxe}

	As illustrated by solutions (c) and (f) in figure \ref{6fig}, within the green region, moving toward the boundary (the black curve) leads to an increasing number of $\f$-bounces, eventually approaching infinity. In contrast, shifting toward the left reveals solutions characterized by a single A-bounce, as shown in figures (a) and (d).

\begin{figure}[!htpb]
		\begin{center}
			\begin{subfigure}{0.31\textwidth}
				\includegraphics[width=\textwidth]{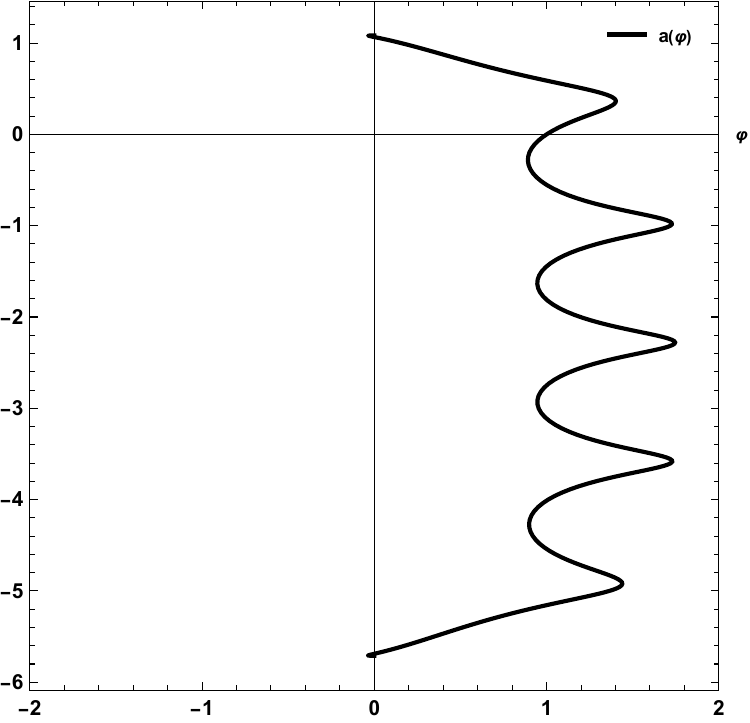}	
				\caption{\footnotesize{$T_0=-19.474, Q=2.08$}}
			\end{subfigure}
			\begin{subfigure}{0.32\textwidth}
				\includegraphics[width=\textwidth]{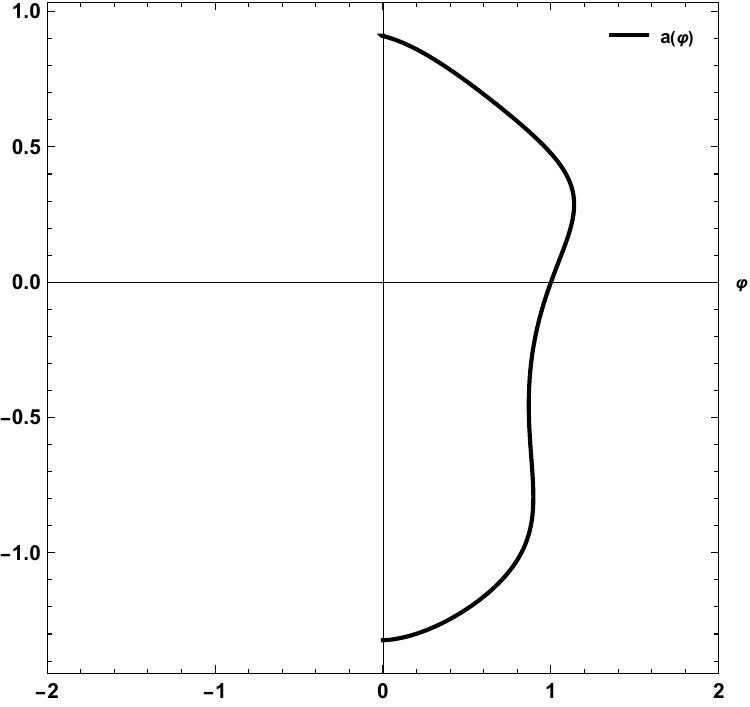}	
				\caption{\footnotesize{$T_0=-17.5, Q=2.08$}}
			\end{subfigure}
			\begin{subfigure}{0.32\textwidth}
				\includegraphics[width=\textwidth]{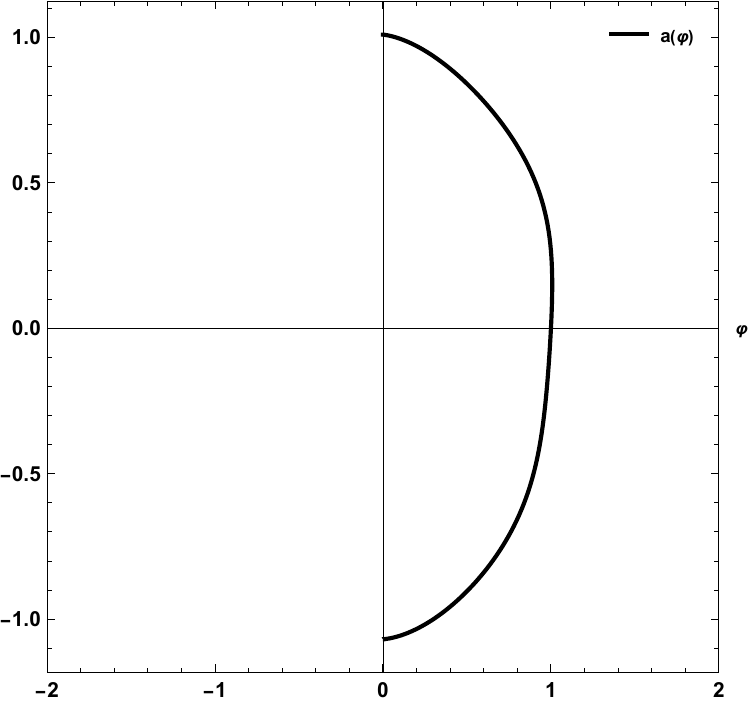}
				\caption{\footnotesize{$T_0=-17, Q=2.08$}}
			\end{subfigure}
		\end{center}
		\begin{center}
			\begin{subfigure}{0.31\textwidth}
				\includegraphics[width=\textwidth]{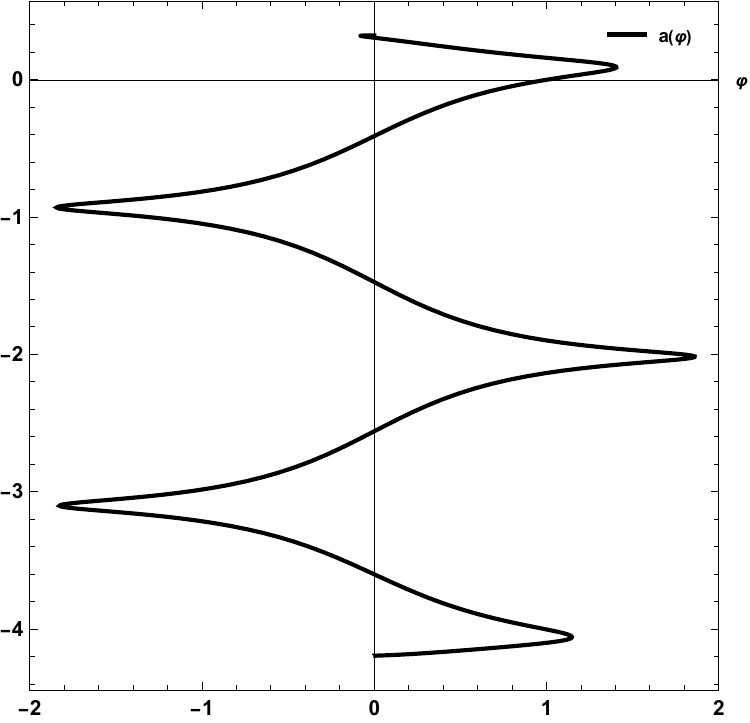}
				\caption{\footnotesize{$T_0=-17.6456, Q=1$}}	
			\end{subfigure}
			\begin{subfigure}{0.32\textwidth}
				\includegraphics[width=\textwidth]{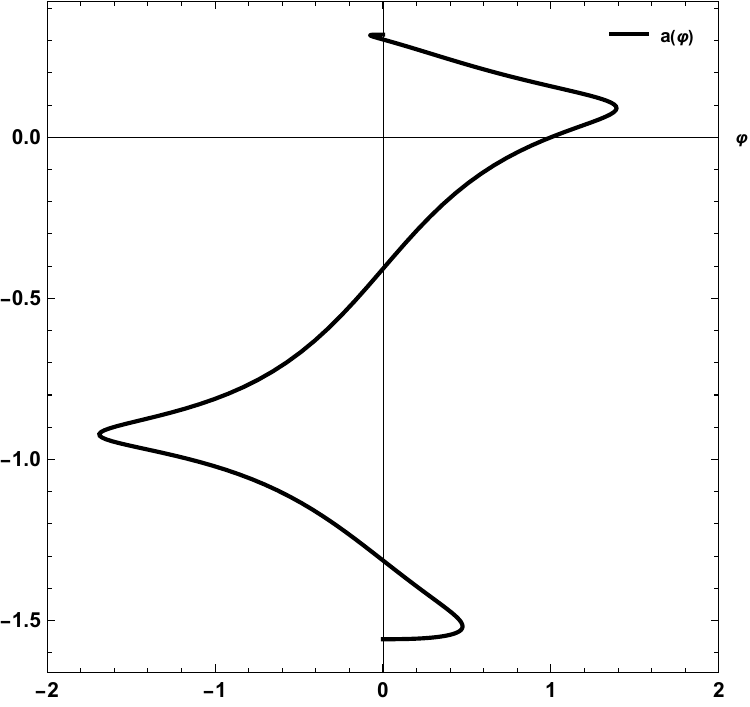}	
				\caption{\footnotesize{$T_0=-17.5, Q=1$}}
			\end{subfigure}
			\begin{subfigure}{0.32\textwidth}
				\includegraphics[width=\textwidth]{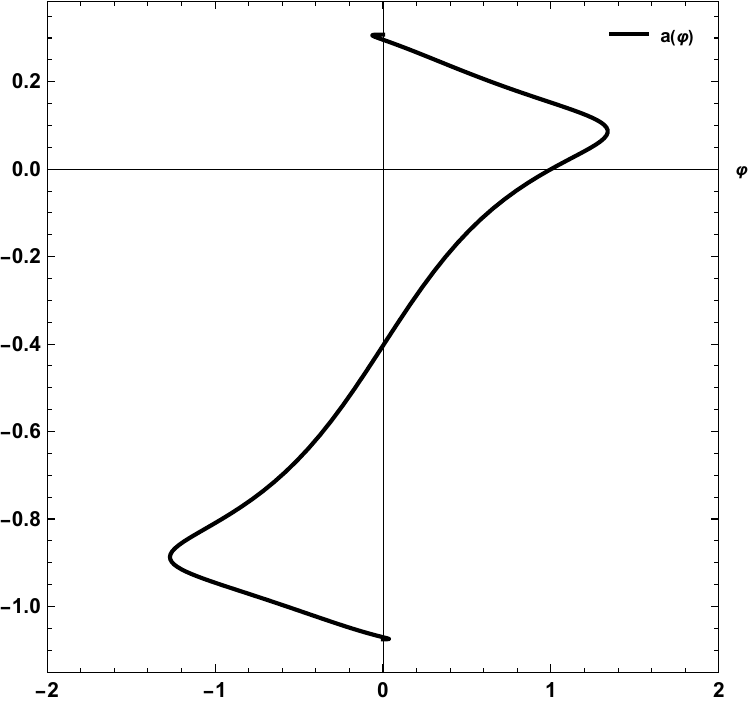}
				\caption{\footnotesize{$T_0=-17, Q=1$}}
			\end{subfigure}
		\end{center}
		\caption{\footnotesize{$a(\f)$ for solutions in figure \ref{6fig}. In all solutions, $a(1)=0$ as an initial value.
		}}\label{6figa}
	\end{figure}

	For the solutions depicted in figure \ref{6fig}, the corresponding axion profiles are illustrated in figure \ref{6figa}. The function $a(\f)$ is derived using equation \eqref{Q} as follows
	\be \label{afi}
	\int_{a_0}^{a} da = \int_{u_0}^{u}  \frac{Q}{ Y e^{4A}} du= \int_{\f_0}^{\f}  \frac{Q}{\dot{\f} Y e^{4A}} d\f\,.
	\ee
	Assuming an A-bounce occurs at $u = u_0$ or equivalently at $\f = \f_0$, we set the initial condition $a_0 = 0$, leading to
	\be \label{afia}
	a(\f) = \int_{\f_0}^{\f}  \frac{Q}{\dot{\f} Y e^{4A}} d\f\,.
	\ee
	As illustrated in figure \ref{6figa}, the values of $a(\f)$ at UV fixed points differ across the solutions. In all cases shown, $a(\f)$ is a monotonic function, satisfying the condition $a(\f=1) = 0$ for every solution.
	
\begin{figure}[!t]
		\begin{center}
			\begin{subfigure}{0.47\textwidth}
				\includegraphics[width=\textwidth]{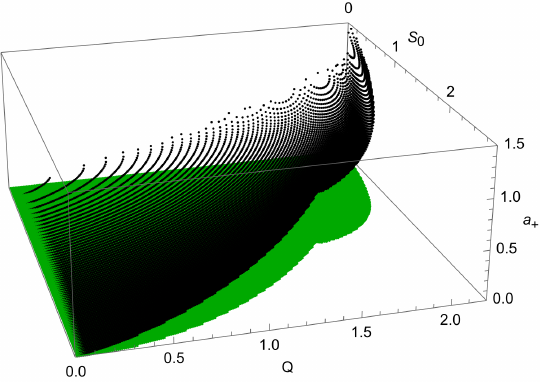}	
				\caption{}\label{axp}
			\end{subfigure}\hspace{0.3cm}
			\begin{subfigure}{0.47\textwidth}
				\includegraphics[width=\textwidth]{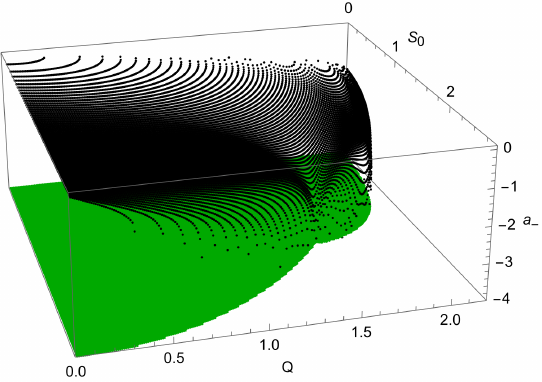}	
				\caption{}\label{axm}
			\end{subfigure}
		\end{center}
		\caption{\footnotesize{The axion field values at the UV fixed points are represented by \( a_+ \) and \( a_- \). In figures (a) and (b), \( a_+ \) and \( a_- \) correspond to the upper UV fixed point at \( u \rightarrow +\infty \) and the lower UV fixed point at \( u \rightarrow -\infty \), respectively. The green shading aligns with the green region in figure \ref{sreg}.}}
	\end{figure}

	To show the possible values of the axion at the UV fixed points, we have presented figures \ref{axp} and \ref{axm}.
	In these illustrations, $a_+$ and $a_-$ represent the upper UV fixed point at $u\rightarrow +\infty$ and the lower
	UV fixed point at $u\rightarrow -\infty$. All values satisfy the conditions $a_+>0$ and $a_-<0$.

	\subsection{IR-UV solutions}\label{IRUV}
	
	In this section, we discuss solutions with one UV boundary. The IR end-point at $\f\to \pm\infty$ is regular (a la Gubser) of Type II asymptotics, where we introduced in the section \ref{SFQ}.
	
\begin{figure}[!t]
		\centering
		\begin{subfigure}{0.32\textwidth}
			\includegraphics[width=\textwidth]{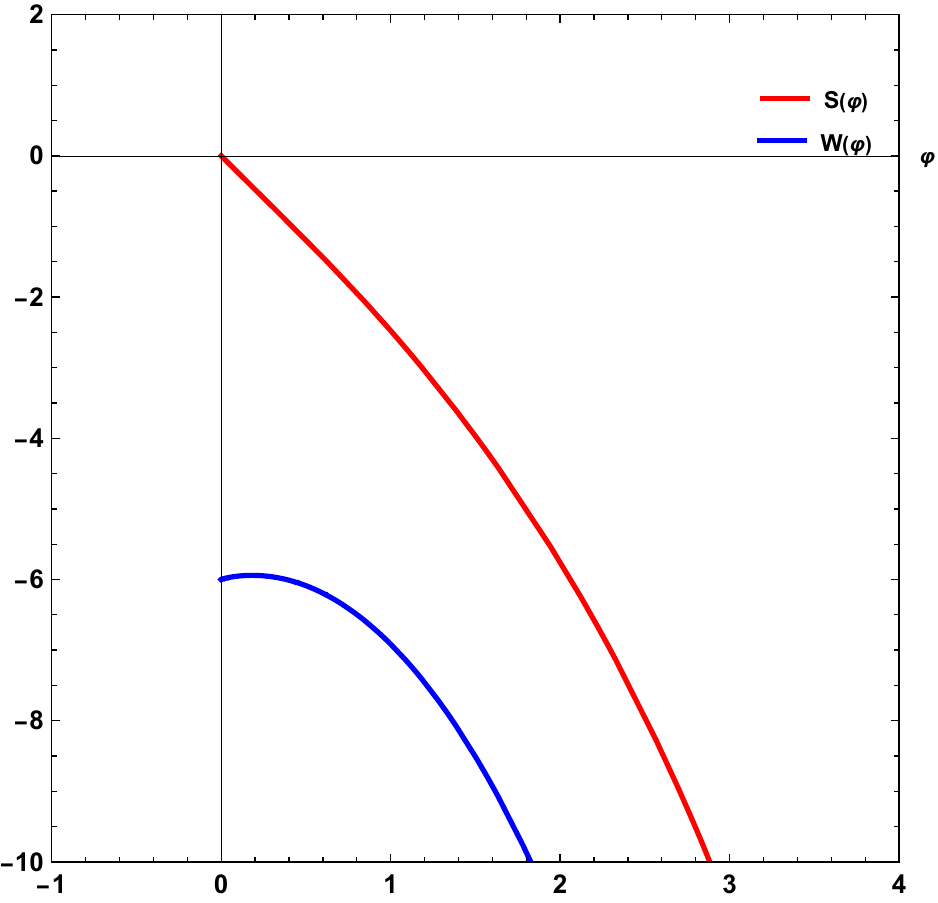}
			\caption{\footnotesize{$\ss=-1.124$}}	\label{p1}
		\end{subfigure}
		\centering
		\begin{subfigure}{0.32\textwidth}
			\includegraphics[width=\textwidth]{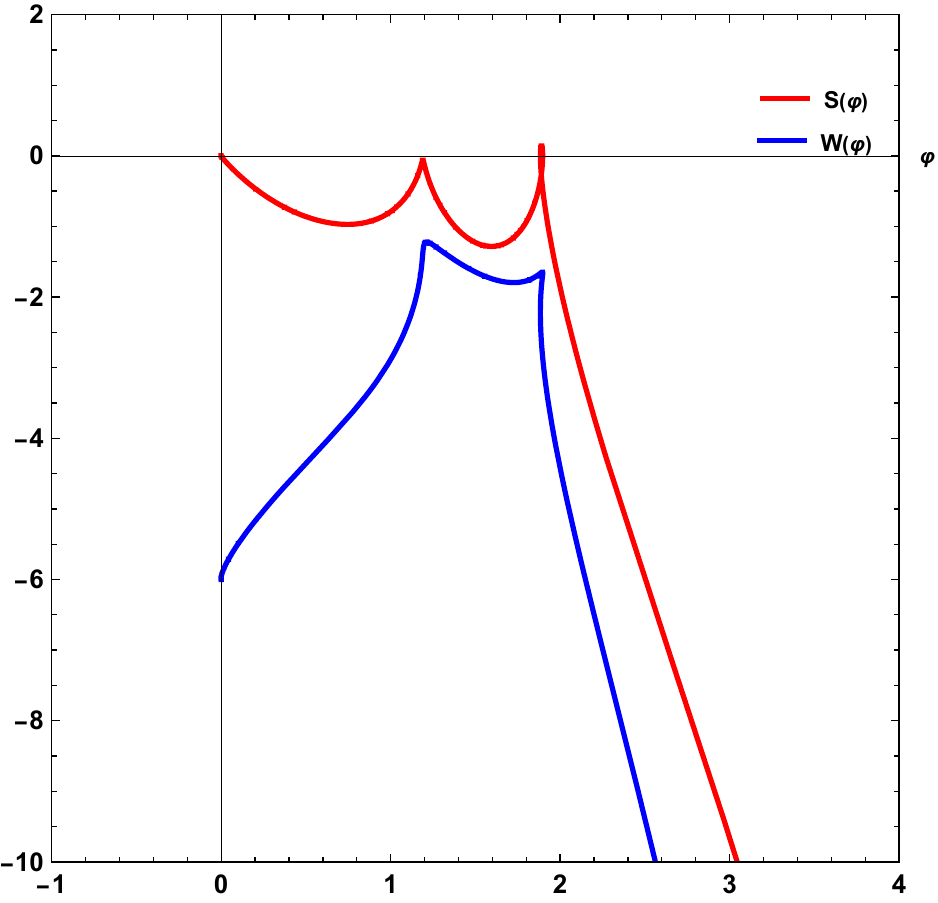}
			\caption{\footnotesize{$\ss=-1.459$}}\label{p2}
		\end{subfigure}
		\centering
		\begin{subfigure}{0.32\textwidth}
			\includegraphics[width=\textwidth]{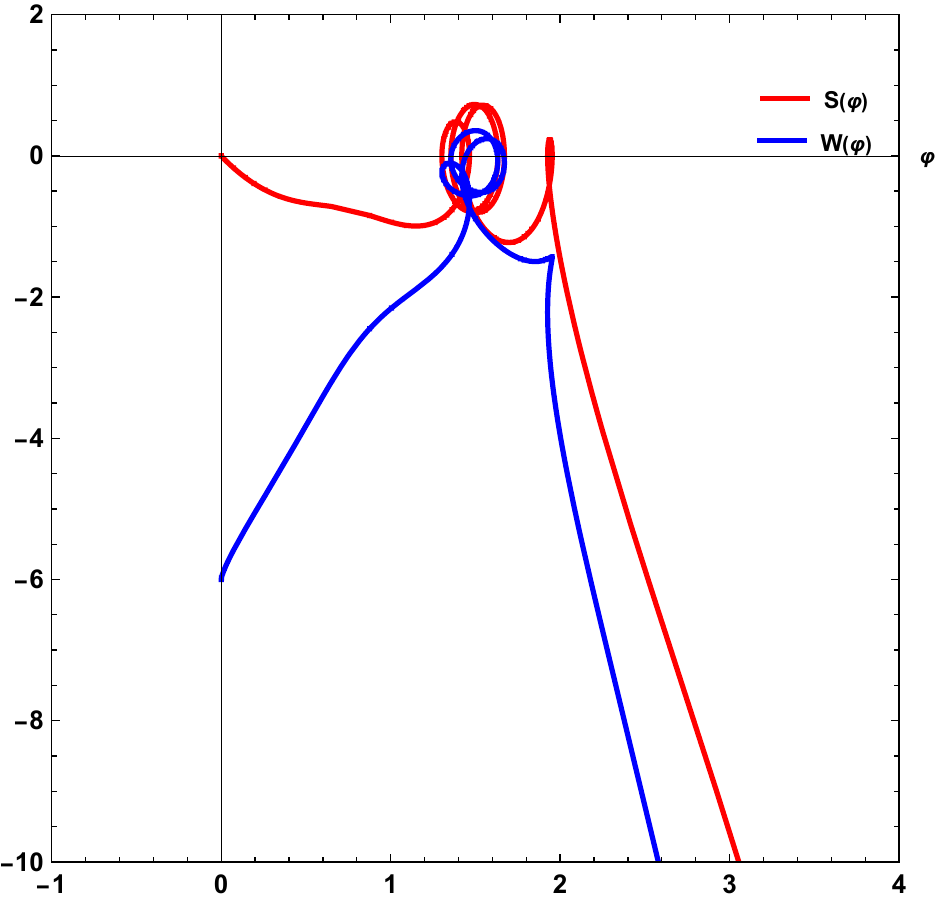}	
			\caption{\footnotesize{$\ss=-1.49013$}}
			\label{p3}
		\end{subfigure}
		\caption{\footnotesize{Regular IR-UV solutions at a fixed $\x=0.1$ for different values of $\ss$. As $\ss$ decreases, the number of bounces (loops) grows, eventually diverging to infinity at a specific threshold.
		}}
	\end{figure}

\begin{figure}[!t]
		\centering
		\begin{subfigure}{0.32\textwidth}
			\includegraphics[width=\textwidth]{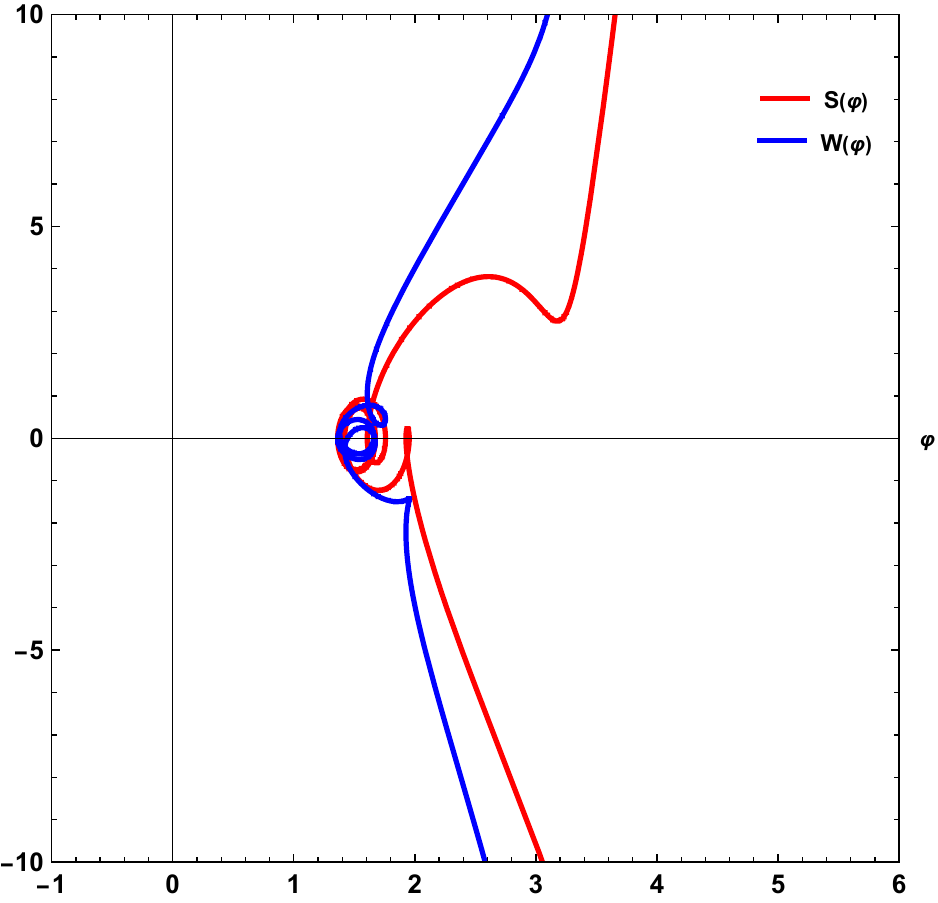}
			\caption{\footnotesize{$\ss=-1.49025$}}
			\label{p4}
		\end{subfigure}
		\centering
		\begin{subfigure}{0.32\textwidth}
			\includegraphics[width=\textwidth]{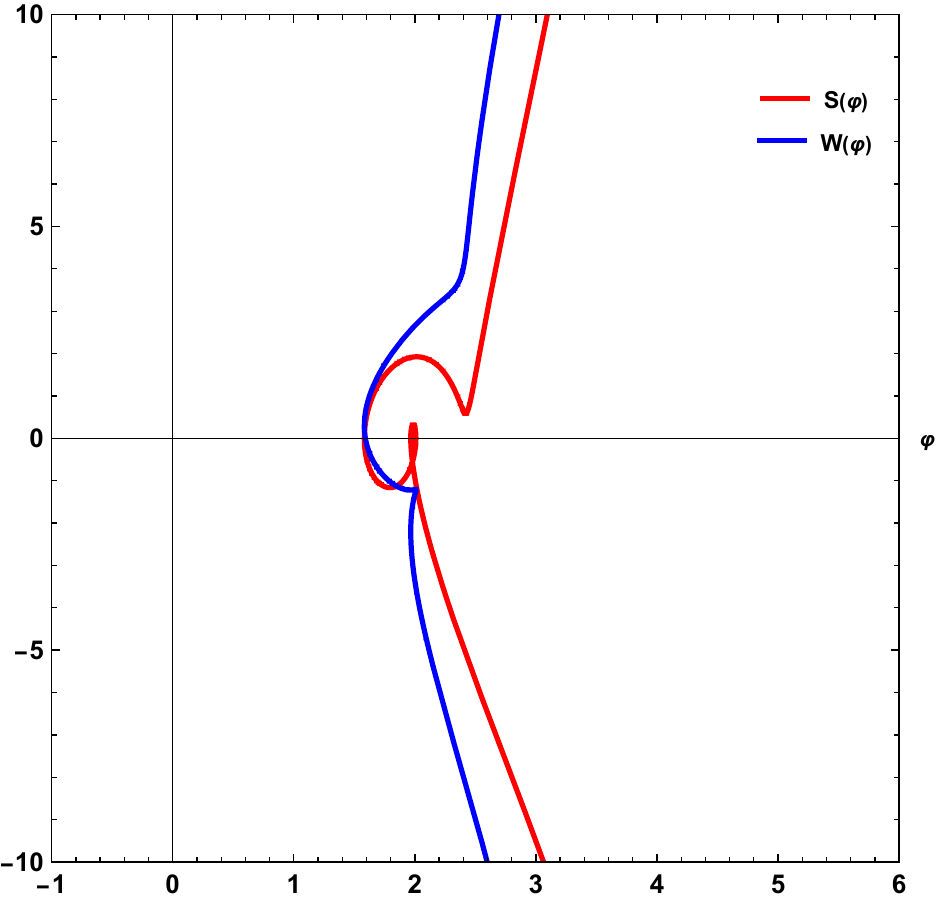}	
			\caption{\footnotesize{$\ss=-1.520$}}
			\label{p5}
		\end{subfigure}
		\centering
		\begin{subfigure}{0.32\textwidth}
			\includegraphics[width=\textwidth]{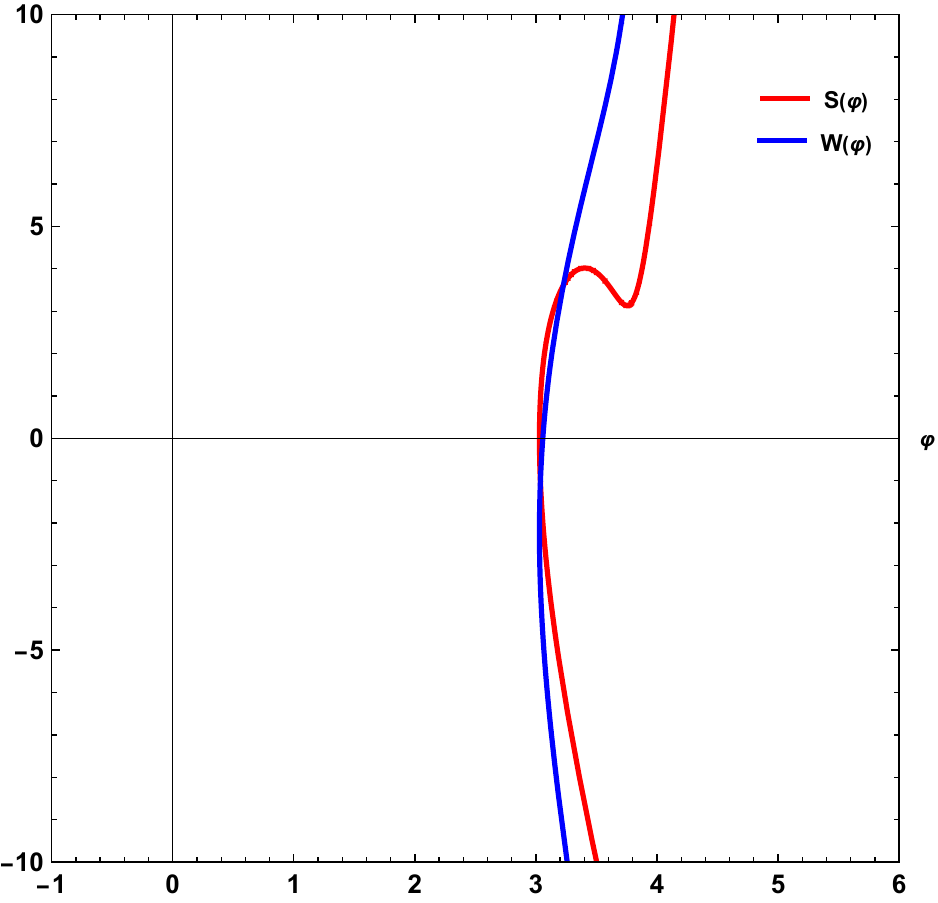}
			\caption{\footnotesize{$\ss=-2.67$}}
			\label{p6}
		\end{subfigure}
		\caption{\footnotesize{Singular solutions at a fixed $\x=0.1$ for different values of $\ss$. Beyond a critical value of $\ss$, approximately $-1.4902$, the UV fixed point vanishes, and the other endpoint extends to infinity. }}
	\end{figure}

	As three numerical examples, we have depicted figures \ref{p1}--\ref{p3}.
	In these figures, we have shown the dependence of the first-order parameters $W$ and $S$ on $\f$.
	The UV fixed point is located at $\f=0$, and the IR end-point is at $\f\to +\infty$.
	As the figures illustrate, reducing the IR parameter $\ss$ leads to an increase in the number of $\f$-bounces or A-bounces, ultimately diverging to infinity at a specific critical value. Beyond this point, the other endpoint no longer connects to the UV fixed point but instead extends to infinity. These types of solutions are depicted in figures \ref{p4}--\ref{p6}.
	
\begin{figure}[!htpb]
		\centering
		\begin{subfigure}{0.32\textwidth}
			\includegraphics[width=\textwidth]{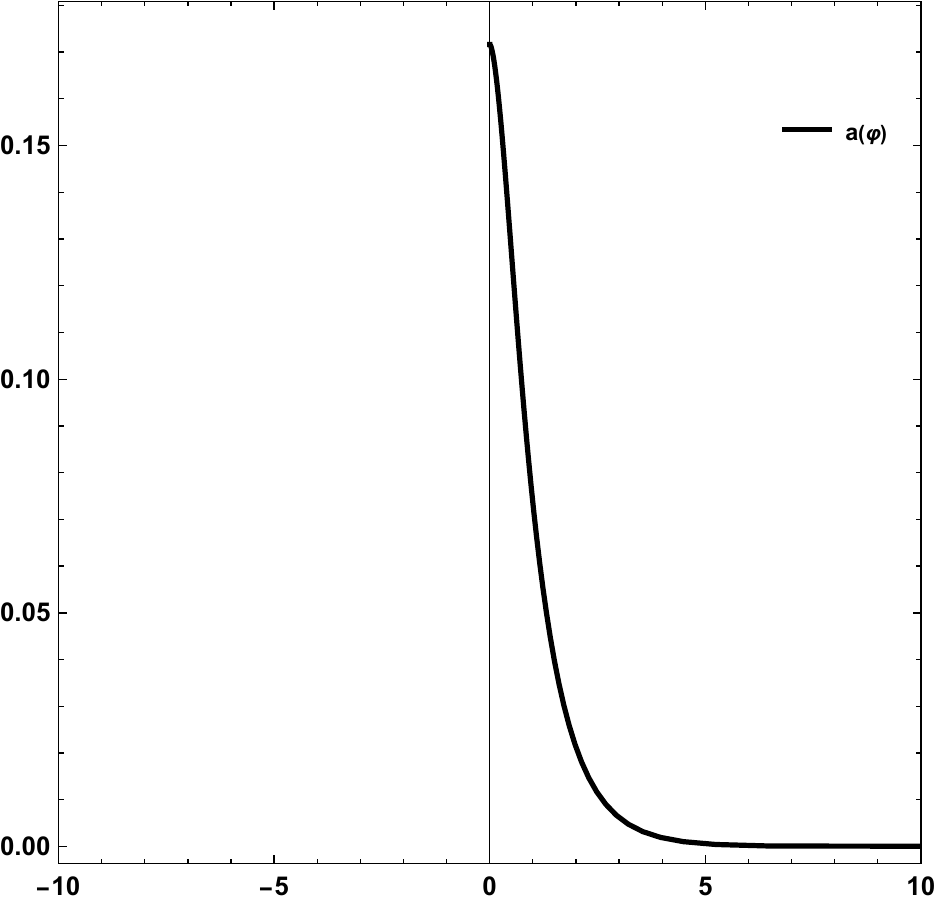}
			\caption{\footnotesize{$\ss=-1.124$}}	\label{xp1}
		\end{subfigure}
		\centering
		\begin{subfigure}{0.32\textwidth}
			\includegraphics[width=\textwidth]{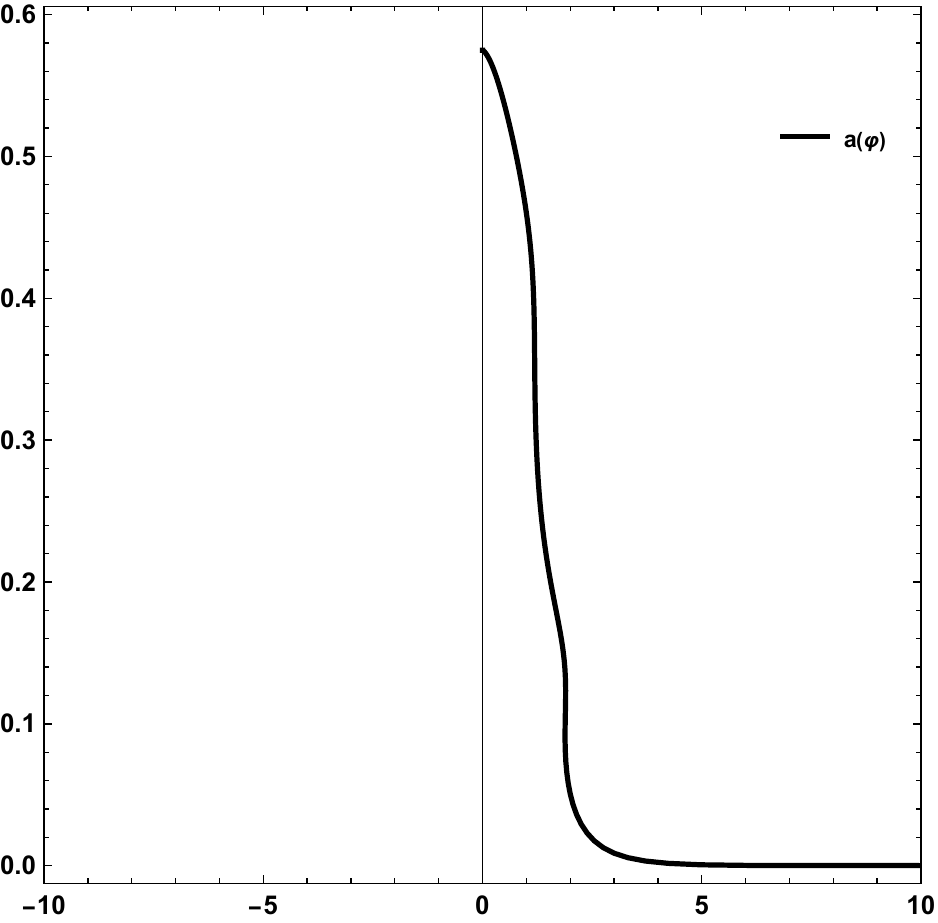}
			\caption{\footnotesize{$\ss=-1.459$}}\label{xp2}
		\end{subfigure}
		\centering
		\begin{subfigure}{0.32\textwidth}
			\includegraphics[width=\textwidth]{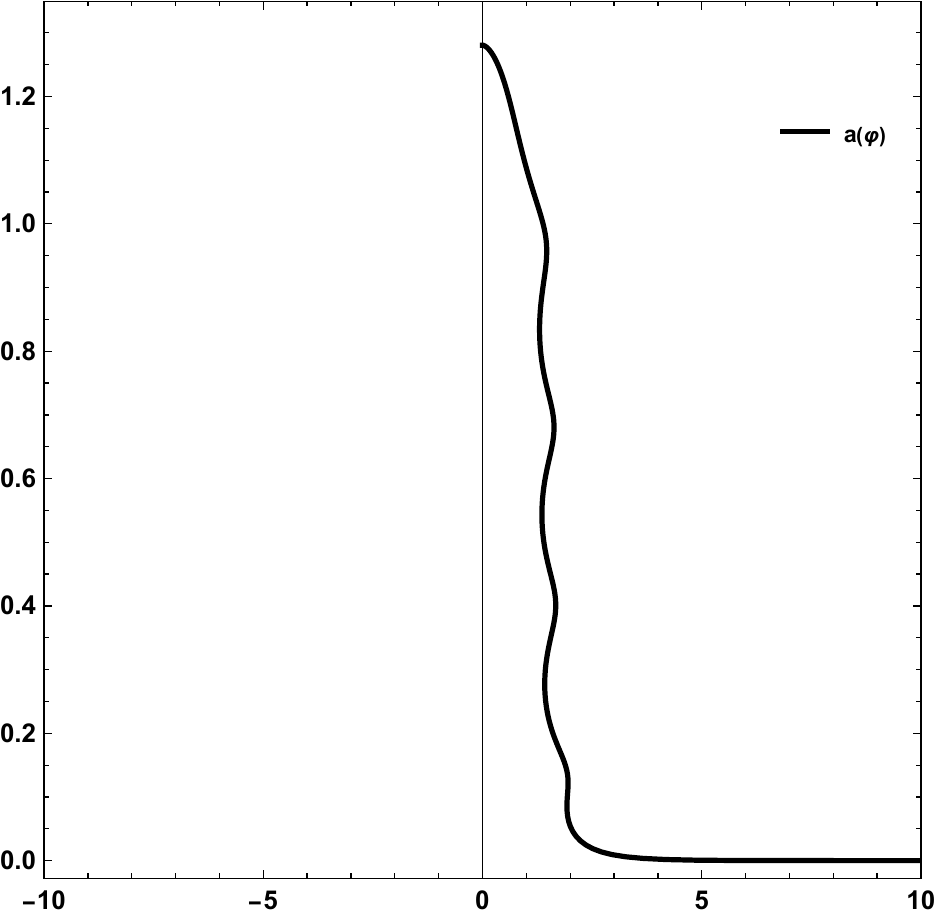}	
			\caption{\footnotesize{$\ss=-1.49013$}}
			\label{xp3}
		\end{subfigure}
		\caption{\footnotesize{The axion as a function of $\f$ for the solutions depicted in figures \ref{p1}--\ref{p3}. At the UV fixed point, where $\f=0$, the axion value attains a finite limit.
		}}
	\end{figure}

	Figures \ref{xp1}--\ref{xp3} illustrate the axion field profile for  figures \ref{p1}--\ref{p3}. At the UV boundary, the axion value approaches a finite limit and increases as $\ss$ decreases. In all solutions within the IR, the axion value is set to zero as the initial condition.
	
	
	In the next section, we shall examine the parameter space of the regular solutions under discussion and establish the relationship between UV and IR parameters.

	\subsubsection{Space of regular IR-UV solutions}\label{IRUVS}
	
	As previously discussed, the space of all regular IR-UV solutions is characterized by two parameters, $\x$ and $\ss$, or equivalently, $\x$ and $Q$. This two-dimensional space of regular IR-UV solutions is depicted in figure \ref{mapsin}, using $(\ss,\x)$ coordinates:
	
	\begin{itemize}
		\item
		Each point within the gray region represents a solution. As one approaches the black curve boundary, the number of loops such as those shown in figure \ref{p3} grows and eventually diverges to infinity. Beyond this boundary, no solution exists with a UV fixed point.
		
	\begin{figure}[!htpb]
		\centering
		\includegraphics[width=0.45\textwidth]{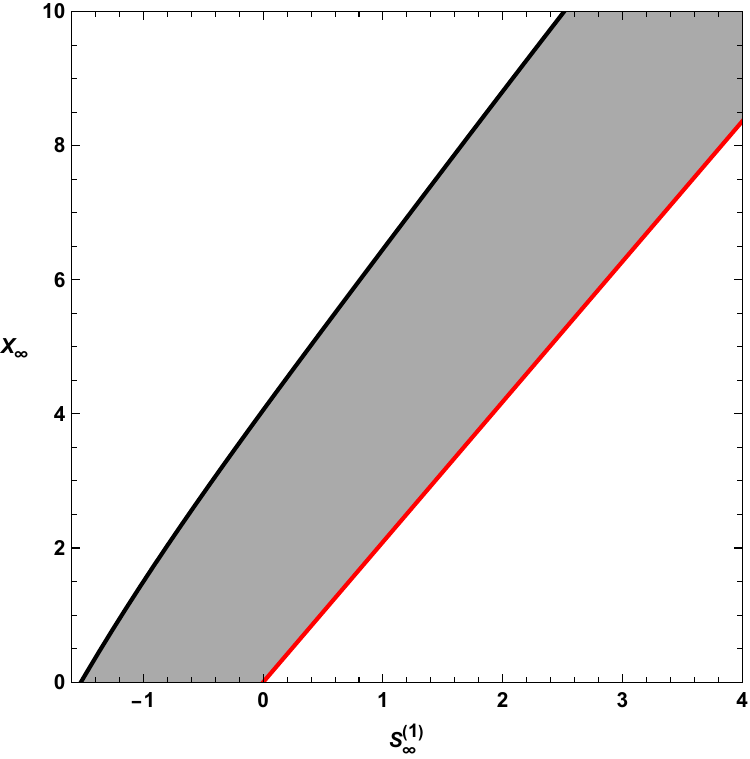}
		\caption{\footnotesize{A corner of the space of regular IR-UV solutions. Each point in the gray region gives the allowed values of $\ss$ and $\x$. Near the black boundary, the number of loops (bounces) increases to infinity. The red bound imposed by requiring a negative slice curvature.}}\label{mapsin}
	\end{figure}
		
		\item
		Obtaining negative curvature slice solutions requires an upper bound on $\ss$, since in this section we have considered $S_\infty<0$ from \eqref{Si1Rz} and \eqref{sstar}, we obtain
		\begin{equation} \label{smaxn}
			\begin{cases}
				\ss < S_\infty^{*}\sp & T_\infty < 0\,, \\[12pt]
				\ss > S_\infty^{*}\sp & T_\infty > 0\,.
			\end{cases}
		\end{equation}
		This upper bound is represented by the red line in figure \ref{mapsin}.
	\end{itemize}

	\subsection{Boundary QFT data}\label{BQFT}

	As we have determined from the expansions of the scalar field, scale factor, and axion field near the UV boundary, i.e., \eqref{fe}--\eqref{ae}, there are five constants of integration: $\f_-$, $A_-$, $C$, $\mathcal{R}$, and $a_{\textrm{UV}}$. On the IR side, the expansions in \eqref{fiir}--\eqref{fAa} involve five other constants: $u_{IR}$, $A_{IR}$, $\x$, $\ss$, and $a_{IR}$. Additionally, there is an extra constant of integration, $Q$, which appears in \eqref{Q} and is present in both the UV and IR expansions.
	
	On the UV side, two sources are present: $a_{\textrm{UV}}$ in \eqref{ae} and the dimensionless curvature $\mathcal{R}$ defined in \eqref{mR}. Furthermore, two parameters related to the vevs, $C$ and $Q$, are given in \eqref{vevf} and \eqref{veva}. The regularity condition of the axion in the UV determines its value on the IR side, i.e., $a_{IR} = 0$. In this section, we examine the behavior of the UV parameters in terms of the IR parameters $\x$ and $\ss$. These two parameters also define the space of solutions.
	
	\begin{figure}[!t]
		\centering
		\begin{subfigure}{0.48\textwidth}
			\includegraphics[width=\textwidth]{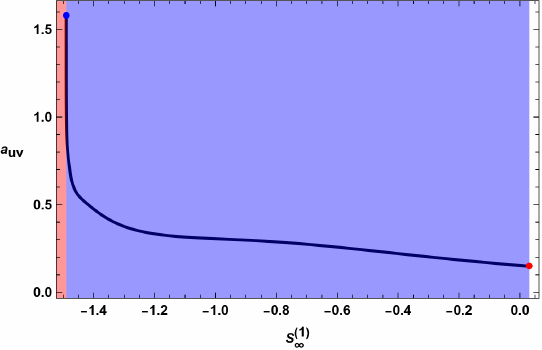}	
			\caption{}
			\label{auvS}
		\end{subfigure}\hspace{0.2cm}
		\begin{subfigure}{0.48\textwidth}
			\includegraphics[width=\textwidth]{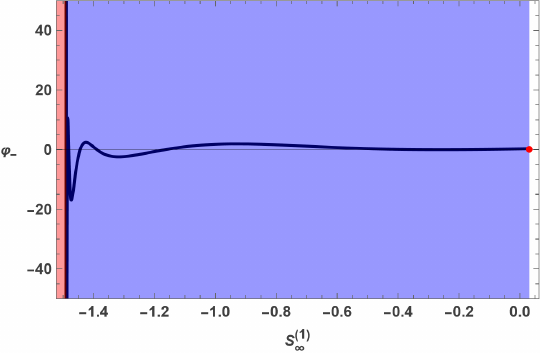}
			\caption{}
			\label{fmS}
		\end{subfigure}
		\begin{subfigure}{0.48\textwidth}
			\includegraphics[width=\textwidth]{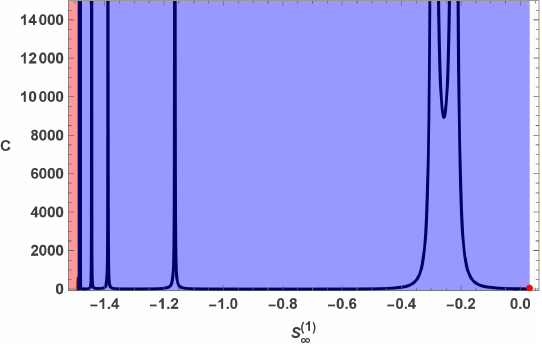}	
			\caption{}
			\label{CS}
		\end{subfigure}\hspace{0.2cm}
		\begin{subfigure}{0.49\textwidth}
			\includegraphics[width=\textwidth]{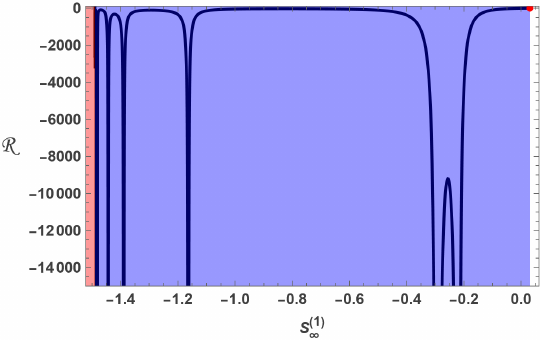}
			\caption{}
			\label{RS}
		\end{subfigure}
		\caption{\footnotesize{UV data ($a_{\textrm{UV}}, \f_-, \mathcal{C}, \mathcal{R}$) as a function of $\ss$ at fixed value of $\x=0.1$. Regular IR-UV solutions exist only in the blue region; beyond that, in the red region, there is no solution with a UV boundary. The red dot represents the maximum number for $\ss$. In figure (a) there is maximum value for $a_{\textrm{UV}}$ pointed by a blue dot.}}
	\end{figure}

	\begin{figure}[!htbp]
		\centering
		\begin{subfigure}{0.49\textwidth}
			\includegraphics[width=\textwidth]{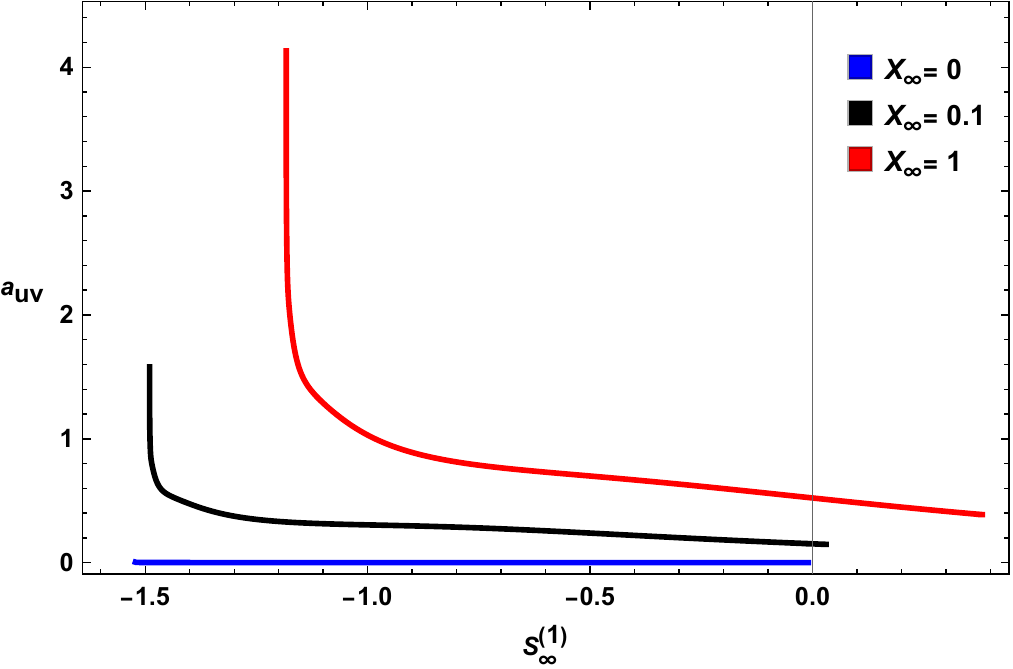}	
			\caption{}
			\label{auvS3}
		\end{subfigure}
		\begin{subfigure}{0.49\textwidth}
			\includegraphics[width=\textwidth]{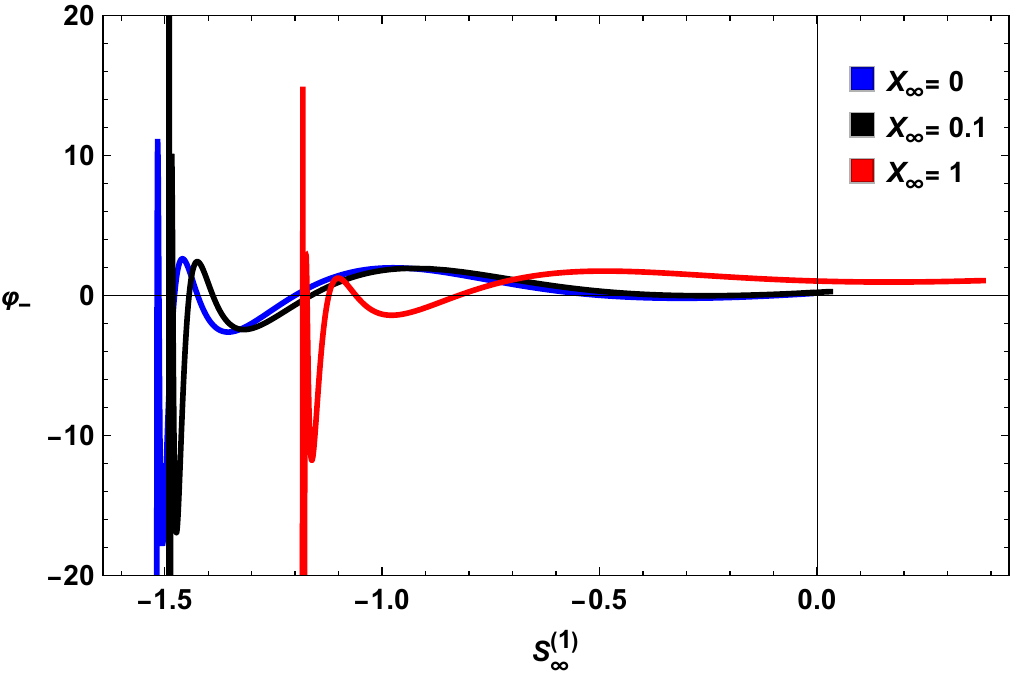}
			\caption{}
			\label{fmS3}
		\end{subfigure}
		\begin{subfigure}{0.48\textwidth}
			\includegraphics[width=\textwidth]{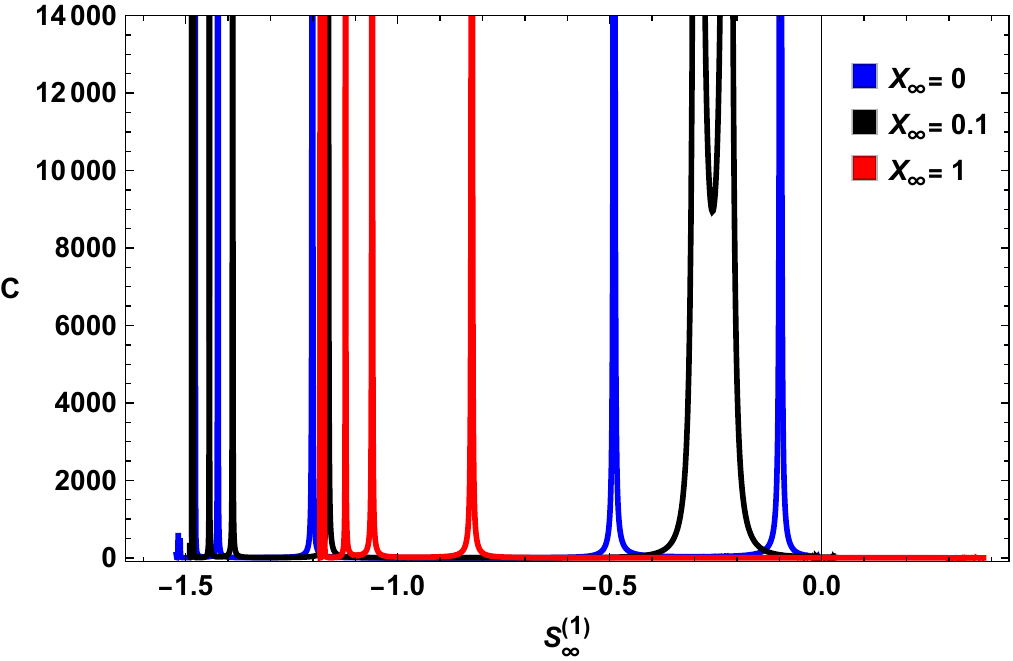}	
			\caption{}
			\label{CS3}
		\end{subfigure}
		\begin{subfigure}{0.49\textwidth}
			\includegraphics[width=\textwidth]{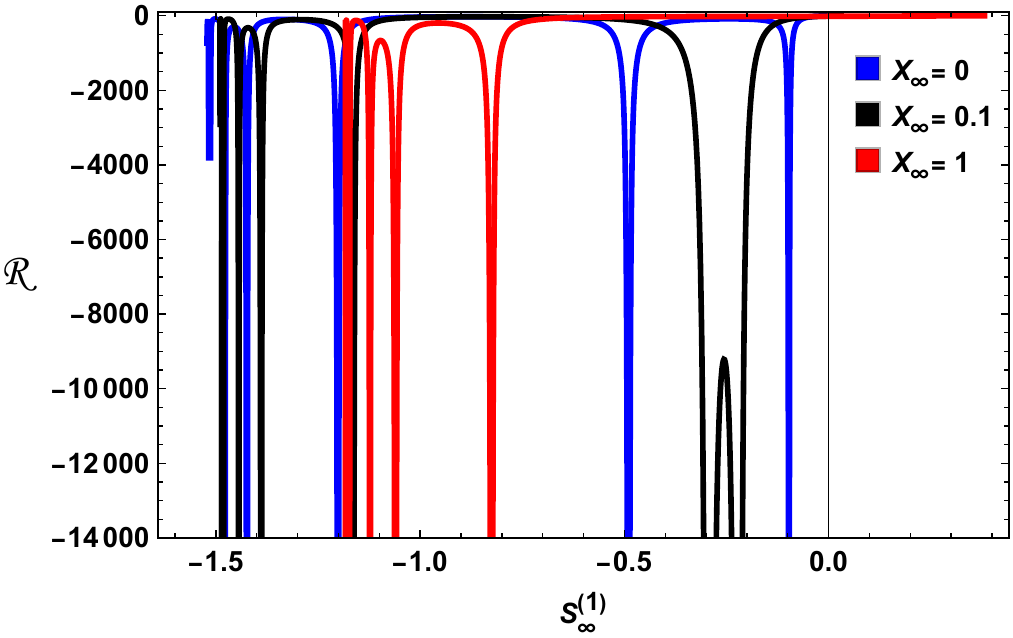}
			\caption{}
			\label{RS3}
		\end{subfigure}
		\caption{\footnotesize{UV data ($a_{\textrm{UV}}, \f_-, \mathcal{C}, \mathcal{R}$) as a function of $\ss$ for three values of $\x=0, 0.1, 1$. All values of $a_{\textrm{UV}}$ are finite.}}
	\end{figure}
	
	Consider the solution space shown in figure \ref{mapsin}, where the value of $\x$ is fixed while $\ss$ varies between the black and red boundaries. This variation reveals the following properties, as illustrated in figures \ref{auvS}--\ref{RS}:

	\begin{itemize}
		\item As illustrated in figure \ref{auvS}, the axion value at the UV fixed point gradually increases as $\ss$ decreases. The minimum value of $a_{\textrm{UV}}$ occurs when the slices are flat, i.e., \( T = 0 \), where $\ss$ attains its maximum value (red dot). Conversely, as one nears the black boundary of figure \ref{mapsin}, the number of A-bounces or $\f$ grows indefinitely. However, the axion value $a_{\textrm{UV}}$ approaches a finite maximum (blue dot). Beyond this boundary, no UV fixed point exists.
		
		\item Figure \ref{fmS} illustrates the coupling $\f_-$ as a function of $\ss$. At points where $\f_- = 0$, the vev $C$ and the dimensionless curvature of the boundary diverge, as shown in figures \ref{CS} and \ref{RS}. These behaviors are governed by equations \eqref{vevf} and \eqref{mR}.
		
	\end{itemize}
	
	Figures \ref{auvS3}--\ref{RS3} present a comparison of three distinct slices within the solution space shown in figure \ref{mapsin}, corresponding to $\x = 0$, $0.1$, and $1$:

	\begin{itemize}
		\item The case of $\x=0$ was previously examined in \cite{Ghodsi:2024jxe}. Here, the axion value remains zero throughout\footnote{The same is true if the value of the axion is constant.}, and $\ss$ is constrained from above by zero, extending downward to a finite negative value near $-1.5$.

		\item As $\x$ increases, both the upper and lower bounds of $\ss$ shift to the right, while the minimum and maximum values of $a_{\textrm{UV}}$ (two end-points of each curve) grow but remain finite.
		
		\item The behavior of the other three parameters, $\f_-$, $C$, and $\mathcal{R}$, for three cases is similar, except for the shifting of the locations of $\f$-bounces.
	\end{itemize}
	To complete the numerical analysis, we have presented two other figures:
	
	\begin{itemize}
		\item
		In figure \eqref{aXS} we have found the value of $a_{\textrm{UV}}$ for each solution in the space of solutions in figure \ref{mapsin}. When $\x\rightarrow 0$, the value of axion goes to zero. By increasing $\x$ eventually $a_{\textrm{UV}}$ increases monotonically at each fixed values of $\ss$.
		
		\begin{figure}[!htbp]
			\centering
			\includegraphics[width=0.55\linewidth]{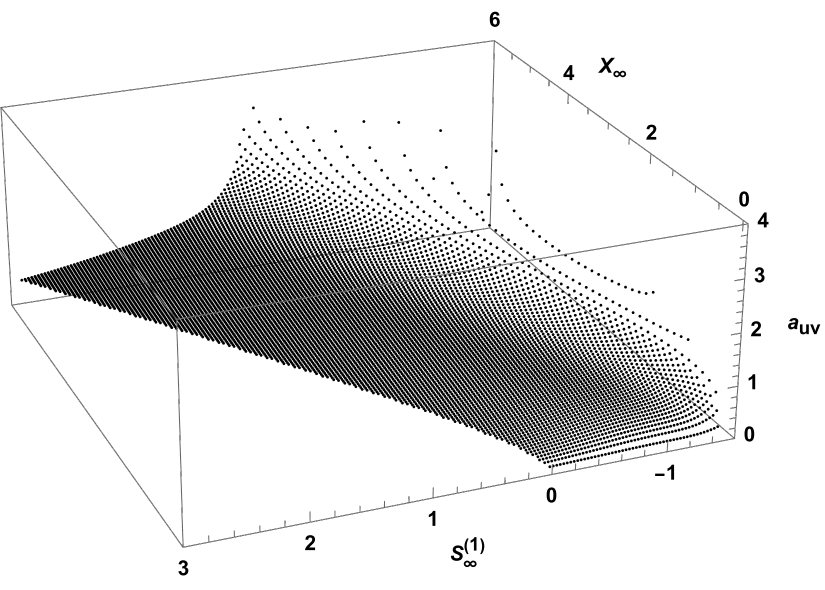}
			\caption{\footnotesize{$a_{\textrm{UV}}$ as a function of $\x$ and $\ss$ for Regular IR-UV solutions.}}
			\label{aXS}
		\end{figure}
		
		\item
		A three-dimensional representation of the possible values of the vev \( Q \) as a function of the IR parameters \( \ss \) and \( \x \) is shown in figure \ref{QXS}. This surface is described by equation \eqref{Si1R}. The red and black lines correspond to the red and black lines in figure \ref{mapsin}. Along the black line, the slice geometries are flat, and the values of \( Q \) remain small but nonzero. Only in the limit \( \x \rightarrow 0 \) does \( Q \) approach zero as well.

		\begin{figure}[!t]
			\centering
			\includegraphics[width=0.6\linewidth]{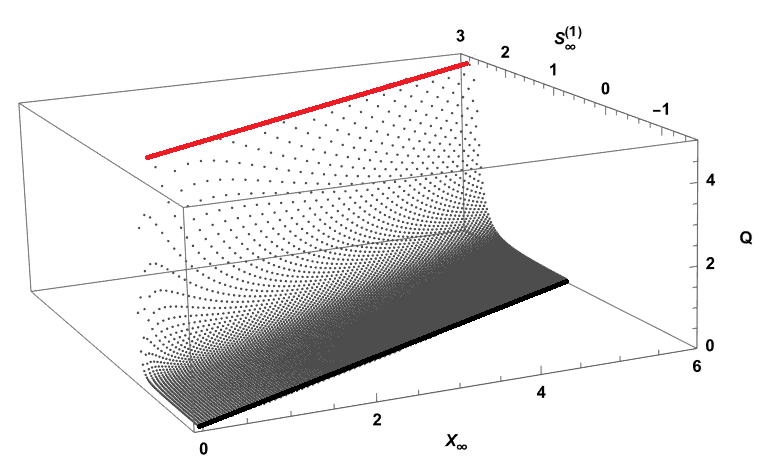}
			\caption{\footnotesize{The vev \( Q \) as a function of \( \x \) and \( \ss \) for regular IR-UV solutions is depicted in figure \ref{mapsin}. The black and red boundaries correspond to those shown in figure \ref{mapsin}.
			}}
			\label{QXS}
		\end{figure}
		
	\end{itemize}

	\subsection{Free Energy}

	In this section, we summarize the computation of the holographic free energy in the framework of \cite{Ghodsi:2024jxe},
	for both one--boundary and two--boundary solutions.
	We also distinguish between the cases where the boundary slices are
	non-compact (so that a side boundary is present)
	and compact (no side boundary).
	All results below refer to the four-dimensional boundary case (\(d=4\)),
	although the generalization to arbitrary \(d\) is straightforward. A brief overview of the free energy for one-boundary solutions is provided in Appendix \ref{FE}.
	
	\subsubsection{One-boundary geometries}
	
	As we discussed earlier, the one--boundary geometries describe a single confining QFT
	living on an AdS\(_4\) boundary.
	The bulk space-time ends smoothly in the interior,
	so that the geometry has a single asymptotic AdS region.
	The free energy in this setup is obtained from the on-shell gravitational action,
	including the Gibbons--Hawking terms and appropriate holographic counter-terms.
	
	For non-compact boundary slices, there exists a side boundary contribution
	which affects the normalization of the on--shell action.
	After holographic renormalization,
	the resulting free energy depends smoothly on the boundary curvature,
	without any indication of a phase transition.
	Among the infinite family of ``pure vev'' solutions,
	only one solution minimizes the free energy and represents the dominant saddle.
	
	When the constant--curvature slices are compact, the geometry no longer has a side boundary.
	This changes the relative weight of curvature terms in the renormalized free energy.
	Nevertheless, the qualitative behavior remains the same:
	The free energy is a smooth function of curvature
	and approaches a finite limit for large negative curvature.
	The absence of the side boundary only modifies numerical coefficients,
	reflecting the different boundary topology.
	
	\subsubsection{Two-boundary geometries}
	
	Two--boundary geometries correspond to bulk configurations with two asymptotic AdS regions,
	interpreted holographically as either an interface or a wormhole between two confining QFTs
	living on curved boundaries with possibly different curvatures.
	The free energy is again computed from the renormalized on-shell action,
	now including contributions from both boundaries and, in the non-compact case, from the side boundary as well.
	
	In this class of solutions, two types of configurations exist:
	a ``connected'' geometry, where the two boundaries are joined through the bulk,
	and a ``disconnected'' configuration,
	which consists of two separate one--boundary solutions.
	By comparing their renormalized free energies,
	it was found that the disconnected configuration always dominates thermodynamically,
	meaning that the interface solutions are subdominant saddles.
	Consequently, the two boundary field theories do not interact at leading order in $N_c$,
	and any cross-correlators between them are exponentially suppressed.
	
	When the slice geometry is compact, the structure of the free energy remains similar,
	but the overall normalization and some scheme-dependent constants change.

	\subsubsection{Summary of results}
	
	The main results of the free energy analysis can be summarized in the following table:
	
	\begin{table}[h!]
		\centering
		\renewcommand{\arraystretch}{1.25}
		\setlength{\tabcolsep}{2pt}
		\begin{tabular}{|p{3.5cm}|p{3.2cm}|p{6.5cm}|}
			\hline
			\multicolumn{1}{|c|}{\textbf{Solution}} &
			\multicolumn{1}{c|}{\textbf{Slice geometry}} &
			\multicolumn{1}{c|}{\textbf{Free energy density $\mathcal{F}$}} \\[2pt]
			\hline
			\multicolumn{1}{|c|}{One--boundary, $u\rightarrow\pm\infty$} &
			\multicolumn{1}{c|}{non--compact} &
			\multicolumn{1}{c|}{$\frac{3}{2}\Big(1 - 96\frac{C}{\mathcal{R}^2} \pm 192 \frac{B}{\mathcal{R}}\Big)$} \\[6pt]
			\hline
			\multicolumn{1}{|c|}{One--boundary, $u\rightarrow\pm\infty$} &
			\multicolumn{1}{c|}{compact} &
			\multicolumn{1}{c|}{$\frac{3}{2}\Big(1 - 96\frac{C}{\mathcal{R}^2} \pm 48 \frac{B}{\mathcal{R}}\Big)$} \\[6pt]
			\hline
			\multicolumn{1}{|c|}{Two--boundary} &
			\multicolumn{1}{c|}{non--compact} &
			\multicolumn{1}{c|}{$3 -144\left(\frac{C_+}{\mathcal{R}_+^2} + \frac{C_-}{\mathcal{R}_-^2}\right) + 288\left(\frac{B_+}{\mathcal{R}_+}-\frac{B_-}{\mathcal{R}_-}\right)$} \\[6pt]
			\hline
			\multicolumn{1}{|c|}{Two--boundary} &
			\multicolumn{1}{c|}{compact} &
			\multicolumn{1}{c|}{$3 -144\left(\frac{C_+}{\mathcal{R}_+^2} + \frac{C_-}{\mathcal{R}_-^2}\right) + 72\left(\frac{B_+}{\mathcal{R}_+}-\frac{B_-}{\mathcal{R}_-}\right)$} \\[6pt]
			\hline
		\end{tabular}
		\caption{\footnotesize{Summary of renormalized free energy for one and two boundary geometries with compact and non-compact slices. $\pm$ subscripts mean UV boundaries at $u\rightarrow \pm \infty $}}
		\label{tabFES}
	\end{table}
	
	When the boundary slices are AdS and an axion is included in the bulk, the general structure of the renormalized free energy remains the same as in the axion-free case. The axion contributes to the UV expansion of the bulk fields only at ``subleading order", so that the leading UV coefficients, which determine the main structure of the free energy in terms of $B$, $C$, and the boundary curvature $\mathcal{R}$ are unaffected. Consequently, the functional form of the free energy formulae is unchanged, but the integration constants $B$ and $C$ acquire a ``weak dependence" on the UV value of the axion, $a_{\rm UV}$, reflecting its next-to-leading-order effect. This ensures that the expressions summarized in table \ref{tabFES} remain valid, with $B$ and $C$ encoding any corrections due to the axion source.

	In conclusion, the presence of a side boundary modifies the renormalized coefficients of the free energy,
	especially the term proportional to $B$.
	For one-boundary solutions, the free energy remains a smooth function of the boundary curvature and the value of the axion at UV
	without phase transitions,
	while for the two-boundary configurations, the connected solutions are always subdominant
	compared to the disconnected ones.

	\section{Conclusion and Open Ends\label{last}}

	In holographic confining theories placed on positively curved backgrounds, the structure of regular bulk solutions leads, as we have shown,  to a rich phase diagram controlled by the dimensionless curvature source $\mathcal{R}$ and, in the axionic case, by the UV value of the axion $a_{\textrm{UV}}$.
	
	For theories above the Efimov bound, where the scalar potential induces oscillatory IR behavior, the families of Type II and Type III solutions generate an Efimov spiral in the $(\mathcal{R}, C)$ plane  (see figures \ref{crall} and \ref{crallz}). This multi-valued structure implies that, for a given set of UV sources, the dual field theory admits more than one saddle. A first-order quantum phase transition occurs when the free energies of the two branches cross: for large curvature, the Type III solutions dominate, while below a critical curvature, the preferred configuration jumps to the Type II branch.
	
	The transition point is characterized by a continuous free energy but a discontinuity in its first derivative with respect to $\mathcal{R}$, confirming the first-order nature of the transition. The axion extends the phase space to $(\mathcal{R}, a_{\textrm{UV}})$, but the basic mechanism remains the same: the oscillatory Efimov regime ensures multi-valuedness of the QFT data, thereby enabling the transition.
	
	For theories below the Efimov bound, the oscillatory structure gradually disappears. As the exponent $b$ decreases past the critical value, the Efimov spiral in the $(\mathcal{R}, C)$ plane opens and finally collapses into a single-valued curve. Consequently, the free energies of Type II and Type III solutions cease to intersect, and the first-order phase transition is washed out and becomes second order. In this regime, the axion does not qualitatively alter the picture: although it introduces an additional UV source $a_{\textrm{UV}}$, the absence of Efimov oscillations eliminates the multi-branch structure needed for a first-order transition. Thus, the existence of the first-order phase transition is intrinsically tied to the presence of the Efimov regime, which is controlled by the large-$\varphi$ asymptotics of the potential. Only above the Efimov bound does the interplay of curvature, scalar dynamics, and axionic deformation produce a genuine first-order quantum phase transition in the dual theory.
	
	\subsection*{Higher-dimensional origins of axions in string/M-theory}
	
	In string and M-theory compactifications, axions arise as Kaluza--Klein zero modes of higher-form gauge fields wrapped on nontrivial internal cycles. Their shift symmetries descend from the gauge invariance of ten or eleven-dimensional $p$-form potentials.
	Concretely, integrating a $p$-form field over a cycle,
	$$
	a(x)\sim \int_{\Sigma_p} C_p(x,y)\,,
	$$
	produces a lower-dimensional scalar with an approximate continuous shift symmetry, broken only by non-perturbative brane or world sheet instantons.
	Although instanton effects generate non-perturbative axion potentials, these vanish in the $g_s \to 0$ limit, or the large $N_c$ limit.

	Such axions appear generically in Calabi-Yau and related compactifications, and similarly in M-theory, where the 3-form generates axions via integration over 3-cycles. These axions often combine with geometric moduli to form complex multiplets in the effective supergravity theory.
	
	In bottom-up holographic models, including the Einstein-axion-dilaton theories studied here, axions typically enjoy an exact shift symmetry at the level of the bulk Lagrangian because the shift symmetry breaking is exponentially small in $N_c$. Their bulk profiles describe RG flows dual to running P-odd couplings or $\theta$-angles in the boundary theory. When such holographic theories are placed on curved backgrounds, the UV value of the axion acts as a tunable source, enlarging the phase space and enabling multi-valued branches of solutions, especially when the scalar potential drives the system into an Efimov-like oscillatory regime. In this setting, the structure of Efimov spirals in the $(\mathcal{R}, a_{\textrm{UV}})$ plane can reflect topological data of the higher-dimensional compactification, since fluxes and wrapped branes induce specific axion couplings and modify IR behavior.
	
	A natural direction is therefore to connect these holographic axion flows to explicit string/M-theory compactifications. Embedding the effective potentials and kinetic terms into a UV-complete construction would clarify how internal geometry, fluxes, and instanton effects determine the allowed axion field ranges and back reaction constraints. This, in turn, may explain which compactifications support the multi-branch structure characteristic of first-order phase transitions in the Efimov regime, and when the dynamics instead collapse to a single smooth branch. Such a link would provide a more complete picture of how confining holographic dynamics and axion-induced phase transitions arise from higher-dimensional geometry and topology.

	\subsection*{Bulk axions, brane-induced phase structures and a Vafa-Witten-like theorem}

	The phase diagrams we have found here, as well as previous works, \cite{Ha1, Ha2}, contain important ``holes". These appear in two contexts, but their significance is similar.
	
	The first, explained in section \ref{FEP}, involves the confining holographic theories without the axion. In that case, we have found that there is a small range of the parameter $b$ controlling
	the large field asymptotics of the potential, around the end of the Efimov regime,  for which, during the transition, the free-energy is discontinuous. This is shown in Figure \ref{below}.  This behavior is unphysical, and in most cases,
	it suggests that other relevant phases that should dominate in this regime are missing.
	
	This can happen in holographic theories if one truncates the low-energy spectrum of operators.
	What can happen in such cases is a scalar that is zero in the initial solution (having zero source), turns on because of a vev, and this new solution may be dominant.
	Multiple examples of this exist in holography.
	
	However, in this particular case, it is difficult to see what can be another field whose dual operator should obtain an expectation value. In YM-like theories, the only other relevant scalar operator is the topological density, but in QFT, there is the Vafa-Witten theorem that forbids such a vev, \cite{vw}.
	
	We have indicated in section \ref{VW} that a similar theorem (the holographic Vafa-Witten theorem) can hold in holographic solutions, modulo some caveats associated with D-instantons.
	It is similarly unlikely that the energy-momentum tensor obtains a vev, as it would break Lorentz invariance in the ground state of the theory.
	
	There is a second possibility: that this range of $b$ is in the ``swampland". The simplest  compactification  from higher dimensions, \cite{GK},
	on a single $S^n$ indicates that one obtains a potential for the scale factor that has
	\be
	b=\sqrt{\frac{n+d-1}{2n(d-1)}}\,.
	\ee
	The Efimov bound corresponds to $n+d=9$. The values of $b$ that are problematic correspond to $n= 6$ ($b=0.5$) and $n=5$ ($b=2/\sqrt{5}$), which would imply a 10 or 11-dimensional origin for the IR limit of the potential.
	These cases are summarized in figures \ref{bzaxc} and \ref{bzax}.
	However, it is not clear that an $S^6$ compactification of M-theory exists that will provide a potential with similar characteristics as the one we are using.
	The previous argument has many loopholes, and further analysis is needed in order to be convinced of its relevance.

	A similar phenomenon happens in the presence of non-trivial axion solutions. It has been observed in  \cite{Ha1,Ha2},
	that the space of UV axion sources ($\theta$-angles) is compact, although in field theory and in the large $N_c$ limit, it can take all real values.
	In \cite{Ha1} the following inequality was shown
	\be
	|a_{\textrm{UV}}|\leq \int_0^{\infty}\frac{d\f }{ \sqrt{Y(\f)}}\,,
	\ee
	which implies $|a_{\textrm{UV}}|\leq \frac{2}{\gamma\sqrt{Y_0}}$ for $Y=Y_0\; e^{\g \f}$.
	This upper bound is also confirmed in all solutions obtained in this paper.
	However, we expect that axionic solutions should not have an upper bound on the source.
	This is certainly linked to the fact that in several situations, we found phase transitions
	at finite axion density, where again the free energy is discontinuous, as discussed in section \ref{FEP}.
	
	However, in the axion case, there is a new possibility that will provide novel solutions.
	A natural extension is the inclusion of branes in the bulk, which act as localized sources for the charge dual to the axion shift symmetry.
	In the higher-dimensional picture where the axion descends from a nontrivial form field, this involves bulk branes that are charged minimally under the form field. These will descend to particles in the lower dimension, and a distribution of them can, in principle, appear in the solutions. We have discussed this occurrence in section \ref{VW}.
	
	In general, we expect to have a lattice of such higher-dimensional branes with a tension $T_n=n T_1$ and charge $Q_n=nQ_1$. In $d$ dimensions, the tension will become a particle mass while the charge will become an axionic charge.  We do expect, however, that results will depend on the choice of $T_1, Q_1$.
	
	The new solutions are expected to
	significantly reshape the $(\mathcal{R}, a_{\textrm{UV}})$ phase space. Existing solution branches may change and give rise to entirely new branches,
	potentially enhancing or suppressing first-order transitions already present in the axion-only setup.
	By modifying both the boundary conditions and the effective IR dynamics, branes enrich the holographic landscape,
	creating new regions of multi-valued vacua and potentially revealing novel Efimov-like structures that reflect the interplay between curvature, axion dynamics, and localized sources.
	We expect to study this in a future publication.

	\subsection*{On the instanton gas for confining holographic theories on AdS}

	It was suggested in \cite{CW} that the nature of the instanton gas in asymptotically free, QCD-like theories is expected to be different than that of flat space.
	In particular, unlike in the standard confining phase of QCD in flat space, where there is an instanton liquid,
	on AdS, they expect that instanton interactions are weak and there is an instanton gas instead, like in the case of flat space above the gluon-plasma transition.
	
	In the holographic case, this would be materialized if at a non-zero $\theta$-angle the vev proportional to $Q$ would vanish\footnote{This is true in large $N_c$ perturbation theory. There would, however, be a non-perturbatively small vev, exponentially small in $N_c$, generated by the instanton gas. This is precisely what happens in the gluon-plasma phase.}.
	
	We do not observe this in our analysis. The reason may be that this is expected in a gauge theory with electric-type boundary conditions, discussed in \cite{CW}. As already observed in \cite{Ghodsi:2024jxe}.
	There is no boundary $SU(N_c)$ global symmetry in our solutions, and most probably the holographic theory resembles more a gauge theory on AdS with magnetic boundary conditions.
	This issue deserves further investigation.

	
	\section*{Acknowledgments}
	\addcontentsline{toc}{section}{Acknowledgments}
	
	We thank  G. Fournodaylos, A. Porfyriadis, and C. Rosen for useful discussions.
	
	This work was partially supported by  the H.F.R.I. call ``Basic research Financing" (Horizontal support of all Sciences)
	under the National Recovery and Resilience Plan ``Greece 2.0" funded by the European Union -NextGenerationEU
	(H.F.R.I. Project Number: 15384), by the In2p3 grant ``Extreme Dynamics", the ANR grant ``XtremeHolo"
	(ANR project n.284452), by the H.F.R.I. Project Number: 23770  of the H.F.R.I call
	``3rd Call for H.F.R.I.'s Research Projects to Support Faculty Members \& Researchers", the ERC starting grant 101078061 SINGinGR, under the European Union's Horizon Europe program for research and innovation"
	and the UoC grant number 12030.


	\appendix
	
	\renewcommand{\theequation}{\thesection.\arabic{equation}}
	\addcontentsline{toc}{section}{Appendix\label{app}}

	\section{IR asymptotic solutions: Details} \label{IRdetails}

	In this appendix, we derive the IR expansion of the solutions to the main equations (\ref{eom2})--(\ref{eom4}). The IR endpoint corresponds to the limit $\f \to +\infty$.
	
	We assume that, as $\f \to +\infty$, the functions take the asymptotic form
	\be \label{VY1}
	V(\varphi)= -V_\infty e^{2b\varphi}+\cdots \sp
	Y(\varphi)= Y_\infty e^{\g\varphi}+\cdots\,,
	\ee
	where $V_\infty$ and $Y_\infty$ are positive constants, and the ellipses denote subleading contributions.
	
	To determine the behavior of $S$ and $X$ near the IR endpoint, we adopt the following exponential ansatz as $\f \to +\infty$:
	\be
	S(\varphi) = S_\infty e^{s\varphi}+ \ss e^{s_1\varphi}+ S_\infty^{(2)} e^{s_2\varphi}+\cdots\,, \nn
	\ee
	\be
	X(\varphi) = X_\infty e^{k\varphi}+ \xx e^{k_1\varphi}+ X_\infty^{(2)} e^{k_2\varphi}+\cdots\,,\label{SXeN}
	\ee
	with the ordering $s>s_1>s_2$ and $k>k_1>k_2$. These expansions are then substituted into the equations derived previously in section \ref{FO}. The first equation is \eqref{SX}, which reads
	\begin{gather}
		S^3S''-\frac{1}{d}S^2{S'}^2+\frac{d+2}{d}\left(V'+\frac{XY'}{2Y^2}\right)SS'-\frac12 S^4-\frac{1}{d}\left(V'+\frac{XY'}{2Y^2}\right)^2\nn \\
		-\left(V''+\frac{V}{d-1}+
		\frac{1}{2 Y^3}\left(Y X' Y'+X Y (Y''+Y)-2 X{Y'}^2\right)\right)S^2=0\,.\label{SXN}
	\end{gather}
	Similarly, the second equation follows from substituting \eqref{1} into \eqref{TX}, namely,
	\be
	R^{(\zeta)}\left(\frac{X}{Q^2}\right)^\frac{1}{d}=\frac{d-1}{d}\left(S'-\frac{V'}{S}-\frac{XY'}{2SY^2}\right)^2-\frac12 S^2+V-\frac{X}{2Y}\,.
	\label{5}
	\ee
	
	By inserting the leading terms of the asymptotic expansions \eqref{VY1} and \eqref{SXeN} into the differential equations \eqref{SXN} and \eqref{5}, we obtain the following algebraic relations:
	\be \label{alg1}
	c_1 + c_2 e^{y \f} + c_3 e^{2y \f}+ c_4 e^{z \f} + c_5 e^{2z \f}+ c_6 e^{(y+z)\f}+\cdots=0\,,
	\ee
	and
	\be \label{alg3}
	b_1 + b_2 e^{y \f} + b_3 e^{2y \f}+ b_4 e^{z \f} + b_5 e^{2z \f}+ b_6 e^{(y+z)\f}+
	R^{(\zeta)}\left(\frac{\x}{Q^2}\right)^{\frac{1}{d}} e^{w\f}+\cdots=0\,,
	\ee
	where we have defined
	\be \label{alg2}
	y=2s-2b \sp z=k-2b-\g \sp w=2s-4b+\frac{k}{d}\,.
	\ee
	
	The coefficients appearing in \eqref{alg1} and \eqref{alg3} are given by
	\begin{gather}
		c_1= 16 b^2 (d-1) V_\infty^2 Y_\infty^2\,,\nn\\
		c_2= -4 \left(d + 4 b^2 (d-1) d - 2 b (d^2+d-2) s\right) S_\infty^2 V_\infty Y_\infty^2\,,\nn\\
		c_3= -2 (d-1) \left(2 (d-1) s^2-d\right) S_\infty^4 Y_\infty^2\,,\nn\\
		c_4=-8 b \gamma  (d-1) V_\infty X_\infty Y_\infty\,,\nn\\
		c_5=(d-1)  \gamma^2 X_\infty^2\,,\nn\\
		c_6= 2 (d-1)  \left(d k \gamma -d \gamma (\gamma +s)-2 \gamma  s +d \right) S_\infty^2 X_\infty Y_\infty\,,\label{c16}
	\end{gather}
	and
	\begin{gather}
		b_1= -\frac{4 b^2 (d-1) \v^2}{d \si^2}\sp
		b_2= \v\Big(1-\frac{4 b (d-1) s}{d}\Big)\,, \nn \\
		b_3=\frac{\si^2 \left(d-2 (d-1) s^2\right)}{2 d}\sp
		b_4=\frac{2 b \gamma  (d-1) \v \x}{d \si^2 \y}\,, \nn \\
		b_5=-\frac{\gamma ^2 (d-1) \x^2}{4 d \si^2 \y^2}\sp
		b_6=\frac{\x (2 \gamma  (d-1) s+d)}{2 d \y}\,.\label{b}
	\end{gather}
	
	\begin{figure}[!b]
		\centering
		\includegraphics[width=0.5\linewidth]{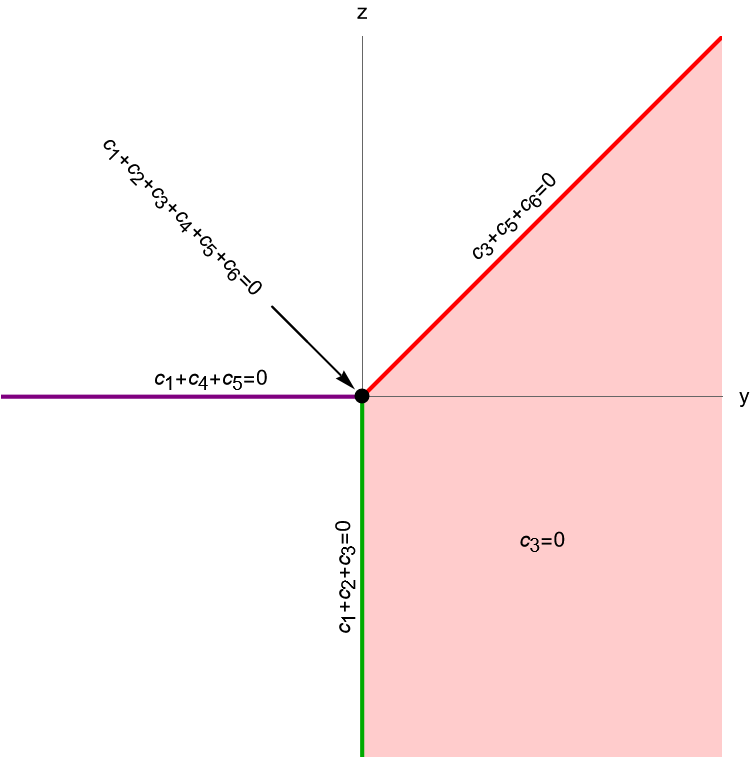}
		\caption{\footnotesize{The solution space for equation \eqref{alg1} is confined to the red region and the colored lines in the corresponding plot. Additionally, the point $y = z = 0$ is also a valid solution.
		}}
		\label{plot1}
	\end{figure}
	
	The set of allowed leading-order solutions of equation \eqref{alg1} is parameterized by the variables $y$ and $z$, defined in \eqref{alg2}. The coefficients $(c_1,\dots,c_6)$ in \eqref{c16} impose nontrivial constraints on this parameter space; in particular, $c_1$, $c_4$, and $c_5$ are non-vanishing. The resulting solution space is illustrated in the $(y,z)$ plane in figure \ref{plot1}, where the allowed regions correspond to the red area and the colored lines. In addition, the point $(y = z = 0)$ is also a valid solution. For each solution of \eqref{alg1}, one must further check whether it satisfies equation \eqref{alg3}.
	
	In the red region, where $y>0$, equation \eqref{alg2} implies that $s>b$. In this case, no regular solutions exist, and the corresponding singularities are classified as ``bad'' according to Gubser's criterion. The same argument applies to the red half-line defined by $z=y>0$, indicating the absence of regular solutions in this region. Therefore, we exclude these cases from further analysis and focus only on the remaining branches, namely the purple and red half-lines, as well as the special point $y=z=0$.
	
	A detailed and systematic analysis of both the leading and subleading terms is presented in the following sections for general values of $\g$.

	\vspace{0.5cm}
	
	\subsection{The half-line $y=0$ and $z<0$, general $\g$ analysis}\label{apa3}
	
	On the green half-line shown in figure \ref{plot1}, where $y=0$, the definition in \eqref{alg2} implies
	\be  \label{gy01}
	s=b\,,
	\ee
	which indicates that the solutions along this line correspond to a ``good'' singularity \cite{good}. Furthermore, since $z<0$, one obtains
	\be \label{gy02}
	k<\g+2b\,.
	\ee
	
	Solving the leading terms in \eqref{alg1}, we find that $\si$ admits four possible values
	\be \label{gy03}
	\si = \pm\sqrt{\frac{2\v}{d-1}} \sp
	\si = \pm 2b \sqrt{\frac{(d-1)\v}{d-2b^2(d-1)}}\,.
	\ee
	These solutions are identified as Type I and Type II. Taking into account the second equation, \eqref{alg3}, one finds that the first (Type I) solution in \eqref{gy03} exists only when the axion field is switched off, $a=0$. The corresponding first-order expansions are presented in section \ref{TY1}.
	
	On the other hand, the second solution in \eqref{gy03} is allowed provided that the last term in \eqref{alg3} is subleading. This follows from the condition $b_1+b_2+b_3=0$, which leads to the constraint
	\be \label{gy04}
	w = -2b + \frac{k}{d}<0\,.
	\ee
	
	Proceeding to the next-to-leading order, we obtain the following relations
	\begin{gather}
		b (d-1)   \left(b (d-1) (\gamma  (b+\gamma -k)-1)+\gamma\right) \x e^{(-2b+k-\g)\f}
		\nn \\
		+  \left(2 b (d-1){s_1}-d\right) \left(b (d-1) (b+{s_1})-1\right) \si  \y \ss e^{(-b + s_1 )\f}+\cdots=0\,,\label{gy05}
	\end{gather}
	\begin{gather}
		(d-1) b \x (b+\gamma ) e^{(-2b+k-\g)\f}	
		-\si (2 b (d-1) s_1 - d) \ss \y e^{(-b + s_1 )\f}
		\nn \\
		+ 2 b^2 (d-1) Q^{-\frac2d} \Rz \x^{\frac1d} \y e^{(-2b+\frac{k}{d})}
		+\cdots=0\,.\label{gy06}
	\end{gather}
	
	Since the first and third terms in \eqref{gy06} are non-vanishing, neither can independently dominate. If the second term were to dominate alone, one would obtain $s_1 = \frac{d}{2b(d-1)}$, which is incompatible with the constraints \eqref{bls}, \eqref{gy02}, \eqref{gy04}, and the ordering $b=s>s_1$. Therefore, in equation \eqref{gy05}, both terms must be of the same order, leading to
	\be \label{gy07}
	s_1 = -b +k -\g \,.
	\ee
	
	From \eqref{gy06}, two distinct cases arise:
	
	\begin{itemize}
		\item Case 1:
		In this case, the third term has the same order as the first two terms, implying
		\be \label{gy08}
		k = \frac{\g d}{d-1}\,,
		\ee
		
		\item Case 2: 	
		Here, the last term in \eqref{gy06} is subleading, so that
		\be \label{gy09}
		k>\frac{\gamma  d}{d-1}\,.
		\ee
	\end{itemize}
	
	In the following, we analyze these two cases separately.
	
	\subsubsection{Case 1}
	
	In this case, all terms in \eqref{gy06} are of the same order, and therefore
	\be  \label{gy010}
	k =  \frac{\gamma  d}{d-1} \sp s_1 = \frac{\gamma }{d-1}-b\,.
	\ee
	From equation \eqref{gy06}, we obtain
	\be  \label{gy011}
	\ss =
	-\sqrt{\frac{(d-1)(d-2 b^2 (d-1))}{\v}}\frac{Q^{-\frac2d} \left(\x (b+\gamma ) Q^{\frac2d}+2 b \Rz \y\x^{\frac1d}\right)}{2  \y \left(2 b^2 (d-1)-2 b \gamma +d\right)}\,.
	\ee
	Solving \eqref{gy05} yields two possibilities:
	
	\begin{itemize}
		\item $\g\neq \frac1b:$
		In this case, one finds
	\end{itemize}
	\be  \label{gy012}
	\x = \left(-\frac{2\Rz \y Q^{-\frac2d}}{d}\right)^{\frac{d}{d-1}}\,.
	\ee
	\begin{itemize}
		\item $\g = \frac1b:$
		In this case, $\x$ remains undetermined, and one must proceed to the next-to-next-to-leading order in the equations of motion
	\end{itemize}
	\begin{gather}
		\bigg(2  \left(2 b^2 (d-1)-d\right) \left(2 b^2 (d-1) d+d-2\right)\si \ss \x \y\nn \\
		+(d-1) \left(2 b^2 (d-1)-d\right) \x^2+4 b^2 \Big(d \big(12 b^4 (d-1)^2\nn \\
		-6 b^2 \left(d^2-1\right)-d+8\big)-4\Big) {\ss}^2 \v \y^2\bigg)
		e^{\frac{2 d}{b (d-1)}\f}\nn \\
		-8 b^2 (d-1) d  \left(b (d-1) {k_1}-d\right)\v \y \xx e^{(2 b+\frac{1}{b}+{k_1})\f}\nn \\
		+8 b^2 d  \left(2 b (d-1) {s_2}-d\right) \left(b (d-1) (b+{s_2})-1\right)\si  \v \y^2 S_\infty^{(2)} e^{(3 b+\frac{2}{b}+{s_2})\f}
		+\cdots=0\,,\label{mm1}
	\end{gather}
	\begin{gather}
		\bigg(
		24 b^4 (d-1) d {\ss}^2 \v \y^2+2 b^2 \Big((d-1)^2 \x^2+4 (d-1)^2 \si \ss \x \y\nn \\
		-2 (d (3 d-4)+4) {\ss}^2 \v \y^2\Big)-(d-1) d \x (4 \si \ss \y+\x)
		\bigg)e^{\frac{2 d}{b (d-1)}\f} \nn \\
		+8 b^2 \left(b^2+1\right) (d-1) d \v  \y \xx e^{(2 b+\frac{1}{b}+{k_1})\f}\nn \\
		+8 b^2 d  \left(d-2 b (d-1) {s_2}\right) \si\v \y^2 S_\infty^{(2)} e^{(3 b+\frac{2}{b}+{s_2})\f}+\cdots=0\,.\label{mm2}
	\end{gather}
	
	The only consistent possibility is that the leading contribution in \eqref{mm1} and \eqref{mm2} corresponds to
	\be \label{vs2}
	s_2=\frac{d}{2b(d-1)}>b\,.
	\ee
	This implies that in the expansion
	\be \label{spaS}
	S= \si e^{b\f} + \ss e^{(-b+\frac{1}{b(d-1)})\f} + S_\infty^{(2)} e^{\frac{d}{2b(d-1)}\f}+\cdots\,,
	\ee
	one must set $S_\infty^{(2)}=0$.
	
	The only solution allowing a free value of $Q$ arises when
	\be \label{k1s2f}
	\frac{2 d}{b (d-1)} = \frac1b + 2 b + k_1 = \frac2b + 3 b + s_2\,,
	\ee
	which gives
	\be \label{k1s2}
	{k_1}= \frac{-2 b^2 (d-1)+d+1}{b (d-1)}\sp
	{s_2}= \frac{2}{b (d-1)}-3 b\,,
	\ee
	and from \eqref{mm1} and \eqref{mm2} one obtains
	\begin{gather} \xx = \frac{\left(2 b^2 (d-1)-d\right)\left(2 b^2 (d-1)-1\right)^{-1} Q^{-\frac4d}}{8 b^4 d \left(2 b^2 (d-1)+d-2\right)^2 \v \y} \times \nn \\
	\bigg(4 b^2 d \left(2 b^2 (d-1)-1\right) \Big(6 b^4 (d-1)+b^2 (7 d-10)-d-2\Big) \Rz \y Q^{\frac2d} \x^{\frac{1}{d}+1}
	\nn \\
	+ 4 b^4 \left(2 b^2 (d-1)-1\right) \Big(d \left(6 b^2 (d-1) -3 d+4\right)-4\Big) {\Rz}^2 \y^2\x^{\frac2d}
	\nn \\
	+ \Big(12 b^8 (d-1)^2 d+2 b^6 (d-1) d (13 d-19)+b^4 \big(d \left(d (16 d-53)+36\right)
	\nn \\
	+4\big)+2 b^2 \left(d (5-6 d)+4\right)+d^2+4\Big) \x^2 Q^{\frac4d}\bigg)\,,\label{Xinf3}
	\end{gather}
	and
	\begin{gather}
		S_\infty^{(2)} =\frac{ -(d-1) \left(2 b^2 (d-1)-d\right)\left(6 b^2 (d-1)+d-4\right)^{-1} Q^{-\frac4d}}{8 b^4 d \si \v \y^2 \left(2 b^2 (d-1)-1\right) \left(2 b^2 (d-1)+d-2\right)^2 }\times
		\nn \\
		\bigg(4 \big(2 b^2 (d-1)
		-1\big) \Big(6 b^4 (d-1)+b^2 (7 d-10)-d-2\Big)b^2 d \Rz \y Q^{\frac2d} \x^{\frac{1}{d}+1}
		\nn \\
		+4 b^4 \big(2 b^2 (d-1)-1\big) \Big(d \big(6 b^2 (d-1)-3 d+4\big)-4\Big){\Rz}^2 \y^2 \x^{\frac2d}
		\nn \\
		+ \Big(12 b^8 (d-1)^2 d+b^6 \big(2 d (d (13 d-30)+15)+4\big)+b^4 \left(d (d (16 d-49)+24)+12\right)
		\nn \\
		+b^2 \left(d (6-11 d)+12\right)+d^2+4\Big)\x^2 Q^{\frac4d}\bigg)\,.\label{Sinf2}
	\end{gather}
	\subsubsection{Case 2}
	
	In this case, solving the leading terms of \eqref{gy05} and \eqref{gy06} yields
	\be \label{C2S1}
	\ss = -\frac{\sqrt{d-1} \x (b+\gamma ) \sqrt{d-2 b^2 (d-1)}}{2 \sqrt{\v} \y (2 b (d-1) (b+\gamma )-d)}\,,
	\ee
	and
	\be \label{C2S2}
	k = \frac{d}{b (d-1)}\,.
	\ee
	Combining this result with the constraint \eqref{gy09} and the other conditions derived above leads to
	\be \label{C2S3}
	\frac{d}{b (d-1)}-2 b<\gamma <\frac{1}{b}\,.
	\ee
	We can proceed one step further by considering the next-to-leading contributions in the equations of motion, which allows us to determine the subleading terms in the expansions.

	\vspace{0.5cm}
	
	\subsection{The point $y=z=0$, general $\g$ analysis}\label{apa4}
	
	We now turn to the case $y=z=0$, which from \eqref{alg2} implies
	\be \label{yz1}
	k=2b+\g \sp s=b\,.
	\ee
	Solving equations \eqref{alg1} and \eqref{alg3} leads to two distinct sets of solutions.
	
	\subsubsection{$\gamma <2 b (d-1)$}
	
	In the first case, the last term in \eqref{alg3} is subleading. One then obtains the following expressions for $\x$ and $\si$:
	\be \label{yza1}
	\si = \frac{\pm 2 d \v}{\sqrt{\gamma  (d-1) d \v (2 b+\gamma )}}
	\sp
	\x = \frac{4 \v \y }{\gamma }\left(b-\frac{d}{(d-1) (2 b+\gamma )}\right) \,.
	\ee
	
	The leading behavior of the fields $S$, $W$, $T$, and $X$ is given by
	\begin{gather}
		S = \si e^{b\f}+\cdots \sp W =\pm 2 \sqrt{\frac{(d-1)(2b+\g)\v}{d\g}} e^{b\f}+\cdots\,, \nn \\
		T = 0 + \cdots \sp X = \x e^{(2b+\g)\f}+\cdots\,. \label{yza2}
	\end{gather}
	
	To proceed, we consider the next-to-leading terms in the equations of motion, which take the form
	\begin{gather}
		-\frac{64 (d-1) d^3 \ss \v^3 \y^2  \left(2 d \left(b^2+{s_1}^2-1\right)+\gamma  (d+2) (b-{s_1})-4 b {s_1}+\gamma ^2\right)}{(\gamma  (d-1) d \v (2 b+\gamma ))^{\frac32}} e^{(s_1-b)\f}
		\nn \\
		-\frac{32 d \v \xx \y (\gamma  (2 b+\gamma )+d (\gamma  (b+\gamma -{k_1})-1)) }{\gamma  (2 b+\gamma )}e^{(-2 b-\gamma +{k_1})\f}+\cdots =0\,,\label{yza3}
	\end{gather}
	\begin{gather}
		\frac{64 \ss \y^2  \left(d (2-(2 b+\gamma ) (2 {s_1}-\gamma ))+(2 b+\gamma ) (2 {s_1}-\gamma )\right) (d \v )^{\frac32}}{(\gamma (d-1)(2 b+\gamma ))^\frac32}e^{({s_1}-b)\f}\nn \\
		+\frac{32 d \v \xx \y \left(2 b \gamma  (d-1)+\gamma ^2 (d-1)+d\right) }{\gamma  (d-1) (2 b+\gamma )}e^{(-2 b-\gamma +{k_1})}\nn \\
		+\frac{64 d^2 \Rz \v \y^2 Q^{-\frac2d} \x^{\frac1d} }{\gamma  (d-1) (2 b+\gamma )}e^{(\frac{2 b}{d}-2 b+\frac{\gamma }{d})\f}+\cdots = 0\,. \label{yza4}
	\end{gather}
	
	From \eqref{yza4}, we observe that the coefficients of the second and third terms are non-zero, and therefore neither term can dominate individually. If the first term were dominant, one could determine $\g$ and $s_1$ using \eqref{yza3}, but a straightforward analysis shows that such values are incompatible with the conditions $s=b>s_1$ and \eqref{bls}. Consequently, the two terms in \eqref{yza3} must be of the same order, which implies
	\be \label{yza5}
	k_1 = b + \g + s_1\,,
	\ee
	and
	\be \label{yza6}
	\xx = \frac{2 d \ss \v \y \left(d \left(2 b^2+b \gamma +2 {s_1}^2-\gamma  {s_1}-2\right)-(2 b+\gamma ) (2 {s_1}-\gamma )\right)}{(-\gamma  (2 b+\gamma )+\gamma  d {s_1}+d) \sqrt{\gamma  (d-1) d \v (2 b+\gamma )}}\,.
	\ee
	
	Using \eqref{yza4}, two possibilities arise depending on whether the last term is subleading or of the same order as the first two.
	
	In the first case, one finds
	\be \label{yza7}
	s_1 = b + \frac{\g}{d}-\frac14 \left(\frac{\g}{d}\right)^\frac12\left(-32 b^2(b+\g) (d-1)+8 b \left(\gamma ^2 (1-d)+2 d\right)+9 \gamma  d\right)^\frac12,
	\ee
	and it can be shown that the leading contribution to $T$ remains zero, requiring an analysis at next-to-next-to-leading order. This solution is excluded when $\g=\frac{1}{b}$ due to the conditions $s=b>s_1$ and \eqref{bls}, and we therefore do not pursue it further.
	
	In the second case, one obtains
	\be \label{yza8}
	s_1 = \frac{2b+\g}{d}-b\,,
	\ee
	and
	\begin{gather}
		\ss =\frac{d}{\chi} \Rz (b \gamma -1) Q^{-\frac2d} \v^{-\frac12}\x^{\frac1d} \sqrt{\gamma  (d-1) d  (2 b+\gamma )}\,,
		\nn \\
		\chi =  8 b^2 (d-1)^2+2 b \left(d \left(2 b^2 (d-1)-5\right)+4\right) \gamma \nn \\
		+ (d-1) \left(\left(4 b^2-1\right) d-2\right)\gamma ^2+b  (d-1) d\gamma ^3\,. \label{yza9}
	\end{gather}
	
	The leading-order solution is then given by
	\begin{gather}
		S =  \frac{\pm 2 d \v}{\sqrt{\gamma  (d-1) d \v (2 b+\gamma )}} e^{b\f}+\cdots\,,  \nn \\
		W =\pm 2 \sqrt{\frac{(d-1)(2b+\g)\v}{d\g}} e^{b\f}+\cdots \,, \nn \\
		T = \Rz Q^{-\frac2d} \left(\frac{4\v \y }{\gamma }\left(b-\frac{d}{(d-1) (2 b+\gamma )}\right)\right)^{\frac1d} e^{\frac{2b+\g}{d}} + \cdots\,, \nn \\
		X =  \frac{4 \v \y }{\gamma }\left(b-\frac{d}{(d-1) (2 b+\gamma )}\right) e^{(2b+\g)\f}+\cdots\,,
		\label{yza10}
	\end{gather}
	
	From \eqref{SWTX}, one also obtains
	\be   \label{yza11}
	e^{-2A} = e^{-2A_{IR}} e^{\,(\frac{2b+\g}{d})\f}+\cdots\,,
	\ee
	where $A_{IR}$ is an integration constant. Substituting this into the definition of $T$ in \eqref{SWTX} and comparing with \eqref{yza10}, we find
	\be  \label{yza12}
	Q^2 =\left({\Rz} e^{2A_{IR}}\right)^d \left(\frac{4\v \y }{\gamma }\left(b-\frac{d}{(d-1) (2 b+\gamma )}\right)\right)\,.
	\ee
	This fixes $Q$ in terms of IR data and shows that it is independent of $a_{UV}$, implying that $a_{IR}=0$ cannot be fixed.
	
	In the flat-slice limit, this solution reduces to the leading-order result found in \cite{Ha1}.
	
	\subsubsection{$\gamma =2 b (d-1)$}
	
	In the second case, the last term in \eqref{alg3} is of the same order as the remaining terms, which implies
	\be \label{yzb1}
	\gamma =2 b (d-1)\,,
	\ee
	and leads to two branches of solutions:
	\be \label{yzb2}
	\si = \pm \sqrt{\frac{2 \v}{d-1}-\frac{\x}{\y}} \sp
	Q^\frac2d = -\frac{(d-1) \Rz \y \x^{\frac1d}}{b^2 (d-1) d ((d-1) \x-2 \v \y)+d \v \y}\,,
	\ee
	and
	\be   \label{yzb3}
	\si = \pm b \sqrt{\frac{2(d-1) (2 \v \y-(d-1) \x)}{\y \left(d-2 b^2 (d-1)\right)}} \sp
	Q^\frac2d = -\frac{2}{d} \Rz \y \x^{\frac{1}{d}-1}\,.
	\ee
	Both branches are valid in the range
	\be  \label{yzb4}
	\Rz<0\sp \x<\frac{2}{d-1} \v \y\,.
	\ee
	For the first solution \eqref{yzb2}, the leading terms are
	\begin{gather}
		S= \si e^{b\f}+\cdots \sp W = \pm 2 b (d-1) \sqrt{\frac{(1-d) \x+2 \v \y}{(d-1) \y}} e^{b\f}+\cdots \,, \nn \\
		T = \left(\frac{b^2 d }{\y}((1-d) \x+2 \v \y)-\frac{d \v}{d-1}\right) e^{2b\f}+\cdots
		\sp
		X = \x e^{2bd\f}+\cdots\,.  \label{yzb5}
	\end{gather}
	From \eqref{SWTX}, one finds
	\be  \label{yzb6}
	e^{-2A} = e^{-2A_{IR}} e^{2b\f}+\cdots\,,
	\ee
	which leads, upon substitution into $T$ and comparison with \eqref{yzb5}, to
	\be  \label{yzb7}
	Q^2 = \frac{\y e^{2 (d-1) A_{IR}}}{b^2 (d-1)^2 d} \left(e^{2 A_{IR}} d \v \left(2 b^2 (d-1)-1\right)+(1-d) \Rz\right)\,.
	\ee
	Again, $Q$ is fixed by IR data and is independent of $a_{UV}$.
	
	For the second solution \eqref{yzb3}, one obtains
	\begin{gather}
		S =  \si e^{b\f}+\cdots \sp W = \pm \sqrt{\frac{2 (d-1) ((d-1) \x-2 \v \y)}{\y \left(2 b^2 (d-1)-d\right)}} e^{b\f}+\cdots\,, \nn \\
		T = -\frac{d \x}{2 \y} e^{2b\f} + \cdots \sp X = \x e^{2bd}\,.
		\label{yzb8}
	\end{gather}
	Repeating the same steps, one finds
	\be  \label{yzb9}
	e^{-2A} = e^{-2A_{IR}} e^{\frac{1}{b(d-1)}\f}+\cdots\,.
	\ee
	Substituting into $T$ and comparing with \eqref{yzb8}, one finds that the only consistent solution within the range \eqref{bls} is
	\be
	\x = 0\,,
	\ee
	which contradicts the presence of a non-trivial axion field.
	
	In conclusion, the solutions given in \eqref{yzb2} and \eqref{yzb3} do not provide acceptable IR endpoints.

	\vspace{0.5cm}
	
	\subsection{The half-line $z=0$ and $y<0$, general $\g$ analysis}\label{apa2}
	
	In this case, we obtain
	\be \label{z01}
	z=0 \rightarrow k=2b+\g\,.
	\ee
	Equation \eqref{alg1} then leads to
	\be \label{z02}
	c_1+c_4+c_5=0 \rightarrow \x = 4\frac{b}{\g}\v \y\,.
	\ee
	Using \eqref{z02}, one finds that $b_1+b_4+b_5=0$. Therefore, according to \eqref{alg3}, the condition \eqref{z02} is consistent provided that the term proportional to $e^{w\f}$ is subleading, namely
	\be \label{z0w}
	w<0 \rightarrow 2s-4b+\frac{k}{d}<0\,.
	\ee
	
	We now proceed to analyze the next-to-leading contributions. These must satisfy the conditions
	\be \label{z03}
	k>k_1 \quad \rightarrow \quad  2b+\frac1b>k_1 \sp y<0 \quad \rightarrow \quad b>s\,.
	\ee
	
	Taking these constraints into account, equation \eqref{SXN} yields
	\be  \label{z04}
	\frac{16}{\g} d \si^2 \v \y^2 (2b(d-1)-\g) e^{2(s-b)\f}+\cdots = 0\,,
	\ee
	where the dots denote subleading terms. This relation fixes the value of $\g$ to
	\be  \label{z05}
	\g = 2b(d-1)\,.
	\ee
	
	We must also examine the next-to-leading terms arising from \eqref{5}. Using the above results, we find
	\be  \label{z06}
	16 d \si^2 \y^2 \left(2^{\frac1d} \Rz Q^{-\frac2d} \left(\frac{\v \y}{d-1}\right)^{\frac1d}+\frac{d \v}{d-1}\right) e^{2(s-b)\f}+\cdots = 0\,.
	\ee
	
	The leading contribution in \eqref{z06} fixes the value of $Q$ in terms of the infrared parameters of the theory, and therefore this solution is not acceptable.
	
	\subsection{Summary of IR asymptotic analysis}
	
	In this appendix, the infrared asymptotic solutions of the system defined by equations \eqref{eom2}--\eqref{eom4} are derived in the limit $\f \to +\infty$. The analysis starts from the exponential behavior of the potentials given in \eqref{VY1}, together with the ansatz for $S$ and $X$ in \eqref{SXeN}. Substituting these into the equations of motion leads to two algebraic constraints, \eqref{alg1} and \eqref{alg3}, controlled by the parameters $(y,z,w)$ defined in \eqref{alg2}.
	
	The allowed solutions are organized in the $(y,z)$ plane (figure \ref{plot1}). Due to the non-vanishing coefficients in \eqref{c16}, not all regions are admissible. In particular, the region with $y>0$ (i.e., $s>b$) leads to bad singularities and is excluded. The analysis, therefore, focuses on three remaining branches: the half-line $y=0$, $z<0$, the point $y=z=0$, and the half-line $z=0$, $y<0$.
	
	\paragraph*{Half-line $y=0$, $z<0$:}
	From \eqref{alg2}, this implies $s=b$, corresponding to good singularities, and $k<\gamma+2b$. Solving \eqref{alg1} yields two classes of solutions for $\si$ (equation \eqref{gy03}), referred to as Type I and Type II. The Type I solution exists only when the axion is turned off ($a=0$), while the Type II solution is allowed when the curvature term in \eqref{alg3} is subleading, leading to the constraint \eqref{gy04}.
	
	At subleading order, consistency of \eqref{gy05}--\eqref{gy06} fixes $s_1$ as in \eqref{gy07}. Two cases arise depending on whether the curvature term contributes at leading order. In Case 1, all terms balance and one finds $k=\frac{\gamma d}{d-1}$ (equation \eqref{gy08}); generically $\x$ is fixed (equation \eqref{gy012}), while the special case $\gamma=\frac{1}{b}$ requires going to higher orders, leading to additional constraints on the expansion coefficients. This special case is called Type II. In Case 2, the curvature term is subleading, and one obtains $k>\frac{\gamma d}{d-1}$ (equation \eqref{gy09}), together with the bound \eqref{C2S3}.
	
	\paragraph*{Point $y=z=0$:}
	This corresponds to $k=2b+\gamma$ and $s=b$ (equation \eqref{yz1}). Two regimes are found. For $\gamma<2b(d-1)$, the leading solution is given by \eqref{yza1}--\eqref{yza2}. Subleading analysis using \eqref{yza3}--\eqref{yza4} imposes relations such as \eqref{yza5} and \eqref{yza6}, leading to two subcases. One branch is largely excluded by consistency conditions, while the other yields a valid asymptotic solution (equation \eqref{yza10}). However, matching with the definition of $T$ fixes the charge $Q$ in terms of IR data (equation \eqref{yza12}), implying that $Q$ is not a free parameter.
	
	For $\gamma=2b(d-1)$, two branches arise (equations \eqref{yzb2} and \eqref{yzb3}), both valid in the range \eqref{yzb4}. In both cases, either $Q$ is fixed by IR data or consistency requires $\x=0$, eliminating the axion field. Therefore, no acceptable IR endpoints are obtained from this case.
	
	\paragraph*{Half-line $z=0$, $y<0$:}
	In this case, $k=2b+\gamma$ (equation \eqref{z01}), and \eqref{alg1} fixes $\x$ as in \eqref{z02}. Consistency of \eqref{alg3} requires the curvature term to be subleading, leading to the condition \eqref{z0w}. Subleading analysis imposes further constraints \eqref{z03} and fixes $\gamma=2b(d-1)$ (equation \eqref{z05}). However, the next equation \eqref{z06} determines $Q$ entirely in terms of IR data, rendering this branch physically unacceptable.
	
	\paragraph*{Conclusion:}
	The IR solution space is strongly constrained by the algebraic relations \eqref{alg1}--\eqref{alg3}. Most branches either correspond to bad singularities, fix the charge $Q$, or eliminate the axion field. The only physically viable and nontrivial IR solutions arise from the half-line $y=0$, $z<0$, where the singularities are good and nontrivial solutions with potentially free parameters exist.

	\subsection*{Summary of results for $\g=\frac1b$}\label{sumA}

	As discussed in section \ref{RCDC}, our focus is on theories arising from higher-dimensional reductions, and in particular on confining backgrounds. This motivates restricting to the class of solutions satisfying
	\be \label{gob}
	\g = \frac1b \,.
	\ee
	
	Using the general analysis developed in the previous sections, together with the structure of the solution space shown in figure \ref{plot1}, we arrive at the following conclusions:
	
	\begin{itemize}
		
		\item Along the green half-line defined by $y=0$ and $z<0$, there exists a single regular solution for which the charge $Q$ remains independent of the IR asymptotic data. This corresponds to the so-called Type II solution. In the flat limit, this solution matches the subleading behavior found in \cite{Ha1}. Additional solutions also exist in this region; however, in those cases, the value of $Q$ is fixed by the IR data.  There is also a Type I solution in which the axion field vanishes.
		
		\item At the special point $y=z=0$, there is one regular solution, but in this case the charge $Q$ is completely determined by the IR asymptotics. In the flat limit, this solution reproduces the leading behavior obtained in \cite{Ha1}.
		
		\item Along the purple half-line defined by $z=0$ and $y<0$, no solutions exist with a free value of $Q$.
		
	\end{itemize}

	\section{IR expansions in specific limits}\label{spl}

	In this appendix, we shall derive the zero axion charge as well as the flat limit of the IR expansions in the previous appendix \ref{IRdetails}. Here we consider the special case of $\g=1/b$.
	
	\subsection{Zero axion limit}\label{zl}
	
	We define
	\be \label{SiRO}
	T_{\infty}\equiv \Rz ~Q^{-\frac2d} \x^{\frac1d}\,,
	\ee
	and rescale $X\rightarrow \y X$ i.e.
	\be
	X =\y \left( \x e^{\frac{d}{b(d-1)}\f}+
	\xx e^{\frac{-2 b^2 (d-1)+d+1}{b (d-1)}\f}+\cdots \right)\,,\label{XRO}
	\ee
	The IR asymptotic expansions in \eqref{SR}--\eqref{TR} can be expressed as follows
	\be
	S= \si e^{b\f} + \ss e^{(\frac{1}{b(d-1)}-b)\f}+\cdots\,,\label{SRO}
	\ee
	\be
	W=\frac{\si}{b} e^{b\f} 	-\left(\frac{d-2}{b d}\ss	+\frac{(d-1) \x}{b d \si}\right)e^{(\frac{1}{b (d-1)}-b)\f}		+\cdots\,,\label{WRO}
	\ee
	\be
	T=T_\infty e^{\frac{1}{b (d-1)}\f}+
	\frac{1}{d} \frac{T_\infty\xx}{\x} e^{-2 \left(b-\frac{1}{b(d-1)}\right)\f}+\cdots\,.
	\label{TRO}
	\ee
	Here we choose
	\be \label{SinfO}
	\si =  - 2b \sqrt{\frac{(d-1)\v}{d-2b^2(d-1)}}\,,
	\ee
	in \eqref{SiR} and from \eqref{Si1R} one reads
	\be \label{Si1ROz}
	T_\infty =  \frac{2\left(2 b^2 (d-1)+d-2\right) }{\left(2 b^2 (d-1)-d\right) }\frac{\v\ss}{\si}-\frac{b^2+1}{2b^2}{\x}\,.
	\ee
	Moreover,
	\begin{gather}
		\xx = \frac{\x \left(2 b^2 (d-1)-d\right)  \left(4 b^2 (d-1)   T_\infty+d \x \right)}{4 b^2 \v  \left(2 b^2 (d-1)-1\right) \left(2 b^2 (d-1)+d-2\right)}\,.
		\label{Xi1RO}
	\end{gather}
	By sending $\x\rightarrow 0$ equations \eqref{SRO}--\eqref{TRO} become
	\be
	S= \si e^{b\f} + \ss e^{(\frac{1}{b(d-1)}-b)\f}+\cdots\,,\label{SROz}
	\ee
	\be
	W=\frac{\si}{b} e^{b\f} 	-\frac{d-2}{b d}\ss e^{(\frac{1}{b (d-1)}-b)\f}		+\cdots\,,\label{WROz}
	\ee
	\be
	T=T_\infty e^{\frac{1}{b (d-1)}\f}+
	\frac{ (d-1) \left(2 b^2 (d-1)-d\right)    T^2_\infty }{ d \left(2 b^2 (d-1)-1\right) \left(2 b^2 (d-1)+d-2\right) \v} e^{-2 \left(b-\frac{1}{b(d-1)}\right)\f}+\cdots\,,
	\label{TROz}
	\ee
	where
	\be \label{Si1RO}
	T_\infty =  \frac{2\left(2 b^2 (d-1)+d-2\right) }{\left(2 b^2 (d-1)-d\right) }\frac{\v\ss}{\si}\,.
	\ee
	As observed, in this limit, we retrieve the Type II regular solution described in \cite{Ghodsi:2024jxe}.

	\subsection{Flat slice limit}\label{FL}

	In the flat limit, as $\Rz\rightarrow 0$, we recover the sub-leading solution previously identified in \cite{Ha1}.
	\begin{gather}
		S= \si e^{b\f} + \frac{\si \left(2 b^2 (d-1)-d\right) \left(b^2+1\right) \x}{4 b^2 \left(2 b^2 (d-1)+d-2\right)\v \y } e^{(\frac{1}{b(d-1)}-b)\f}+\cdots\,,\nn \\
		W=\frac{\si}{b} e^{b\f}
		-\frac{b (d-1) \x}{\left(2 b^2 (d-1)+d-2\right)\si \y }e^{(\frac{1}{b (d-1)}-b)\f}		+\cdots\,, \nn \\
		T=0\,, \nn \\
		X = \x e^{\frac{d}{b(d-1)}\f}+
		\frac{d\x^2 \left(2 b^2 (d-1)-d\right)\left(2 b^2 (d-1)-1\right)^{-1}}{4 b^2 \v \y  \left(2 b^2 (d-1)+d-2\right)} e^{\frac{-2 b^2 (d-1)+d+1}{b (d-1)}\f}+\cdots\,.\label{FLM1}
	\end{gather}

	\section{Expansions near the A-bounces}\label{EB}

	In this appendix, we analyze the expansions of the fields near a specific point known as the A-bounce, where the scale factor turns around. This bounce is characterized by the condition $\dot{A} = 0$, indicating that the scale factor reaches a minimum at this point.
	
	Next, we solve the equations of motion in the vicinity of this point. This is a regular point of the equations.
	We utilize the equations for $S(\f)$ and $X(\f)$ given in \eqref{SXN} and \eqref{5}, and subsequently derive the expansions for both fields. These expansions are formulated in terms of the parameter $x = \f - \f_0$, where $\f_0$ denotes the bounce point. Furthermore, the bulk functions are expanded as follows:
	
	\be \label{EXVY}
	V = V_0 + V_1 x + V_2 x^2 + \cdots \sp Y = Y_0 + Y_1 x + Y_2 x^2 + \cdots\,.
	\ee
	Keeping in mind the symmetry present in the equations of motion under the transformations $S\rightarrow -S$ and $W \rightarrow -W$ we find
	\be \label{AbS}
	S = S_0 + \frac{1}{2 {S_0}}\left(2 {V_1}+\frac{{X_0} {Y_1}}{{Y_0}^2}\right) x + \mathcal{O}(x^2)\,,
	\ee
	\be \label{AbW}
	W = \frac{d-1}{d {S_0} {Y_0}} \left( S_0^2 {Y_0}+ {X_0}+2 \frac{V_0 Y_0}{d-1}\right) x  + \mathcal{O}(x^2)\,,
	\ee
	\be \label{AbT}
	T = -\frac{S_0^2}{2}+{V_0}-\frac{{X_0}}{2 {Y_0}}
	+ \mathcal{O}(x)\,,
	\ee
	Here, $S_0$ is treated as an independent parameter, while $X_0$ is determined in terms of it through the relation
	\be \label{AbXS}
	2 \left({V_0}-\Rz \left(\frac{{X_0}}{Q^2}\right)^{1/d}\right)-\frac{{X_0}}{{Y_0}} = S_0^2 \,.
	\ee
	
	The reality condition imposed by the equation above places constraints on the parameters, requiring them to satisfy the following condition
	\be \label{Abc1}
	2 \left({V_0}-\Rz \left(\frac{{X_0}}{Q^2}\right)^{1/d}\right)-\frac{{X_0}}{{Y_0}}>0\,.
	\ee
	Each A-bounce is specified by three parameters: $Q$, $\f_0$, and $S_0$.
	It is also important to note that the sign of the slice curvature is determined by the following expression
	\be
	T_0\equiv -\frac{S_0^2}{2}+{V_0}-\frac{{X_0}}{2 {Y_0}} = \Rz \left(\frac{X_0}{Q^2}\right)^\frac1d\,.
	\ee

	\section{Free energy and entropy of one boundary solutions} \label{FE}
	
	In this appendix, we summarize the evaluation of the free energy and thermal entropy for one-boundary configurations characterized by positively curved slices. These space-time geometries exhibit a UV boundary, whereas their IR endpoints are governed by the specific classifications outlined in section \ref{IRAS}. Because the spatial slices possess positive curvature, taking the form of either $dS_d$ or $S^d$, these solutions inherently lack a side boundary. In what follows, we explicitly perform holographic renormalization to derive the finite free energy and analyze the associated conformal trace anomalies. Subsequently, we compute the thermal entropy and ADM mass, demonstrating that these quantities perfectly satisfy the first law of thermodynamics and can be formulated entirely in terms of the vacuum expectation values (vevs) of the dual boundary stress tensor.
	
	\subsection{The free energy}\label{FES}
	
	In the context of holography, the free energy of a classical saddle point corresponds to minus the Euclidean on-shell action. For our specific model, this is given by
	\begin{gather}
		F = - S_E = - M_p^{d-1}\Bigg(\!\int du d^d x \sqrt{g}\left[R-\frac{1}{2}g^{AB}\partial_A\varphi\partial_B\varphi-\frac{1}{2}Y(\varphi)g^{AB}\partial_A a\partial_B a \!-\! V(\varphi)\right]
		\nn \\
		+2\int_{UV-boundary}\!\!\!\!\!\!\!\!\! d^d x \sqrt{\g} K \Bigg)\,,\label{FE1}
	\end{gather}
	where the second line represents the Gibbons-Hawking-York boundary term evaluated at the UV boundary, with $\g$ denoting the induced metric and $K$ being the trace of its extrinsic curvature.
	
	Assuming the metric ansatz
	\be \label{smet}
	ds^2 = du^2 +e^{2A(u)} \a^2 ds_{S^d}^2\,,
	\ee
	where $\a$ represents the $S^d$ radius, the free energy expression in \eqref{FE1} reduces to (refer to \cite{Ghodsi:2024jxe} for full details)
	\be \label{FE2}
	F = -2 \frac{M_p^{d-1}}{d} V_{S^d} \a^d \left(\Rz \int^{u_0}_{u_-} du\,  e^{(d-2)A}+d(d-1) e^{dA}\dot{A}\Big|^{u_0}_{u_-}\right)\,.
	\ee
	In this formula, $u_0$ marks the location of the IR endpoint, and $u_-$ acts as the UV regulator for the boundary as $u \rightarrow -\infty$. The constant $V_{S^d}$ is the volume of a $d$-dimensional sphere of unit radius.
	
	By defining an auxiliary scalar function $U(u)$ such that
	\be \label{defU}
	(d-2) \dot{A} U + \dot{U} = -1\,,
	\ee
	we can recast equation \eqref{FE2} into the form
	\be \label{FE3}
	F = -2 \frac{M_p^{d-1}}{d} V_{S^d} \a^d \left(-\Rz  U e^{(d-2)A}+d(d-1) e^{dA}\dot{A}\right)\Big|^{u_0}_{u_-}\,.
	\ee
	By imposing a suitable boundary condition, it can be shown that the IR contribution to this expression is effectively driven to zero.
	
	For $d=4$, the asymptotic behavior of $U(u)$ near the UV boundary (in the limit $u \rightarrow - \infty$) can be deduced from \eqref{defU}, yielding
	\be \label{FEU}
	U(u) =  \frac{\ell}{2} + \mathcal{B}\ell^3 |\varphi_-|^{2/\Delta} e^{ 2u/\ell} + \frac{\ell^3\mathcal{R} |\varphi_-|^{2/\Delta} }{24}\frac{u}{\ell} e^{ 2u/\ell} + \cdots\,,
	\ee
	where $\mathcal{B}$ serves as a constant of integration.
	Inserting the UV asymptotic expansions for both $U$ and $A$, given by \eqref{FEU} and \eqref{Ae}, respectively, into \eqref{FE3} yields the bare free energy, which requires appropriate renormalization. In four dimensions, this is accomplished by adding the following local counter-terms at the UV cutoff surface $u_- = \ell \log \epsilon$
	\be \label{Sct}
	S_{ct} = -M_p^3 \int d^4x \sqrt{\gamma} \left( \frac{6}{\ell} + \frac{\Delta}{2\ell} \varphi^2 + \frac{\ell}{2} R^{(\gamma)} + \frac{\ell^3}{48} \left( R^{(\gamma)} \right)^2 \log \omega \epsilon \right)_{u_- = \ell \log \epsilon}\,.
	\ee
	The parameter $\omega$ determines the renormalization scheme. For simplicity, we set $\omega = 1$ in the subsequent analysis.
	
	With these counter-terms in place, we extract the following expression for the renormalized free-energy density
	\be \label{F}
	\mathcal{F}(\mathcal{R},a_{\textrm{UV}}) \equiv \frac{F_{{ren}}}{M_p^3 \ell^3  V_{S^4} } = \frac{3}{2} \left( 1- 96 \frac{C(\mathcal{R},a_{\textrm{UV}})}{\mathcal{R}^2} - 48 \frac{\mathcal{B}(\mathcal{R},a_{\textrm{UV}})}{\mathcal{R}} \right)\,.
	\ee
	Here, both the scalar vacuum expectation value $C$, and the curvature vev $\mathcal{B}$ are functions of the UV source parameters $\mathcal{R}$ and $a_{\textrm{UV}}$.
	
	\subsection{Anomalies}
	
	We should note that the preceding counter-term calculation was performed using domain-wall coordinates. Transforming this into the Fefferman-Graham coordinate system gives
	\be
	\frac{d\rho^2}{4\rho^2} = \frac{du^2}{\ell^2} \rightarrow \rho = e^{\frac{2u}{\ell}}\,.
	\ee
	If we define a corresponding cutoff $\tilde{\epsilon}$ in the Fefferman-Graham frame, it relates to our original cutoff via
	\be
	\tilde{\epsilon} = \epsilon^2\,,
	\ee
	which rewrites the counter-term action as
	\be \label{SctFG}
	S_{ct} = -M_p^3 \int d^4x \sqrt{\gamma} \left( \frac{6}{\ell} + \frac{\Delta}{2\ell} \varphi^2 + \frac{\ell}{2} R^{(\gamma)} + \frac{\ell^3}{96} \left( R^{(\gamma)} \right)^2 \log \omega \tilde{\epsilon} \right)_{\rho = \tilde{\epsilon}}\,.
	\ee
	This expression perfectly aligns with the standard counter-terms documented in the holographic literature, inherently ensuring that the renormalized free energy in \eqref{F} remains unaltered.
	
	Following from this, the boundary trace of the energy-momentum tensor is governed by twice the coefficient of the logarithmic divergence
	\be \label{trem}
	\langle  T^\mu_\mu  \rangle = -\frac{1}{48}(M_p \ell)^3 {R^{(\gamma)}}^2\,.
	\ee
	For a generic four-dimensional conformal field theory (CFT), the trace anomaly is typically dictated by the central charges $c$ and $a$, which couple to the square of the Weyl tensor $W^2$ and the Euler density $E_4$, respectively
	\begin{equation} \label{anmCFT}
		\langle T^\mu_\mu \rangle = \mathcal{A}_c + \mathcal{A}_a = \frac{c}{16} W^2 - \frac{a}{16} E_4 \,.
	\end{equation}
	Since the background geometry is an $S^4$, the Weyl tensor $W$ vanishes identically, meaning $W^2 = 0$.\footnote{Also, the scheme-dependent part vanishes for maximally symmetric spaces.} Conversely, the Euler density $E_4$ in four dimensions and on an $S^4$ background evaluates to
	\be \label{E4}
	E_4 = R_{\mu\nu\rho\sigma}R^{\mu\nu\rho\sigma} - 4 R_{\mu\nu}R^{\mu\nu} + R^2 =\frac{1}{6} R^2 \,.
	\ee
	Substituting  these geometric invariants back into the generic CFT anomaly formula \eqref{anmCFT}, we find
	\begin{equation} \label{tanom}
		\langle T^\mu_\mu \rangle  = - \frac{a}{96\pi^2} R^2 \, .
	\end{equation}
	Comparing \eqref{tanom} with our holographic result \eqref{trem} allows us to extract the central charge $a$
	\be \label{av}
	a = 2 M_p^3 \ell^3\,.
	\ee
	Because the $S^4$ geometry dictates $W^2=0$, the central charge $c$ does not appear in the trace anomaly and cannot be fixed purely algebraically from this geometric setup. Nevertheless, since the bulk dual is described by a standard two-derivative Einstein gravity theory, well-established holographic dictionaries guarantee that $c = a = 2  M_p^3 \ell^3$.

	\subsection{Thermal entropy}\label{TES}

	For geometries featuring de Sitter slices, the metric takes the form
	\be \label{te0}
	ds^2=du^2+e^{2A(u)} \left[-d\t^2+ \a^2 \cosh^2\!\left(\frac{\t}{\a}\right)\left(d\theta^2+\sin^2\theta ds_{S^{d-2}}^2\right)\right]\,.
	\ee
	By applying a coordinate transformation, the geometry of the de Sitter slice can be rewritten in static patch coordinates \cite{Ghosh:2018qtg}. In this frame, the boundary cosmological horizon extends into the bulk space-time, yielding the metric
	\be \label{te1}
	ds^2=du^2+e^{2A(u)} \left[-\left(1-\frac{r^2}{\a^2}\right)dt^2+
	\left(1-\frac{r^2}{\a^2}\right)^{-1}dr^2+r^2 ds_{S^{d-2}}^2\right]\,.
	\ee
	The horizon of this geometry is located at $r = \alpha$, and its associated Hawking temperature is given by \cite{Ghosh:2018qtg}
	\be \label{te2}
	T\equiv\frac{1}{\b}=\frac{1}{2\pi\a}\,.
	\ee
	For static metrics like \eqref{te1}, the ADM mass is computed as
	\be
	M_{ADM}
	= 2 M_p^{d-1} \int_{UV} d^{d-1}x \, \sqrt{h}\, N K_{ADM}\,,
	\label{mass1}
	\ee
	where $h$ represents the induced metric at the UV boundary, defined as
	\begin{equation}
		h_{ab} dx^a dx^b
		= e^{2A(u)} \Big|_{UV}
		\left[
		\left(1 - \frac{r^2}{\alpha^2} \right)^{-1} dr^2
		+ r^2 d\Omega_{d-2}^2
		\right].
		\label{mass2}
	\end{equation}
	Additionally, the ADM extrinsic curvature $K_{ADM}$ and the lapse function $N$ evaluate to
	\begin{gather}
		K_{ADM} = -(d-1)\frac{dA(u)}{du}\Big|_{UV}\sp
		N = e^{A(u)} \left(1 - \frac{r^2}{\alpha^2} \right)^{\frac12} \,.
		\label{mass3}
	\end{gather}
	Utilizing the geometric identity $V_{S^d} = \frac{2\pi}{d-1}V_{S^{d-2}}$ within the mass definition \eqref{mass1}, we find
	\begin{gather}
		M_{ADM}
		=- 2(d-1) M_p^{d-1} e^{dA(u)} \frac{dA(u)}{du} \Big|_{UV}
		V_{S^{d-2}} \int_0^{\alpha} dr \, r^{d-2} \nn \\
		= - 2(d-1) M_p^{d-1} e^{dA(u)} \frac{dA(u)}{du} \Big|_{UV}
		\, V_{S^d} \, \frac{\alpha^{d-1}}{2\pi} .
		\label{mass4}
	\end{gather}
	Incorporating the temperature relation from \eqref{te2}, we obtain
	\begin{equation}
		\beta M_{ADM}
		= -2(d-1) M_p^{d-1} \, V_{S^d} \a^d e^{dA(u)} \frac{dA(u)}{du} \Big|_{UV}
		\,.
		\label{mass5}
	\end{equation}
	Next, the thermal entropy\footnote{As demonstrated in \cite{Ghosh:2018qtg}, if the de Sitter slice metric is expressed in global coordinates, the entanglement entropy between two halves of the spatial sphere exactly matches the thermal entropy computed in static patch coordinates.} is deduced via the Bekenstein-Hawking area law
	\begin{gather}
		S_{\text{th}} = \frac{\text{Area}}{4 G_{d+1}}
		= 4\pi M_p^{d-1} V_{S^{d-2}}\a^{d-2}
		\int_{UV}^{IR} du\, e^{(d-2)A(u)} \nn \\
		= 2M_{p}^{d-1}\frac{R}{d} V_{S^d}\a^d \int_{UV}^{IR} du\, e^{(d-2)A(u)}\,, \label{te3}
	\end{gather}
	where
	\be \label{RS4}
	R=\frac{d(d-1)}{\a^2}\,,
	\ee
	denotes the scalar curvature of $S^{d}$.
	Because the ADM mass represents the total energy evaluated at asymptotic infinity (the UV boundary), it is naturally identified with the thermodynamic internal energy of the system
	\be  \label{te3a}
	U_{th}= M_{ADM}\,.
	\ee
	Combining the free energy relation \eqref{FE2} with equations \eqref{te3} and \eqref{te3a} simply reproduces the first law of thermodynamics
	\be  \label{te4}
	S_{th} = \b (U_{th}-F_{th})\,,
	\ee
	with the thermodynamic free energy identified as
	\be  \label{te5}
	\b F_{th}= F\,.
	\ee
	
	To rigorously link the internal energy and entropy to the vacuum expectation values of the dual field theory, we compute the holographically renormalized stress tensor
	\be  \label{te6}
	\langle T^{(ren)}_{\m\n}\rangle =
	-\frac{2}{\sqrt{\g}}\frac{\delta S^{(ren)}_{on-shell}}{\delta \g^{\m\n}}\,,
	\ee
	Here, $\g_{\m\n}$ denotes the metric on $S^d$. Focusing on the $d=4$ case and utilizing the relationship $-S^{(ren)}_{on-shell} = F_{ren}$, alongside the renormalized free energy derived in \eqref{F}, we find
	\be \label{te7}
	\langle T^{(ren)}_{\m\n}\rangle =  \frac{1}{48}(M_p\ell)^3  R^2 \mathcal{C} \g_{\m\n}\,,
	\ee
	where $R$ was introduced in \eqref{RS4}, and the quantity $\mathcal{C}$ is defined as
	\be \label{te8}
	\mathcal{C} =48  \frac{C}{\mathcal{R}^2}+12\frac{\mathcal{B}}{\mathcal{R}}
	-24\frac{1}{\mathcal{R}}\frac{\pa C}{\pa \mathcal{R}}-12  \frac{\pa \mathcal{B}}{\pa \mathcal{R}}\,.
	\ee
	The thermodynamic internal energy can then be extracted from the temporal component of the stress tensor
	\be \label{te9}
	\b U_{th} = \int d^4x \sqrt{\g}\langle {T^{\,0}}_{0} \rangle =
	3 (M_p\ell)^3 V_{S^4} \mathcal{C}\,,
	\ee
	where $V_{S^4}$ is the volume of a unit 4-sphere.
	Inserting this into the first law \eqref{te4}, the system's thermal entropy becomes
	\be \label{te10}
	S_{th} = -\frac32 (M_p\ell)^3 V_{S^4} \left(1 -24\frac{\mathcal{B}}{\mathcal{R}}-48\frac{1}{\mathcal{R}}\frac{\pa C}{\pa \mathcal{R}}-24 \frac{\pa \mathcal{B}}{\pa \mathcal{R}}\right)\,.
	\ee
	One can easily verify that this expression is equivalent to the following differential operation on the free energy.
	\be \label{te11}
	S_{th} = -\left(\frac12\mathcal{R}\frac{\partial}{\partial\mathcal{R}}+1\right) F\,.
	\ee
	In terms of thermodynamic densities (analogous to \eqref{F}), this relationship is written as
	\be \label{Sth}
	\mathcal{S}_{th}(\mathcal{R},a_{\textrm{UV}})
	= -\left(\frac12\mathcal{R}\frac{\partial}{\partial\mathcal{R}}+1\right)\mathcal{F}(\mathcal{R},a_{\textrm{UV}})\,.
	\ee
	Alternatively, evaluating the thermal entropy integral \eqref{te3} directly in terms of the auxiliary scalar functions yields
	\be \label{te12}
	S_{th} =  72 (M_p\ell)^3 V_{S^4} U(u) T(u)^{-\frac{d-2}{2}}\Big|^{UV}\,.
	\ee
	Substituting the UV asymptotic expansions for $A(u)$ and $U(u)$ from \eqref{Ae} and \eqref{FEU}, and carefully subtracting the appropriate counter-terms, we arrive at the renormalized entropy
	\be \label{te13}
	S_{th} = 72 (M_p\ell)^3 V_{S^4} \frac{\mathcal{B}}{\mathcal{R}}\,.
	\ee
	By equating the two distinct entropy formulations \eqref{te10} and \eqref{te12}, we derive the following differential constraint\footnote{For $d=4$, this specific result was previously obtained in \cite{Ghosh:2020qsx}. However, our definition of $\mathcal{B}$ differs by a factor of 2 compared to \cite{Ghosh:2020qsx}.}
	\be \label{Cp}
	\frac{\pa C}{\pa \mathcal{R}}=\frac{\mathcal{B}}{2}-\frac{\mathcal{R}}{2}\frac{\pa \mathcal{B}}{\pa \mathcal{R}}+\frac{\mathcal{R}}{48}\,,
	\ee
	Combining this constraint with equation \eqref{F} results in an expression for the derivative of the free energy density
	\be \label{dfdr}
	\frac{\partial\mathcal{F}(\mathcal{R},a_{\textrm{UV}})}{\partial\mathcal{R}} = 144\left(
	\frac{2C(\mathcal{R},a_{\textrm{UV}})}{\mathcal{R}^3}-\frac{1}{48\mathcal{R}}\right)\,.
	\ee
	
	Finally, the conformal anomaly can be verified by computing the trace of the renormalized boundary stress tensor \eqref{te7}
	\be
	\langle T^{\mu(ren)}_{\mu}\rangle =\g^{\m\n}\langle T^{(ren)}_{\m\n}\rangle =
	\frac{1}{12}(M_p\ell)^3  R^2 \mathcal{C} =
	(M_p\ell)^3 R^2 \left(\frac{4C(\mathcal{R},a_{UV})}{\mathcal{R}^2}-\frac{1}{24}\right)\,,
	\label{anom}
	\ee
	where the last equality utilizes the relations \eqref{te8} and \eqref{Cp}. The fact that this trace is non-zero confirms the presence of a conformal anomaly. Because $R$ acts as the scalar curvature of $S^4$, the overall dependence on $R^2$ perfectly matches the universal structure of the trace anomaly in four dimensions.

	The trace of energy-momentum in \eqref{anom} can be written as
	\be
	\langle {T^{\mu}}_{\mu}\rangle^{(ren)} = \beta \langle O_{\f} \rangle - \frac{a}{48} R^2\sp \beta = (d-\Delta)\, \f_-\,, \label{anomv}
	\ee
	where $\beta$ is the $\beta$-function of the relevant
	operator  $O_{\f}$  deforming the UV CFT and $a$ is the anomaly given in \eqref{av}.
	In \eqref{anomv}, $\Delta$ is the scaling dimension of the scalar operator dual to $\f$, $\f_-$ is the source, and $\langle O_{\f} \rangle$ is the vev defined in \eqref{vevf}.


	\addcontentsline{toc}{section}{References\label{refs}}
	


\begin{thebibliography}{100}
	
		\bibitem{Ghodsi:2024jxe}
		A.~Ghodsi, E.~Kiritsis and F.~Nitti,
		{\em ``On holographic confining QFTs on AdS,''}
		\hrj{10.1007/JHEP04(2025)154}{JHEP \textbf{04} (2025), 154};
		\hri{2409.02879}{[hep-th]}.
		
		
		\bibitem{Kastikainen:2025eys}
		J.~Kastikainen, E.~Kiritsis and F.~Nitti,
		``Holographic confining theories on space-times with constant positive curvature,''
		\hrj{10.1007/JHEP09(2025)139}{JHEP \textbf{09} (2025), 139};
		\hri{2502.04036}{hep-th]}.
		
		
		\bibitem{Gursoy:2007er}
		U.~Gursoy, E.~Kiritsis and F.~Nitti,
		{\em ``Exploring improved holographic theories for QCD: Part II,''}
		\hrj{10.1088/1126-6708/2008/02/019}{JHEP \textbf{02} (2008), 019};
		\hri{0707.1349}{[hep-th]}.
		
		\bibitem{Gursoy:2008za}
		U.~Gursoy, E.~Kiritsis, L.~Mazzanti and F.~Nitti,
		``Holography and Thermodynamics of 5D Dilaton-gravity,''
		\hrj{10.1088/1126-6708/2009/05/033}{JHEP \textbf{05} (2009), 033};
		\hri{0812.0792}{[hep-th]}.
		
		
		\bibitem{Witten:1998uka}
		E.~Witten,
		{\em ``$\theta$-dependence in the large N limit of four-dimensional gauge theories,''}
		\hrj{10.1103/PhysRevLett.81.2862}{Phys. Rev. Lett. \textbf{81} (1998), 2862-2865};
		\hre{hep-th}{9807109}
		
		
		\bibitem{relaxion}
		P.~W.~Graham, D.~E.~Kaplan and S.~Rajendran,
		{\em ``Cosmological Relaxation of the Electroweak Scale,''}
		\hrj{10.1103/PhysRevLett.115.221801}{Phys. Rev. Lett. \textbf{115} (2015) no.22, 221801};
		\hri{1504.07551}{ [hep-ph]};\\
		
		A.~Nelson and C.~Prescod-Weinstein,
		{\em ``Relaxion: A Landscape Without Anthropics,''}
		\hrj{10.1103/PhysRevD.96.113007}{Phys. Rev. D \textbf{96} (2017) no.11, 113007};
		\hri{1708.00010}{[hep-ph]}.
		
		\bibitem{Hamada:2020bbf}
		Y.~Hamada, E.~Kiritsis, F.~Nitti and L.~T.~Witkowski,
		``The Self-Tuning of the Cosmological Constant and the Holographic Relaxion,''
		\hrj{10.1002/prop.202000098}{Fortsch. Phys. \textbf{69}, no.2, 2000098 (2021)};
		\hri{2001.05510}{[hep-th]}.
		
		
		
		\bibitem{Gursoy:2012bt}
		U.~G{\"u}rsoy, I.~Iatrakis, E.~Kiritsis, F.~Nitti and A.~O'Bannon,
		{\em ``The Chern-Simons Diffusion Rate in Improved Holographic QCD,''}
		\hrj{10.1007/JHEP02(2013)119}{JHEP \textbf{02} (2013), 119};
		\hri{1212.3894}{hep-th]}.
		
		
		
		\bibitem{Ha1}
		Y.~Hamada, E.~Kiritsis, F.~Nitti and L.~T.~Witkowski,
		``Axion RG flows and the holographic dynamics of instanton densities,''
		\hrj{doi:10.1088/1751-8121/ab4712}{J. Phys. A \textbf{52} (2019) no.45, 454003};
		\hri{1905.03663}{ [hep-th]}.
		
		\bibitem{Ha2}
		Y.~Hamada, E.~Kiritsis and F.~Nitti,
		``Holographic Theories at Finite \ensuremath{\theta}-Angle, CP-Violation, Glueball Spectra and Strong-Coupling Instabilities,''
		\hrj{doi:10.1002/prop.202000111}{Fortsch. Phys. \textbf{69} (2021) no.2, 2000111};
		\hri{2007.13535}{ [hep-th]}.
		
		
		\bibitem{Ghodsi:2022umc}
		A.~Ghodsi, J.~K.~Ghosh, E.~Kiritsis, F.~Nitti and V.~Nourry,
		``Holographic QFTs on AdS$_{d}$, wormholes and holographic interfaces,''
		\hrj{10.1007/JHEP01(2023)121}{JHEP \textbf{01} (2023), 121}
		\hri{2209.12094}{[hep-th]}.
		
		
		
		\bibitem{Ghosh:2017big}
		J.~K.~Ghosh, E.~Kiritsis, F.~Nitti and L.~T.~Witkowski,
		{\em ``Holographic RG flows on curved manifolds and quantum phase transitions,''}
		\hrj{10.1007/JHEP05(2018)034}{JHEP \textbf{05} (2018), 034};
		\hri{1711.08462}{[hep-th]}.
		
		\bibitem{Jani}
		J.~Kastikainen, E.~Kiritsis and F.~Nitti,
		{\em ``Holographic confining theories on space-times with constant positive curvature,''}
		\hrj{10.1007/JHEP09(2025)139}{JHEP \textbf{09} (2025), 139};
		\hri{2502.04036}{ [hep-th]}.
		
		
		\bibitem{Raymond}
		E.~Kiritsis, F.~Nitti and J.~L.~Raymond,
		``Improved Holographic QCD on a Curved Background: An Application of Dynamical System Theory in Holography,''
		\hrj{10.1002/prop.70046}{Fortsch. Phys. \textbf{73} (2025) no.12, e70046};
		\hri{2505.10703}{[hep-th]}.
		
		
		\bibitem{good}
		S.~S.~Gubser,
		{\em ``Curvature singularities: The Good, the bad, and the naked,''}
		\hrj{10.4310/ATMP.2000.v4.n3.a6}{Adv. Theor. Math. Phys. \textbf{4} (2000), 679-745};
		\hre{hep-th}{0002160}.
		
		
		
		\bibitem{GK}
		B.~Gouteraux   and E.~Kiritsis,
		{\em ``Generalized Holographic Quantum Criticality at Finite Density,''}
		\hrj{10.1007/JHEP12(2011)036}{JHEP \textbf{12} (2011), 036};
		\hri{1107.2116 }{[hep-th]}.
		
		
		
		\bibitem{Marolf:2010tg}
		D.~Marolf, M.~Rangamani and M.~Van Raamsdonk,
		{\em ``Holographic models of de Sitter QFTs,''}
		\hrj{10.1088/0264-9381/28/10/105015}{Class. Quant. Grav. \textbf{28} (2011), 105015};
		\hri{1007.3996}{[hep-th]}.
		
		\bibitem{Blackman:2011in}
		J.~Blackman, M.~B.~McDermott and M.~Van Raamsdonk,
		{\em ``Acceleration-Induced Deconfinement Transitions in de Sitter Space-time,''}
		\hrj{10.1007/JHEP08(2011)064}{JHEP \textbf{08} (2011), 064};
		\hri{1105.0440}{[hep-th]}.
		
		\bibitem{vw}
		C.~Vafa and E.~Witten,
		{\em ``Parity Conservation in QCD,''}
		\hrj{10.1103/PhysRevLett.53.535}{Phys. Rev. Lett. \textbf{53} (1984), 535}.
		
		
		\bibitem{CW}
		C.~G.~Callan, Jr. and F.~Wilczek,
		{\em ``Infrared behavior at negative curvature,''}
		\hrj{10.1016/0550-3213(90)90451-I}{Nucl. Phys. B \textbf{340} (1990), 366-386}.
		
		\bibitem{vw2}
		C.~Vafa and E.~Witten,
		{\em ``Restrictions on Symmetry Breaking in Vector-Like Gauge Theories,''}
		\hrj{doi:10.1016/0550-3213(84)90230-X}{Nucl. Phys. B \textbf{234} (1984), 173-188}.


		\bibitem{Aharony}
		O.~Aharony, E.~Y.~Urbach and M.~Weiss,
		{\em ``Generalized Hawking-Page transitions,''}
		\hrj{10.1007/JHEP08(2019)018}{JHEP \textbf{08} (2019), 018};
		\hri{1904.07502 }{[hep-th]}.
		
		\bibitem{Edwan}
		E.~Kiritsis, F.~Nitti and E.~Pr{\'e}au,
		{\em ``Holographic QFTs on S$^{2}${\texttimes}S$^{2}$, spontaneous symmetry breaking and Efimov saddle points,''}
		\hrj{10.1007/JHEP08(2020)138}{JHEP \textbf{08} (2020), 138};
		\hri{2005.09054}{ [hep-th]}.
		
		
		\bibitem{Ghosh:2018qtg}
		J.~K.~Ghosh, E.~Kiritsis, F.~Nitti and L.~T.~Witkowski,
		``Holographic RG flows on curved manifolds and the $F$-theorem,''
		\hri{10.1007/JHEP02(2019)055}{JHEP \textbf{02} (2019), 055};
		\hri{1810.12318}{[hep-th]}.
		
		
		
		\bibitem{Ghosh:2020qsx}
		J.~K.~Ghosh, E.~Kiritsis, F.~Nitti and L.~T.~Witkowski,
		``Back-reaction in massless de Sitter QFTs: holography, gravitational DBI action and f(R) gravity,''
		\hrj{10.1088/1475-7516/2020/07/040}{JCAP \textbf{07} (2020), 040};
		\hri{2003.09435}{[hep-th]}.
		
		
	\end{thebibliography}
\end{document}